\documentclass{aa}
\bibpunct{(}{)}{;}{a}{}{,}
\usepackage{graphicx}
\usepackage{txfonts}
\usepackage{subcaption}         % necessary for continued figures, example in section 3
\usepackage{lscape}             % to rotate a single page table, example in appendix.
\usepackage{placeins}           % useful with \FloatBarrier, to keep 
\usepackage{makecell}
\usepackage{tikz}
\usetikzlibrary{arrows.meta, positioning}
\tikzset{
  box/.style={
    rectangle, rounded corners=3mm,
    draw=black, fill=blue!40, text=white,
    minimum width=2.5cm, minimum height=1cm,
    align=center
  },
  gcbox/.style={
    rectangle, rounded corners=3mm,
    draw=black, fill=blue!70!black, text=white,
    minimum width=2.5cm, minimum height=1cm,
    align=center
  },
  textonly/.style={
    align=left,
    text=black,
    minimum width=2cm, minimum height=1cm,
  },
  arrow/.style={
    -{Latex[length=3mm]}, thick
  },
  dashedarrow/.style={
    -{Latex[length=3mm]}, thick, dashed
  }
}

\usepackage{natbib}
\usepackage{hyperref}
\hypersetup{
    colorlinks=true,
    linkcolor=blue,
    citecolor=blue,
    urlcolor=blue,
    filecolor=blue,
    pdftitle={AuriGLOBES: Auriga GLOBular clustErs Simulations},
    pdfauthor={Pablo Contreras Guerra},
}

\usepackage{orcidlink}

\begin{document}
   \titlerunning{Introducing AuriGLOBES}
   \title{Introducing AuriGLOBES: The effects of compressive tides, compact object-induced mass loss, and size evolution on modelling globular clusters}

   \authorrunning{Contreras Guerra, Grand, Reina-Campos and Dalla Vecchia}
   \author{Pablo Contreras Guerra\inst{1}\fnmsep\inst{2}\orcidlink{0009-0001-6155-6745},
        Robert J. J. Grand\inst{3}\orcidlink{0000-0001-9667-1340},
        Marta Reina-Campos\inst{4}\fnmsep\inst{5}\fnmsep\inst{6}\orcidlink{0000-0002-8556-4280},
        \and
        Claudio Dalla Vecchia\inst{1}\fnmsep\inst{2}\orcidlink{0000-0002-2620-7056}
        }

   \institute{Instituto de Astrof\'isica de Canarias, c/ Via Lactea s/n, 38205 La Laguna, Spain
             \and Departamento de Astrof\'isica, Universidad de La Laguna, 38206 La Laguna, Spain
             \and Astrophysics Research Institute, Liverpool John Moores University, 146 Brownlow Hill, Liverpool L3 5RF, UK
             \and Instituto Galego de F\'isica de Altas Enerx\'ias, Universidade de Santiago de Compostela, 15782 Santiago de Compostela, Galicia, Spain
             \and Canadian Institute for Theoretical Astrophysics (CITA), University of Toronto, 60 St. George Street, Toronto, Ontario M5S 3H8, Canada
             \and Department of Physics \& Astronomy, McMaster University, 1280 Main Street West, Hamilton, L8S 4M1, Canada
             }

   \date{Received, 20XX}

  \abstract
  {Globular clusters (GCs) are long-lasting survivors of galaxy assembly and evolution, yet the processes behind their emergence from an initial cluster population are still poorly constrained. Here, we present Auriga GLOBular clustEr Simulations (AuriGLOBES) a physically motivated subgrid model for star cluster (SC) formation and evolution that includes enhanced mass loss from compact object remnants. With this model, implemented in the Auriga cosmological galaxy formation model, we ran a suite of zoom-in cosmological simulations comprising nine Milky Way (MW) mass and five lower mass galaxies. We demonstrate that our model produces plausible GC populations, compared to the MW/M31 systems, and reproduces the empirical GC system-mass-halo-mass relation within a 2$\sigma$ scatter. We show that the formation of SCs in tidally compressive, high-pressure gas in addition to enhanced mass loss from compact object remnants heating is required to capture the transformation of an initial Schechter mass function to the characteristic observed GC mass function in the MW/M31 systems. The resulting GC populations exhibit spatial and metallicity distributions that are qualitatively similar to the MW/M31 systems, as well as a variety of age distributions that correlate with the star formation history of the simulated galaxies. However, the peak of the age distribution of MW GCs is older than any of our simulated MW-mass galaxies, which is attributed to unrepresented star formation and galaxy assembly histories. AuriGLOBES represents a reliable framework for the study of GC populations through cosmic history and offers a robust foundation for future applications in modelling stellar streams arising from GC disruption.
  }
  \keywords{galaxies: star clusters: general - globular clusters: general - galaxies: formation - galaxies: evolution - methods: numerical}

\maketitle
\nolinenumbers

%%%%%%%%%%%%%%%%%%%%%%%%%%%%%%%%%%%%%%%%%%%%%%%%%%%%%%%%%%%%%%
\section{Introduction}
Star clusters (SCs) make up one of the fundamental building blocks of galaxies and are powerful tracers of their formation and assembly history. Among them, globular clusters (GCs) are shown to be especially puzzling due to their old ages, $\gtrsim 6$~Gyr \citep[e.g.][]{Strader_2005, Forbes_2010, VandenBerg_2013}, and near-universal presence in galaxies across a wide mass range \citep{Harris_1991, Brodie_2006, Harris_2017, Forbes_2018, Dornan_2025, Forbes_2025}. The old ages of many GCs mean they formed early in the universe and survived until low redshift, offering a unique window into the physical conditions of high-redshift star formation and the hierarchical growth of galaxies. Their metallicities, ages, kinematics, and spatial distributions encode information about their host galaxy assembly, making them valuable tools for studying both early galaxy formation and late-time galactic dynamics \citep[e.g.][]{Searle_1978, Ibata_1994, Forbes_2010}.

How the present-day GC population emerged from its initial cluster population remains an open question. Observations of GC-like young massive clusters (YMCs) forming in the Local Universe \citep[e.g.][]{Holtzman_1992, Whitmore_1995, Zepf_1999, Adamo_2017} have suggested that SC formation is a byproduct of star formation in high-pressure regions \citep[e.g.][]{Harris_1994, Elmegreen_1997, Kruijssen_2014}. In this picture, GCs are the long-term survivors of YMCs formed early in the universe. This has been reinforced by recent JWST observations of compact, dense, clusters and clumps at high redshift (i.e. $0.5<z<10$) in the context of sites of proto-GC formation \citep[e.g.][]{Adamo_2024, Mowla_2024, Claeyssens_2025, Claeyssens_2026}.
The mass functions (MFs) of YMCs follow a power law with an exponential cutoff at high masses \citep{Portegies_Zwart_2010, Johnson_2017, Adamo_2020, Wainer_2022}. Furthermore, this MF seems to be already in place from the early universe throughout cosmic history \citep{Claeyssens_2026}.
In contrast, the GC mass function (GCMF) across galaxies is remarkably uniform and log-normal-like, with the peak, at $M\sim 10^5\mathrm{M}_\odot$, weakly depending on the host galaxy stellar mass and galactocentric radius \citep{Harris_1991, Jordan_2007}. This suggests a transformation driven by dynamical evolution, including mass loss from stellar evolution, internal two-body relaxation, and tidal stripping, which preferentially affects low-density clusters \citep{Fall_2001, Vesperini_2001, Elmegreen_2010a, Kruijssen_2015, Gieles_2023}.

Capturing this transformation within a full cosmological framework is a major challenge due the vast dynamic range involved. The formation and evolution of SCs depend on parsec-scale processes, whereas they must be modelled within their host galaxies evolving on kiloparsec and gigayear scales. Different approaches have been used to tackle this issue, with coarser methods, allowing for a more extended cosmic time and a larger number of galaxies to be explored. 

First, high-resolution simulations with improved physics prescriptions have enabled us to resolve the formation of clumps, proto-clusters, and star clusters at high redshift \citep[e.g.][]{Mandelker_2014, Kim_2018,Ma_2020, Williams_2025}, in galaxy mergers \citep[e.g.][]{Lahen_2020, Li_2022}, and in dwarf galaxy systems \citep[e.g.][]{Gutcke_2024, Deng_2024, Taylor_2025}. However, these simulations are computationally extremely expensive and are usually limited to high redshifts and/or small volumes. Second, semi-analytic models offer computationally efficient frameworks to follow SCs formation and disruption across cosmic time by populating N-body, or full galaxy simulations, with SCs via post-processing \citep[e.g.][]{Renaud_2017, Creasey_2019, Phipps_2020, Halbesma_2020, Valenzuela_2021, Rodriguez_2023, Chen_2024, De_Lucia_2024}. As semi-analytic models do not require complete reruns, they enable the simulation of SC populations in large samples of galaxies and an efficient exploration of model parameters. Nevertheless, these could represent an incomplete SC model from not having the full information of the tidal environment shaping their disruption at all times.

Third, subgrid models embed simplified prescription for SCs formation and evolution directly within cosmological galaxy simulations enabling a consistent interaction between the SCs and their host galaxies. The results, however, can be sensitive to the specific galaxy model of choice, interstellar medium (ISM) structure, or disruption mechanisms and prescriptions. Important contributions under this approach include the models of \citet{Li_2017, Li_2018, Li_2019}, which use sink particle techniques to model SC formation in giant molecular clouds (GMC), the E-MOSAICS project \citep{Pfeffer_2018, Kruijssen_2019b} including the MOSAICS\footnote{MOdelling Star cluster population Assembly in Cosmological Simulations} model for cluster formation and evolution \citep{Kruijssen_2008, Kruijssen_2009, Kruijssen_2011} in a rerun of the EAGLE\footnote{Evolution and Assembly of GaLaxies and their Environments} simulations \citep{Crain_2015, Schaye_2015}, and the EMP-Pathfinder simulations \citep{Reina_Campos_2022, Reina_Campos_2023} implementing an updated MOSAICS prescription under a new galaxy formation model that includes a cold, dense phase of the ISM.

Despite a significantly different numerical setup and physical processes included, previous efforts are demonstrating that the process of regular cluster formation, combined with that of hierarchical galaxy assembly, produce diverse populations of SCs that broadly resemble the observed clusters in the Local Universe. However, as highlighted in \citet{Reina_Campos_2022}, uncertainties in the formation and evolution of SCs produce degeneracies in several of their properties, such as their mass distribution (see their Fig.~8). In particular, matching any observed GCMF involves considerations on the ability to capture tidal disruption \citep[e.g.][]{Pfeffer_2018, Li_2019, Rodriguez_2023, Reina_Campos_2022}, assumed SCs size and density prescriptions \citep[e.g.][]{Reina_Campos_2023}, as well as their compatible assembly histories \citep[e.g][]{Chen_2024}.
These uncertainties motivate further studies of the formation and evolution of GCs under different galaxy formation models.

To this end, we present AuriGLOBES: Auriga GLOBular clustEr Simulations, a physically motivated subgrid model for the formation and evolution of star clusters implemented within the Auriga suite of cosmological zoom-in simulations. These are evolved in a lambda cold dark matter ($\Lambda$CDM) cosmology including a comprehensive galaxy formation physics model giving predictions that offer a good comparison with many observed scaling relations, such as the Tully-Fisher relation, the star-forming main sequence, HI gas fraction, and HI disc thickness \citep{Grand_2017, Grand_2024}. Our model incorporates four updates to the SC modelling used in previous works:  

First, \textit{(i)} it includes a novel criterion for constraining SC formation to tidally compressive gas environments only and \textit{(ii)} implements an empirically motivated initial mass-radius relation (MRR). Such compressive regions conductive to the formation of gravitationally bound clusters \citep{Renaud_2014, Renaud_2015} are consistent with theoretical expectations and observations linking cluster formation to dynamically driven, high-pressure environments \citep{Renaud_2009, Whitmore_2010, Johnson_2015}. The SCs are then dynamically evolved in mass through two-body relaxation and tidal shocks. In particular, \textit{(iii)} our two-body relaxation mass loss prescription accounts for the enhanced two-body scattering expected from clusters that retain a fraction of their compact stellar remnants. The inclusion of this enhancement is motivated by recent studies showing that SCs could retain populations of black holes (BHs) and their role as a source of energy for the clusters dynamical evolution \citep{Gieles_2023, DellaCroce_2024}. This is further motivated by increasing detections of BHs in MW GCs \citep[e.g.][]{Giesers_2018, Kremer_2020, Weatherford_2020, Dickson_2024} and extragalactic GCs \citep[e.g.][]{Maccarone_2007, Maccarone_2011, Barnard_2012, Saracino_2022}. Finally, \textit{(iv)} we implement a size evolution prescription that is coupled to the mass evolution of the SCs. Together, these ingredients provide a reliable framework to model the formation and long-term evolution of SC populations in a cosmological context.

This paper is structured as follows. In Sect.~\ref{sec:simdescription}, we describe the simulation set-up including the description of the Auriga simulations and of our AuriGLOBES model. In Sect.~\ref{sec:simulationsuite}, we present the suite of simulations run with our model and in Sect.~\ref{sec:scpopulations}, we show the results of the simulated SC and GC populations. Finally, we give our discussion and conclusions in Sects. \ref{sec:discussion} and \ref{sec:conclusions}.

%%%%%%%%%%%%%%%%%%%%%%%%%%%%%%%%%%%%%%%%%%%%%%%%%%%%%%%%%%%%%%
\section{Simulation set-up}\label{sec:simdescription}
\begin{figure*}
\begin{tikzpicture}[node distance=1.5cm]

\node[box] (mhd) {Magnetohydrodynamics\\ + Gravity (AREPO)};
\node[box, below=1.4cm of mhd] (auriga) {Auriga Galaxy\\ Formation Model};
\node[gcbox, right=2cm of mhd] (eq) {$T_{ij} = -\partial_i \partial_j \Phi = \sum_{n=0}^{N} -\partial_i \partial_j \phi_n (x - x_n)$};

\node[box, right=2.5cm of auriga, yshift=1cm] (sf) {Star Formation};
\node[box, right=2.5cm of auriga, yshift=-1cm] (se) {Star Evolution};
\node[gcbox, right=1.5cm of sf] (cf) {Cluster Formation};
\node[gcbox, right=1.5cm of se] (ce) {Cluster Evolution};
\node[gcbox, below=0.8cm of ce] (cd) {Star Cluster \\ population at $z=0$};
\node[gcbox, right=2.5cm of cd] (gcs) {GC Population};

\node[gcbox, right=3cm of eq] (globes) {AuriGLOBES};
\node[textonly, right=0.5cm of cf, yshift=-0.1cm, text=black] (cfinfo) {%
  $\bullet$ \textbf{Tidally compressive regions} \\ 
  $\bullet$ Environmental CFE \\ 
  $\bullet$ Environmental CIMF \\
  $\bullet$ \textbf{Empirical initial MRR} \\
};
\node[textonly, right=0.5cm of ce, text=black] (ceinfo) {%
  $\bullet$ Two-body relaxation with \\ \textbf{compact object remnants heating} \\ 
  $\bullet$ Tidal shocks \\ 
  $\bullet$ \textbf{Half-mass radius evolution} \\
  $\bullet$ DF at post-processing
};
\node[textonly, right=0.25cm of cd.east, yshift=0.4cm, text=black, fill=white, align=center](gcsel){GCs \\ selection};

% Arrows
\draw[arrow] (mhd) -- (auriga);
\draw[arrow] (auriga.east) -- ++(1,0) |- (sf.west);
\draw[arrow] (auriga.east) -- ++(1,0) |- (se.west);

\draw[arrow] (sf) -- (cf);
\draw[dashedarrow] (se) -- (ce);
\draw[arrow] (cf) -- (ce);
\draw[arrow] (ce) -- (cd);
\draw[arrow] (cd) -- (gcs);

\draw[arrow] (eq.west) -- (mhd.east);
\draw[dashedarrow] (eq.east) .. controls +(1,0) and +(0.5,0.5) .. (cf.north);
\draw[dashedarrow] (eq.east) .. controls +(1.5,0) and +(0.5,1.5) .. (ce.north east);
\draw[dashedarrow] (auriga.east) .. controls +(0.,0.) and +(-0.5,-0.5) .. (cf.south west);
\node[textonly, right=2.9cm of auriga, align=center, fill=white](gas){Local Gas \\ Properties};

\end{tikzpicture}

\caption{The AuriGLOBES star cluster formation and evolution model. The base simulation code and the Auriga galaxy formation model, with star formation and evolution stages, are depicted in the light blue boxes. The AuriGLOBES extension to the base simulation is depicted in the dark blue boxes. Continuous arrows represent a direct relation, while dashed arrows represent a dependency relation. Bold fonts highlight extensions with respect to previous models. The used acronyms are: cluster formation efficiency (CFE), cluster initial mass function (CIMF), mass radius relation (MRR), dynamical friction (DF), and globular clusters (GCs).}
\label{fig:auriglobes}
\end{figure*}
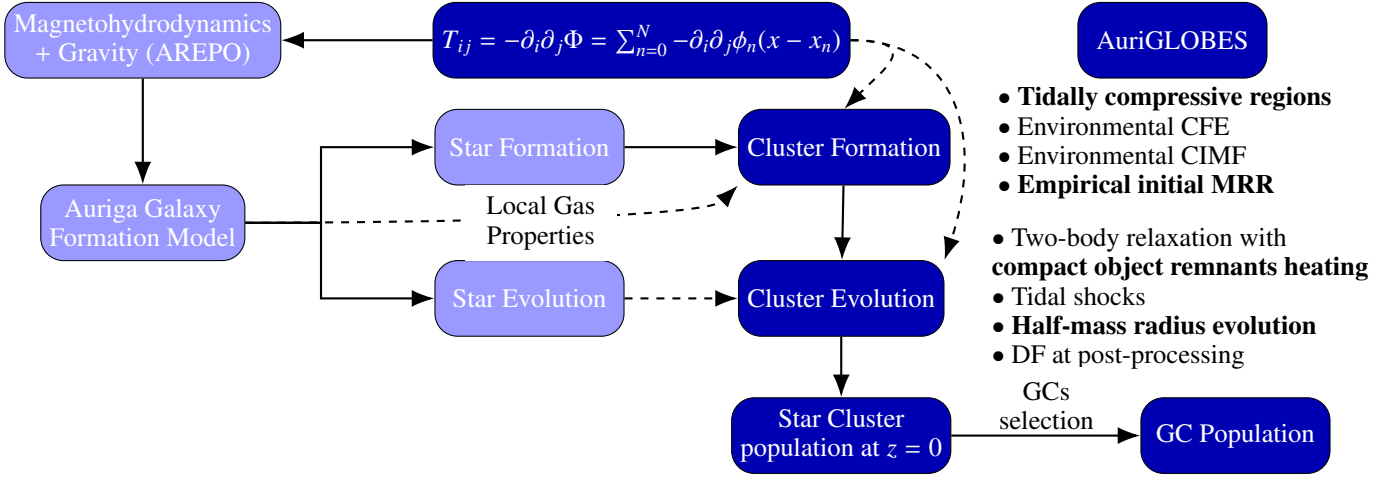

In this section, we describe our AuriGLOBES sub-grid SC model, which is implemented into the Auriga galaxy formation model \citep{Grand_2017}. We first describe the Auriga model in Sect.~\ref{sec2:auriga} and then give a full description of AuriGLOBES in Sect.~\ref{sec2:globes}. The complete schematic of the model workflow, and its relation with Auriga, is shown in Fig. \ref{fig:auriglobes}.

\subsection{The Auriga galaxy formation model}
\label{sec2:auriga}
The Auriga formation model is implemented in the second-order accurate gravo-magnetohydrodynamic (gravo-MHD) code AREPO \citep{Springel_2010}, which follows MHD and collisionless dynamics in a cosmological context. AREPO calculates gravitational forces with a standard tree particle-mesh method \citep{Springel_2005} comprising a hierarchical oct-tree algorithm \citep{Barnes_1986} for short range forces and a fast Fourier treatment for long range forces. The timestep is adaptive and MHD is solved in an unstructured Voronoi mesh that moves approximately with the fluid. This way, AREPO performs as hybrid Eulerian and Langrangian code reducing errors accumulating from large relative bulk velocities inherent to galaxy formation \citep{Wadsley_2008}, while retaining the ability to capture shocks to a good accuracy \citep{Schaal_2015, Schaal_2016}.

The simulations follow a comprehensive galaxy formation model including the evolution of gas, dark matter, stars, and BHs. It is described in detail in \citet{Grand_2017}, and builds on top of models presented in \citet{Vogelsberger_2013} and \citet{Marinacci_2014}. The model implements:
\begin{itemize}
    \item a pressurised two phase ISM model consisting of cold, dense, clouds embedded in a hot, ambient medium \citet{Springel_2003};
    \item stochastic star formation for dense ISM gas above a threshold density of $n_{\mathrm{th}}=0.13~\mathrm{cm}^{-3}$ with a \citet{Chabrier_2003} initial mass function (IMF). This threshold is derived from the parameters of the two gas phases and a desired star formation time-scale of $t_\mathrm{SF}=2.2$~Gyr;
    \item primordial and metal-line cooling and heating from a uniform ultraviolet background radiation field \citep{Faucher_2009};
    \item stellar evolution from asymptotic giant branch stars and supernovae type Ia and II with a stellar feedback scheme for thermal and kinetic energy injection;
    \item supermassive black holes seeding and growth;
    \item active galactic nuclei thermal feedback in quasar and radio modes;
    \item magnetic fields seeded uniformly at the starting redshift of the simulation at a comoving strength of $10^{-14}~\mathrm{G}$ \citep{Pakmor_2017}.
\end{itemize}

The Auriga galaxy formation model has been employed to run large suites of cosmological MHD simulations for the formation and evolution of Milky Way (MW) mass and dwarf-mass galaxies in the $\Lambda$CDM cosmogony \citep{Grand_2024}. The simulations use the zoom-in technique, which resimulates galaxy halos from the Eagle Dark Matter Only simulation of comoving side length 100~cMpc \citep[L100N1504 in][$1.15\times 10^7 \mathrm{M}_\odot$ mass resolution]{Schaye_2015} at high resolution. 
The initial conditions of these resimulations are created with the public Gaussian white noise field realisation Panphasia \citep{Jenkins_2013} adopting the cosmological parameters from \citet{Planck_2013}, namely:  $\Omega_\mathrm{m}=0.307$, $\Omega_\mathrm{b}=0.048$, $\Omega_\Lambda=0.693$, $\sigma_8=0.8288$, and Hubble constant of $H_0=100h~\mathrm{km \, s}^{-1}\,\mathrm{Mpc}^{-1}$, where $h=0.6777$. For each identified halo, the particles inside four times its virial radius at $z=0$ where traced back to their initial position at $z=127$ and resampled to a higher mass resolution while lowering the resolution of particles at larger distances. This reduces computational cost while maintaining the correct cosmological tidal field. For more details on the Auriga simulation suite, we refer to \citet{Grand_2017,Grand_2024}. In this paper, we describe how we applied our SCs formation and evolution model by re-running some of these cosmological simulations (see Section~\ref{sec:simulationsuite}).

\subsection{The AuriGLOBES sub-grid star cluster model}
\label{sec2:globes}
The SC formation and evolution model largely follows sub-grid prescriptions available in the literature \citep[e.g.][]{Pfeffer_2018, Reina_Campos_2022} adapting them to work with our galaxy formation setting and extending the covered physical processes to cover compressive natal environments, empirical initial sizes, enhanced mass loss from compact object heating, and size evolution. In the following subsections, we describe in detail the physics and numerical implementation of the AuriGLOBES SC model.

\subsubsection{Cluster formation}
In our fiducial model, a newly created star particle is eligible to host SCs based on the properties of the natal gas cell. In addition to high gas density and pressure leading to bound systems to likely form, we also require the gas environment to be tidally compressive.
The latter is determined from calculating the tidal tensor, $\mathbf{T}$. We do it in a similar way as in \citet{Grudic_2020}, by using the linearity of $\mathbf{T}$ to sum the contributions of each particle in the simulation,
\begin{equation}\label{tidal_field}
    T_{ij}=\sum_{n=0}^{N} -\partial_i\partial_j \Phi_n(\mathbf{x}-\mathbf{x}_n) = \sum_{n=0}^{N} \partial_i \mathbf{g}_{n,j}(\mathbf{x}-\mathbf{x}_n),
\end{equation}
where $\Phi_n$ is the gravitational potential, and $\mathbf{g}_{n, j}$ the  acceleration ($j$-th component) contribution of the particle $n$ as a function of the separation $\mathbf{x} - \mathbf{x}_n$, and the partial derivatives are with respect to the target position $\mathbf{x}$. The evaluation of the tidal tensor is done in the same tree walk as the evaluation of $\mathbf{g}$.

Once $\mathbf{T}$ is defined\footnote{Noting that long range, Fourier based, gravitational contributions are not taken into account for the tidal tensor calculus in favour of decreasing the computational time. We tested its inclusion and found negligible differences in the resulting tidal tensor. Both the test of the correct calculation of $\mathbf{T}$ and how negligible is the error when ignoring long range contributions are shown in the appendices \ref{app:tidal_tensor} and \ref{app:long_range_tensor}.}, its eigenvalues $\lambda_i$ describe whether the gravity field stretches or squeezes a region in each of the three spatial directions: negative values mean that gravity is compressive in a given direction, whereas positive values reflect the opposite. Therefore, whether the medium is compressive or not is ensured by having all three negative eigenvalues. We further strengthen this condition to having all three eigenvalues less than a certain negative threshold, i.e. $\forall \lambda_i \leq -1 \; \mathrm{Gyr}^{-2} $. This threshold was chosen by visually inspecting the tidal tensor eigenvalues distribution and finding that the negative eigenvalues are generally well below this value (see Appendix \ref{app:long_range_tensor}).

Once a star particle is identified to be born in a tidally compressive region, the formation of SCs is determined by three ingredients: \textit{(i)} the cluster formation efficiency (CFE, $\Gamma$), namely, the fraction of star formation happening in bound structures \citep{Bastian_2008}, estimated from the physical properties of the parent gas cell; \textit{(ii)} a locally defined cluster initial mass function (CIMF) from which the individual SC masses are drawn; and \textit{(iii)} an empirical initial MRR from which the individual SC sizes are determined. These are explained in the following paragraphs. As previous papers have reported, the numerical details of how star formation is modelled affect the resulting cluster population \citep[e.g.][]{Pfeffer_2018, Reina_Campos_2022}. Thus, we perform additional simulation tests to assess the influence of including or not this compressive criterion (see Sect.~\ref{sec3.1}).

We estimated the CFE by implementing the \citet{Kruijssen_2012} model, which estimates the fraction of star formation happening in bound structures from integrating the density spectrum of the ISM. In particular, we used the model formulation based on the local properties of the gas, namely, $f_\mathrm{bound}=\Gamma(\rho_\mathrm{g}, \sigma_\mathrm{loc},c_s)$, with $\rho_\mathrm{g}$ the gas density, $\sigma_\mathrm{loc}$ the gas velocity dispersion, and $c_s$ its sound speed. Equivalent to E-MOSAICS \citep{Pfeffer_2018}, we estimated the one-dimensional velocity dispersion from $\sigma_\mathrm{loc}=\sqrt{P_\mathrm{g}/\rho_\mathrm{g}}$, with $P_\mathrm{g}$ the local gas pressure, and assumed a thermal sound speed corresponding to a cold interstellar gas at $\sim 10\,\mathrm{K}$, i.e. $c_\mathrm{s}=0.3\, \mathrm{km \, s}^{-1}$. The implementation of the local formulation of \citet{Kruijssen_2012} model is justified given that the local gas properties are a good description of the global state of the gas, mainly from the ISM model not explicitly modelling a cold, clumpy, structure.

The fraction, $\Gamma$, sets the mass budget for SC masses to be drawn from the CIMF, given by a \citet{Schechter_1976} function,
\begin{equation}\label{cimf}
    \mathrm{d}N \propto m^{-\alpha} \exp\left( -m / m_\mathrm{cl,max}\right)\mathrm{d}m,
\end{equation}
where $\alpha \simeq 2$ \citep{Zhang_1999} is the power-law index, $m_\mathrm{cl,max}$ the CIMF truncation mass, and $m$ the individual SC mass. This CIMF has been shown to be consistent with populations of Young Star Clusters (YSC) in the local Universe \citep{Portegies_Zwart_2010, Johnson_2017, Adamo_2020, Wainer_2022}. High-resolution simulations following star-by-star compact SC formation agree with this power law, whereas the exponential cut-off has been difficult to characterise due the size-of-sampling effect \citep[e.g.][]{Lahen_2020, Andersson_2024, Pascale_2025}.

The truncation mass is set by the mass of the GMC from which clusters originate \citep{Kruijssen_2014}, $M_\mathrm{GMC,max}$. To estimate this, we used the environment dependent model of \citet{Reina_Campos_2017}, namely,
\begin{equation}
    m_\mathrm{cl,max} = \epsilon \Gamma M_\mathrm{GMC,max} \; ; \; M_\mathrm{GMC,max} = f_\mathrm{coll} M_\mathrm{T},
\end{equation}
where $\epsilon = 0.1$ is the integrated star formation efficiency for an entire molecular cloud\footnote{As noted by \citet{Pfeffer_2018}, the variation of $\epsilon$ estimates is much smaller than the dynamical range of the product $\Gamma M_\mathrm{GMC,max}$ making it reasonable to set as a constant.} \citep{Duerr_1982, Murray_2011, Kruijssen_2019a, Sun_2023}, $M_\mathrm{T}$ is the \citet{Toomre_1964} mass, i.e. the highest mass on a region that can collapse under its own gravity, and $f_\mathrm{coll}$ is the collapse fraction before stellar feedback stops this process,
\begin{equation}
    f_\mathrm{coll} = \min \left(1,\frac{t_\mathrm{fb,g}}{t_\mathrm{ff,2D}}\right)^4,
\end{equation}
where $t_\mathrm{fb,g}$ is the gas feedback time-scale, and $t_\mathrm{ff,2D}$ the two-dimensional cloud free-fall time. Following \citet{Kruijssen_2012}, the gas feedback time-scale is
\begin{equation}
    t_\mathrm{fb,g} = \frac{t_\mathrm{sn}}{2}\left( 1+ \sqrt{1 + 
    \frac{4t_\mathrm{ff}\sigma_\mathrm{loc}^2}{\phi_\mathrm{fb} \epsilon_\mathrm{ff} t_\mathrm{sn}^2}} \right),
    \label{eq:fb_time}
\end{equation}
where $t_\mathrm{ff}=\sqrt{3\pi/32G\rho_\mathrm{g}}$ is the gas free-fall time, $\epsilon_\mathrm{ff}=0.012$ the star formation efficiency per free-fall time \citep{Elmegreen_2002, Utomo_2018, Grisdale_2019}, $t_\mathrm{sn} = 3 \mathrm{Myr}$ is the typical time of the first supernova \citep{Ekstrom_2012}, and $\phi_\mathrm{fb} = 0.16 \; \mathrm{cm}^2 \mathrm{s}^{-3}$ is the rate of energy injection per unit stellar mass into the ISM from feedback of a simple stellar population with a normal IMF \citep{Kruijssen_2012}. Lastly, the cloud free-fall time is defined as
\begin{equation}
    t_\mathrm{ff,2D} = \frac{\sqrt{2\pi}}{\kappa},
\end{equation}
where $\kappa$ is the epicyclic frequency that is computed from the tidal tensor at the location of the particle (see Appendix \ref{app:tidal_tensor}), as demonstrated by \citet{Pfeffer_2018} and \citet{Reina_Campos_2022}.

The Toomre mass is obtained from the parent gas cell for each new star particle through
\begin{equation}
    M_\mathrm{T} = 4\pi^5 G^2  \frac{\Sigma_\mathrm{g}^3}{\kappa^4},
    \label{eq:toomre_mass}
\end{equation}
where $\Sigma_\mathrm{g}$ is the local surface gas density estimated assuming hydrostatic equilibrium; namely, equating the mid-plane pressure of an equilibrium disc to the pressure of the parent gas cell \citep{Krumholz_2005} via
\begin{equation}
    \Sigma_\mathrm{g} = \sqrt{\frac{2 P_\mathrm{g}}{\pi G \phi_P}},
\end{equation}
with $\phi_P$ being the contribution of the gravity of stars to the mid-plane pressure. Following \citet{Pfeffer_2018}, this contribution is expressed as
\begin{equation}
    \phi_P = 1+\frac{\sigma_\mathrm{g}}{\sigma_\star}\left(\frac{1}{f_\mathrm{gas}} -1\right),
\end{equation}
with $f_\mathrm{gas}$ the gas mass fraction, and $\sigma_\mathrm{g}$ ($\sigma_\star$) the velocity dispersion of the gas (stars) in the vicinity region where $\phi_P$ is determined. This region is up to three times the parent gas cell smoothing length until a minimum of 64 gas cells are enclosed.

Clusters are formed by stochastically drawing from the CIMF. The number of clusters expected to form in a new star particle of mass $m_\star$ is given by the mass budget defined by $\Gamma$ over the mean cluster mass, i.e. $ N_\mathrm{clus} = \Gamma m_\star / \bar{m}_\mathrm{cl}$. This mean cluster mass is defined by integrating the normalised CIMF over the mass range for cluster formation, expressed as
\begin{equation}
    \bar{m}_\mathrm{cl} = \int_{10^2 \mathrm{M}_\odot}^{10^8 \mathrm{M}_\odot} m \frac{\mathrm{d}N}{\mathrm{d}m}\mathrm{d}m.
\end{equation}
We note that the mass range is set to be consistent with YSCs in the Local Universe \citep[e.g.][]{Brown_2021}. Finally, the actual number of cluster masses to be drawn from the CIMF is defined from a Poisson distribution with the mean given by $N_\mathrm{clus}$. While the low limit to draw masses is $10^2 \, \mathrm{M}_\odot$, clusters below $5 \times 10^3 \, \mathrm{M}_\odot$ are discarded to lower the memory consumption. This is justified given that all of these low mass clusters are expected to disrupt shortly after their formation \citep[$<<1$~Gyr, ][]{Kruijssen_2012a}. Due to the stochasticity of cluster formation, and depending on the mass resolution of the simulation, some of the star particles may contain cluster populations with a total mass exceeding their actual mass while others may contain no cluster populations. At the Auriga resolution level~4, this accounts for $\sim 10 \%$ of the star particles that host SCs having a total mass in SCs higher than its mass; for most of the cases it is well bellow the parent star particle mass and close to the expected $\Gamma m_\star$ mass budget.

Finally, the created SCs are assigned an effective radius, $R_\mathrm{eff}$, determining its half-mass radius, $r_\mathrm{h,cl}=4R_\mathrm{eff}/3$. This is done by using the empirical mass-radius relation (MRR) for YSCs reported by \citet{Brown_2021}. This relation corresponds to SCs in the LEGUS sample in the $1\textrm{--}10 \; \mathrm{Myr}$ age range with its intrinsic scatter,
\begin{equation}
    R_\mathrm{eff} = 2.365 \mathrm{pc} \; \left(\frac{m_\mathrm{cl}}{10^4 \mathrm{M}_\odot}\right)^{0.18} \; \mathrm{with}\; \sigma_{R_\mathrm{eff}} = 0.319 \; \mathrm{dex},
\end{equation}
where $m_\mathrm{cl}$ is the individual SC mass.

\subsubsection{Cluster evolution}
\label{sec2:clevo}
Once SCs are formed, their dynamical evolution is immediately followed determining their mass loss coupled with their size evolution. The mass loss is determined by stellar evolution, two-body relaxation, and tidal shocks via
\begin{equation}
    \frac{\mathrm{d}m}{\mathrm{d}t} = \left(\frac{\mathrm{d}m}{\mathrm{d}t}\right)_\mathrm{sev}
    + \left(\frac{\mathrm{d}m}{\mathrm{d}t}\right)_\mathrm{rlx} + \left(\frac{\mathrm{d}m}{\mathrm{d}t}\right)_\mathrm{sh}.
\end{equation}
The stellar evolution mass loss is directly inherited from the parent star particle evolution given by the Auriga model. For the dynamical mass loss mechanisms, we follow, or modify, available models applicable to sub-grid prescriptions \citep[e.g.][]{Gnedin_1997, Prieto_2008, Kruijssen_2011, Alexander_2012}. Both two body relaxation and tidal shocks mass loss strongly depend on the underlying tidal field, eq.~\ref{tidal_field}, as described below.

\paragraph{Two-body relaxation mass loss enhanced by compact object heating}\mbox{}\\
\label{sec2:relaxation}
The mass loss driven by two-body interactions leading to the evaporation of the cluster is estimated by\footnote{The canonical mass loss rate is given by $\left(\mathrm{d}m/\mathrm{d}t\right)_\mathrm{rlx} = - \xi m/t_\mathrm{rh}$}
\begin{equation}\label{rlx_massloss}
    \left(\frac{\mathrm{d}m}{\mathrm{d}t}\right)_\mathrm{rlx} = - a\xi\frac{m}{t_\mathrm{rh}} \left( \frac{m}{m_i}\right)^{-2/3},
\end{equation}
where $\xi$ is the fraction of stars escaping the SC potential per relaxation time, $a$ is a boosting parameter, $t_\mathrm{rh}$ is the relaxation time-scale at the half-mass radius, $m$ is the current mass of the cluster, and $m_i~\equiv~\mu_\mathrm{sev}m_0$ is the initial mass after the stellar evolution mass loss has occurred \citep{Gieles_2023}. For Auriga, $\mu_\mathrm{sev}$ covers a range of $\mu_\mathrm{sev} \approx 0.53\textrm{--}1$.

The relaxation time-scale, $t_\mathrm{rh}$, is defined following \citet{Spitzer_1987} via
\begin{equation}
    t_\mathrm{rh}=\frac{0.138 \sqrt{N} r_\mathrm{h}^{3/2}}{\sqrt{G \bar{m}}\ln(\gamma N)},
\end{equation}
with $N=m/\bar{m}$ the mean number of stars in the cluster, $\bar{m}$ the mean stellar mass in the cluster, and $\ln(\gamma N)$ the Coulomb logarithm with $\gamma\approx0.11$\citep{Giersz_1994} for clusters of single-mass stars.
For Auriga, which adopts a \citet{Chabrier_2003} IMF in the $0.1 \leq m/\mathrm{M}_\odot \leq 100$ mass range, the mean stellar mass is $\bar{m}= 0.62\, \mathrm{M}_\odot$. The escapers fraction $\xi$ is estimated following the model described in the EMACSS prescription \citep{Alexander_2012, Gieles_2014, Alexander_2014} that fits the evaporation of N-body simulations of SCs submerged in a tidal field\footnote{\citet{Alexander_2014} model revision provides a parametrisation to include the effect of an evolving stellar mass spectrum making up the SCs where the relaxation timescale $t_\mathrm{rh}$ is modified to account for the effect of such a mass spectrum. It provides a recalibration of the parameters for multi-mass SCs accounting for the evolution of the mass spectrum and should, up to some degree, include the effect of the retention of compact object remnants. We performed tests using this model prescription with results showing an inconsistent behaviour. For simplicity, we adopt the single-mass SCs prescription, with fewer parameters, and apply the compact objects correction afterwards.} via
\begin{equation}
    \label{escapers_rate}
    \xi = \xi_0 (1-\mathcal{P}) + \frac{3}{5}\zeta \mathcal{P},
\end{equation}
where $\xi_0=0.0142$ is the fraction of stars escaping a SC in isolation \citep{Alexander_2012}, $\zeta =0.1$ is a dimensionless constant for the fractional energy change per relaxation time-scale \citep{Henon_1961, Henon_1965, Gieles_2011}, and the effect of the tidal field is captured in the dimensionless evaporation rate $\mathcal{P}$ that is a function of the half-mass radius, the tidal radius, and the number of stars in the cluster (eq. 25 in \citet{Alexander_2012} with the revised parameter values from \citet{Gieles_2014}). The tidal radius is the boundary at which the gravitational acceleration due to the galactic potential equals the acceleration from the potential of the cluster and is given by
\begin{equation}
    r_\mathrm{t} = \left( \frac{G m}{T}\right)^{1/3}.
    \label{eq:tidal_radius}
\end{equation}
where $T = -\partial^2 \Phi / \partial r^2 + \Omega^2$ is the tidal field strength \citep{King_1962}, with the circular frequency, $\Omega$. This is estimated from the tidal tensor eigenvalues as $T = \max(\lambda_i) + \Omega^2 = \max(\lambda_i) + 1/3 \left| \sum_{i=1}^3 \lambda_i \right|$, following \citet{Pfeffer_2018} and \citet{Reina_Campos_2022}.

The boosting parameter, $a=1.5$, and the last factor in eq.~\ref{rlx_massloss}, $\left( m/m_i\right)^{-2/3}$, account for the effect of compact object remnants in the mass loss rates of SCs, following the parametrisation given by  \citet{Gieles_2023}. These authors found that SCs retaining a fraction of their compact object remnant population experience an accelerated mass loss rate from two-body interactions and that this effect is fundamental in evolving an initial Schechter-like CIMF into a peaked GCMF consistent with observations (targeting the MW GC population, in particular). In addition, studies have pointed to BH populations being present in MW and extragalactic GCs \citep[e.g.][]{Maccarone_2007, Maccarone_2011, Barnard_2012, Kremer_2020, Weatherford_2020, Saracino_2022, Dickson_2024} supporting the inclusion of this effect. Their parametrisation follows from comparing N-body models of SCs forming and retaining compact object remnants in a galactic tidal field to the baseline N-body models of \cite{Baumgardt_2003}, where SCs do not form BHs. We tested the effect of this enhancement by comparing our results from running simulations with and without this effect (see Sect.~\ref{sec3.1}).

\paragraph{Tidal shocks mass loss}\mbox{}\\
Tidal shocks are perturbations of the gravitational potential exerted by the passage through the disc or close encounters with dense structures. As a result, they increase the kinetic energy of the stars in the clusters, reducing the binding energy of the cluster, and could cause stars to escape its potential \citep{Ostriker_1972, Spitzer_1987, Kundic_1995, Gnedin_1997, Murali_1997a, Murali_1997c, Murali_1997b, Gnedin_1999a, Gnedin_1999b, Prieto_2008}. The mass loss is given by the time-scale for tidal heating, $t_\mathrm{sh}$, to change the energy of the stars at the half-mass radius by an order of itself \citep{Prieto_2008} via
\begin{equation}
    \left( \frac{\mathrm{d}m}{\mathrm{d}t}\right)_\mathrm{sh} = -\frac{m}{t_\mathrm{sh}}.
\end{equation}
Following \citet{Prieto_2008}, this time-scale is
\begin{equation}
    t_\mathrm{sh} \equiv \frac{|E_\mathrm{h}|}{(\mathrm{d}E_\mathrm{h}/\mathrm{d}t)_\mathrm{sh}}
    \simeq \frac{|E_\mathrm{h}| P}{2 \langle \Delta E_\mathrm{h}\rangle} = \frac{3}{5}\frac{Gm}{r_\mathrm{h}^3}\frac{P}{I_\mathrm{tid}},
\end{equation}
where $P$ is the time interval between shocks, $E_\mathrm{h}$ is the energy per unit mass at the half-mass radius, and $\langle \Delta E_\mathrm{h}\rangle$ its ensemble average. This energy is approximated as 
\begin{equation}
|E_\mathrm{h}| = \frac{v_\mathrm{h}^2}{2} \approx ~ \eta \frac{Gm}{2r_\mathrm{h}},
\end{equation}
 with the proportionality constant $\eta \simeq 0.4$ \citep{Spitzer_1987}. Its ensemble average is given by 
\begin{equation}
    \langle \Delta E_\mathrm{h}\rangle = \frac{1}{2}\langle \Delta v\rangle^2 = \frac{1}{6}I_\mathrm{tid}r_\mathrm{h}^2,
\end{equation}
where the tidal heating parameter, $I_\mathrm{tid}$, is defined by integrating the tidal tensor components through the duration of the shocks and summing over all its components:
\begin{equation}\label{heating_param}
    I_\mathrm{tid} \equiv \sum_{i,j} \left( \int T_{ij} \mathrm{d}t\right)^2 \left(1+\frac{\tau_{ij}^2}{t_\mathrm{dyn}^2} \right)^{-3/2}.
\end{equation}
For each pair of coordinates, $\tau_{ij}$ is the shock duration\footnote{For each component, the start (end) of a shock event is identified by finding a contrast of at least 88\% with respect to the last recorded (last maximum) component value, and the integral runs over the full duration of the shock. The contrast is defined by the height ratio of a Gaussian distribution at $0.5\sigma$ and at the mean, defining a full width of $1\sigma$.}, and $t_\mathrm{dyn} = \sqrt{8\pi r_\mathrm{h}^3/Gm}$ is the dynamical time-scale at the half-mass radius. The second term in eq.~\ref{heating_param} accounts for the \citet{Weinberg_1994c, Weinberg_1994b, Weinberg_1994a} adiabatic correction for extended shocks describing the absorption of energy by adiabatically expanding the cluster. We note that, although we do not explicitly limit the simulation timestep for tidal shocks capture, their typical duration is one to two orders of magnitude shorter than the typical shock duration implying that most shock disruption is correctly resolved (see Appendix~\ref{app:shocks_capture}). 

\subsubsection{Half-mass radius evolution}
Following \citet{Reina_Campos_2022}, we evolve the half-mass radius of SCs according to their mass evolution. As the stellar populations in SCs lose mass through stellar evolution, SCs experience an adiabatic expansion given by the inverse ratio of the parent star particle mass at the current and previous timestep, giving $(\mathrm{d}r_\mathrm{h}/r_\mathrm{h})_\mathrm{sev} = |\mathrm{d}m_\mathrm{\star,sev}|/m_\star$. In addition to the adiabatic expansion, a dynamical evolution is also implemented where, having $\mathrm{d}m_\mathrm{rlx}$ and $\mathrm{d}m_\mathrm{sh}$, the size of the cluster is evolved accordingly. This evolution follows the density evolution derived by \citet{Gieles_2016} extended to account for mass loss from two-body interactions \citep[first derived in Appendix E in][]{Reina_Campos_2023} via
\begin{equation}\label{eq:size_evo}
    \left(\frac{\mathrm{d}r_\mathrm{h}}{r_\mathrm{h}}\right)_\mathrm{dyn} = \left(2-\frac{1}{f}\right)\frac{\mathrm{d}m_\mathrm{sh}}{m}
    + \left(2-\frac{\zeta}{\xi}\right)\frac{\mathrm{d}m_\mathrm{rlx}}{m},
\end{equation}
where $f=3/5$ is the fraction of the energy gain of the remaining bound stars\footnote{Not to be confused with the fraction of the energy gain translated into a mass loss following a shock event introduced by \citet{Gieles_2006} using the same terminology but corresponding to all stars, i.e. including the ones that become unbound after the shock.} translated into a change in density \citep{Gieles_2016}. The choice of $f=3/5$ means that the density remains constant after the mass loss, adjusting the size accordingly. We tried different parameters and the results show no strong dependence on this particular selection (see Appendix \ref{app:rh_evolution}). Eq.~\ref{eq:size_evo} describes the shrinking of clusters due to shocks and their overall expansion from two-body interactions. We note, however, that in tidally limited scenarios, $\xi \simeq 3/5 \zeta$ (from eq.~\ref{escapers_rate}), also gives cluster contraction from two-body interactions. Implementing a half-mass radius evolution model corresponding to the mass evolution enables an exploration of the different cluster disruption processes. As both two-body relaxation and tidal shocks mass loss models are dependent on $r_\mathrm{h}$, i.e. $t_\mathrm{rh}\propto r_\mathrm{h}^{3/2}$ and $t_\mathrm{sh}\propto r_\mathrm{h}^{-3}$, the size evolution of the SCs  affects their survivability \citep[e.g.][]{Reina_Campos_2023}. To explore this effect, we compare simulations with and without evolving the half-mass radius (see Sect.~\ref{sec3.1}). 

\subsubsection{Dynamical friction}
Finally, the effect of dynamical friction is estimated at post-processing with the same approach of \citet{Pfeffer_2018}, and \citet{Reina_Campos_2022}. For a star particle hosting more than one SC, each one with its own mass, applying an on-the-fly dynamical friction treatment would result in star particles experiencing different forces from those of its SC population. This is specially relevant when the subgrid population has an important mass contrast between the clusters. Therefore, the effect is estimated by comparing the dynamical friction time-scale with the age of the cluster and assume SCs are disrupted whenever this time-scale is shorter. The dynamical friction time-scale is defined following \citet{Lacey_1993} via
\begin{equation}
    t_\mathrm{df} = \frac{f(\epsilon)}{2 B(v_c/\sqrt{2\sigma})}\frac{\sqrt{2}\sigma \, r_c^2}{Gm \ln \Lambda},
\end{equation}
with $r_c$ being the radius of a circular orbit with the same energy of the particle, $\sigma$ the stellar velocity dispersion inside $r_c$, and $v_c$ the circular velocity at $r_c$. The Coulomb logarithm is $\ln\Lambda = \ln(1+M(r_c)/m)$, with $M(r_c)$ the total mass inside $r_c$, and $B(x)\equiv \mathrm{erf}(x) - 2x\,\exp(-x^2)/\sqrt{\pi}$. Finally, $\epsilon = J/J_c$ is the circularity parameter defined as the angular momentum relative to the angular momentum of a circular orbit of the same energy, while $f(\epsilon)=\epsilon^{0.78}$ defines the orbital eccentricity of the SC \citep{Lacey_1993}. This time-scale is estimated at every snapshot of the simulation data products and identified disrupted clusters are assigned a mass of zero from that snapshot onward.

\section{The AuriGLOBES suite of simulations}\label{sec:simulationsuite}
In this work, we applied the AuriGLOBES sub-grid star cluster formation and evolution model to a subset of the simulated halos from the Auriga suite. The model is run at the baryonic mass resolution of $M_\mathrm{bary}\sim 5 \times 10^4 \mathrm{M}_\odot$ and comoving softening length of $\epsilon_\mathrm{comov}\sim750$~cpc up to a maximum physical softening length of $\epsilon_\mathrm{phys}\sim375$~pc, referred as the level~4 resolution.

\subsection{Auriga re-simulations with AuriGLOBES}\label{sec3.1}
Firstly, we explore the effect of model choices on the resulting SC populations. For this we introduce six different model variations that we run in one of the halos, labelled as halo 6 in Table 3 in \citet{Grand_2024}. This is a disc galaxy with a stellar mass of $\sim 5 \times 10^{10} \, \mathrm{M}_\odot$, similar to the estimated stellar mass of the MW, i.e. $\sim 2-6\times 10^{10} \, \mathrm{M}_\odot$ \citep{Licquia_2015, Lian_2025}. Starting from our \texttt{fiducial} AuriGLOBES (see Sect.~\ref{sec2:globes}), we ``turn off'' part of the model choices and analyse their effect on the SCs formation and evolution.

All the model variations, summarised in Table~\ref{tab:modelvariations}, are: \textit{(i)} \texttt{fiducial} which includes all the physics described in Sect.~\ref{sec2:globes}; \textit{(ii)} \texttt{no BH heating} where we do not apply the mass loss enhancement due compact object heating; \textit{(iii)} \texttt{no compressive criterion} where cluster formation is not constrained to tidally compressive regions only; and a \textit{(iv)} \texttt{no added physics} model turning off both the BH heating and tidally compressive formation.
In addition, we repeated the \texttt{fiducial} model with two final variations: \textit{(v)} \texttt{enhanced shocks} arbitrarily enhancing the mass loss from tidal shocks by a factor of 10 during the full evolution of SCs; and \textit{(vi)} \texttt{no size evolution} keeping the initial MRR but turning the half-mass radius evolution off.
The purpose of these variations is to further test how the model is capturing the tidal disruption of SCs with an emphasis on the early disruption from their natal environments \citep{Kruijssen_2012a} and the role of their half-mass density in their dynamical evolution.

Secondly, we ran a suite of simulations comprising 9 halos with our \texttt{fiducial} model. These simulated halos, in the $1\times10^{12} \mathrm{M}_\odot <M_{200}(z=0)<2\times10^{12} \mathrm{M}_\odot$ mass range, are summarised in Table \ref{tab:fiducialruns}. Thus covering a mass range similar to the MW and M31 systems with a variety of accretion and assembly histories that can impact the produced SC and GC populations. We note that the overall halo properties are slightly different compared to those listed in \citet{Grand_2024}, which is attributed to phenomena such as the Butterfly effect \citep{Genel13, Keller_2019, Pakmor_2025a}.

\begin{table}
\centering
\caption{Summary table for the halo 6 model variations. }
\label{tab:modelvariations}
\begin{tabular}{ccc}
\hline \hline
Model  & Formation & Mass loss \\
\hline
Fiducial  & \makecell{Compressive \\ environments} & BH enhanced \\
No BH heating & \makecell{Compressive \\ environments} & \makecell{Canonical \\ evaporation} \\
\makecell{No compressive \\ criterion} & All environments & BH enhanced \\
No added physics  & All environments & \makecell{Canonical \\ evaporation} \\
\hline
\multicolumn{3}{c}{Additional variations} \\
Model & \multicolumn{2}{c}{Change description} \\
\hline
Enhanced shocks & \multicolumn{2}{c}{Tidal shocks boosted by 10} \\
No size evolution  & \multicolumn{2}{c}{Keep initial MRR, $\mathrm{d}r_\mathrm{h}/\mathrm{d}t = 0$} \\
\hline
\end{tabular}

\raggedright
\textit{Notes.} The columns are the model variation identifier (i), and the
model variation description in (ii) and (iii).
\end{table}

\begin{table}
\centering
\caption{Summary table for the rerun halos with the AuriGLOBES model.}
\label{tab:fiducialruns}
\begin{tabular}{ccccc}
\hline \hline
Halo & \makecell{$M_{200}$ \\ $(10^{10} \mathrm{M}_\odot)$} & \makecell{$R_{200}$ \\ (kpc)} & \makecell{$M_\ast$ \\ $(10^{10} \mathrm{M}_\odot)$} & \makecell{$M_{\mathrm{GCs}}$ \\ $(10^{6} \mathrm{M}_\odot)$} \\
\hline
 6  &  104.7 & 214.1 &  5.1 & 197.6 \\
14  &  157.6 & 245.3 & 11.8 & 248.3 \\
18  &  120.8 & 224.5 &  9.4 & 266.6 \\
22  &  100.9 & 211.5 &  7.0 &  66.3 \\
23  &  156.7 & 244.9 &  9.3 & 258.5 \\
24  &  155.9 & 244.4 &  9.5 & 316.3 \\
26  &  153.6 & 243.2 & 10.4 & 455.8 \\
27  &  178.3 & 255.6 & 10.7 & 389.9 \\
28  &  156.2 & 244.6 & 10.6 & 273.8 \\
\hline
Milky Way & $108^{+20}_{-14}$ & -- & $2.6\pm0.4$ & 29.7 \\
M31 & $250\pm30$ & -- & $10.3^{+2.3}_{-1.7}$ & 228.9 \\
\hline
\end{tabular}

\raggedright
\textit{Notes.} 
The columns are (i) halo identifier, 
(ii) total dynamical mass, (iii) halo radius, (iv) stellar mass, 
(v) total, cumulative, mass in GCs (see Sect.~\ref{sec3.2:gcselection}) older than 8~Gyr.

The simulations are run at the resolution level~4, i.e. $M_\mathrm{bary}\sim 5 \times 10^4 \mathrm{M}_\odot$ and $\epsilon_\mathrm{phys}\sim375$~pc. We include the properties of the MW and M31 at the bottom for comparison. The values quoted for the MW are: $M_{200}$ from \citet{Cantun_2020}, $M_\star$ from \citet{Lian_2025}, and $M_\mathrm{GCs}$ as calculated from \citet{Harris_1996, Harris_2010} catalogue; and for M31: $M_{200}$ from \citet{Villanueva_Domingo_2023}, $M_\star$ from \citet{Sick_2015}, and $M_\mathrm{GCs}$ as calculated from \citet{Caldwell_2016} catalogue.
\end{table}

\subsection{Globular cluster selection}
\label{sec3.2:gcselection}
\begin{figure}
    \centering
    \resizebox{\hsize}{!}{\includegraphics{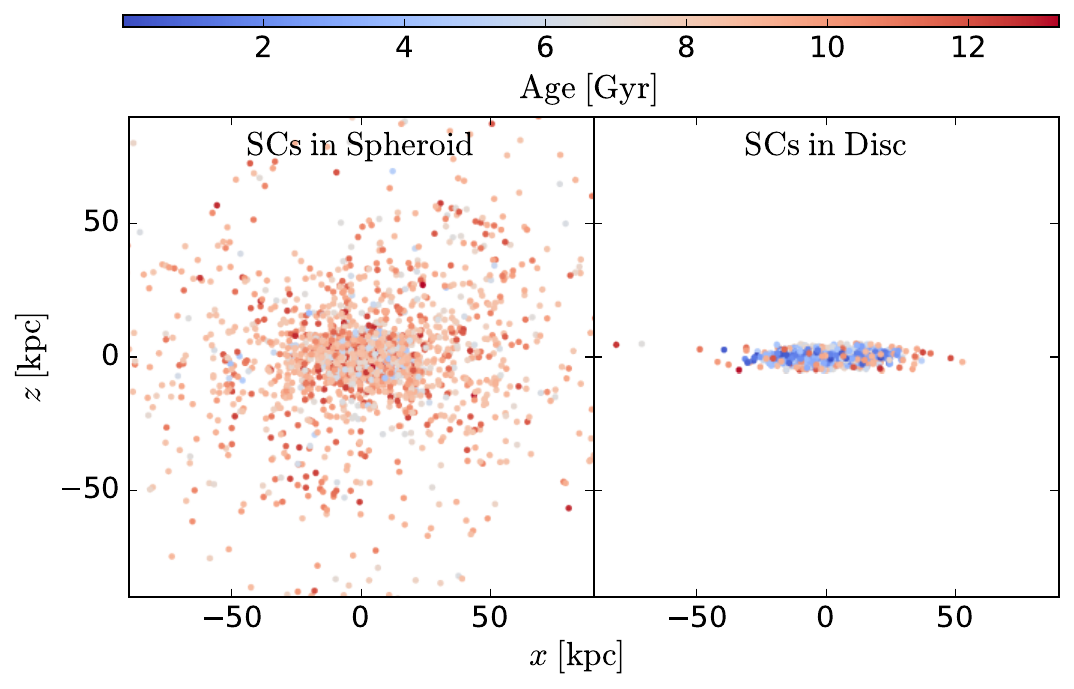}}
    \caption{Spatial distribution of the modelled SCs in the halo 6 colour coded by their age in spheroidal and disc kinematic components. \textit{Left}: SCs identified to be in the spheroidal, i.e. bulge and halo, component. \textit{Right}: SCs identified to be in the disc component. SCs compatible with old GCs are generally scattered around the bulge and halo while younger SCs are generally concentrated towards the disc.}
    \label{fig:spatial_age}
\end{figure}

The described formation and evolution model corresponds to SCs in general and so, to turn our attention to GCs specifically, we needed to perform a physically motivated candidate selection. Based on the fact that GCs are products of long term evolution and survival of SCs that escaped disruptive environments and now populate the bulge and halo of galaxies, we can restrict our GC selection to SCs with kinematic signatures resulting in a spheroidal spatial distribution.

For this purpose, we performed a kinematic decomposition through a Gaussian mixture model (GMM) fit for a two component (spheroidal and disc) system over a three-dimensional space defined by the circularity parameter\footnote{In this case, the circularity parameter is defined as the $z$-component of the angular momentum over the maximum angular momentum allowed for the orbital energy of the star particle, $\epsilon = L_z/L_{z,\mathrm{max}}(E)$, such that circular orbits in the direction of galactic rotation have $\epsilon=1$, counter-rotating orbits have $\epsilon<0$, and non-circular orbits have $\epsilon \sim 0$.}, $\epsilon$, orbital energy, $E$, and logarithm of the height coordinate of the star particles, $\log |z|$.
We plot the spatial distribution resulting from the decomposition in Fig.~\ref{fig:spatial_age}. The distribution corresponds to the simulated SCs of halo 6 at redshift $z=0$ and is colour coded by age.
We show in the left and right panels the SCs corresponding to the spheroidal and disc components, respectively. The SCs located in the disc component (right panel in Fig.~\ref{fig:spatial_age}) are generally young, consistent with the observed distribution of YSCs and open clusters. In contrast, the spheroidal components (left panel in Fig.~\ref{fig:spatial_age}) host SCs with generally older ages, in line with GCs being located mainly in the bulges and halos of galaxies.

Finally, we further constrain our selection to SCs with current masses $m>10^4 \, \mathrm{M}_\odot$, well below completeness limits of extragalactic studies \citep{Jordan_2007}, and comparable to the MW and M31 GCMFs completeness limit \citep{Harris_1996, Harris_2010, Caldwell_2016}. This limit also restricts our selection to SCs less affected by a potential underestimation of their tidal shocks disruption.

\section{Simulated star clusters in AuriGLOBES}\label{sec:scpopulations}

In the following subsections, we describe the general properties of our simulated SC populations. We also discuss the effects resulting from the different model choices.

\subsection{Mass functions under different model variations}
\label{results:variation_mfs}
\begin{figure}
    \centering
    \resizebox{0.95\hsize}{!}{\includegraphics{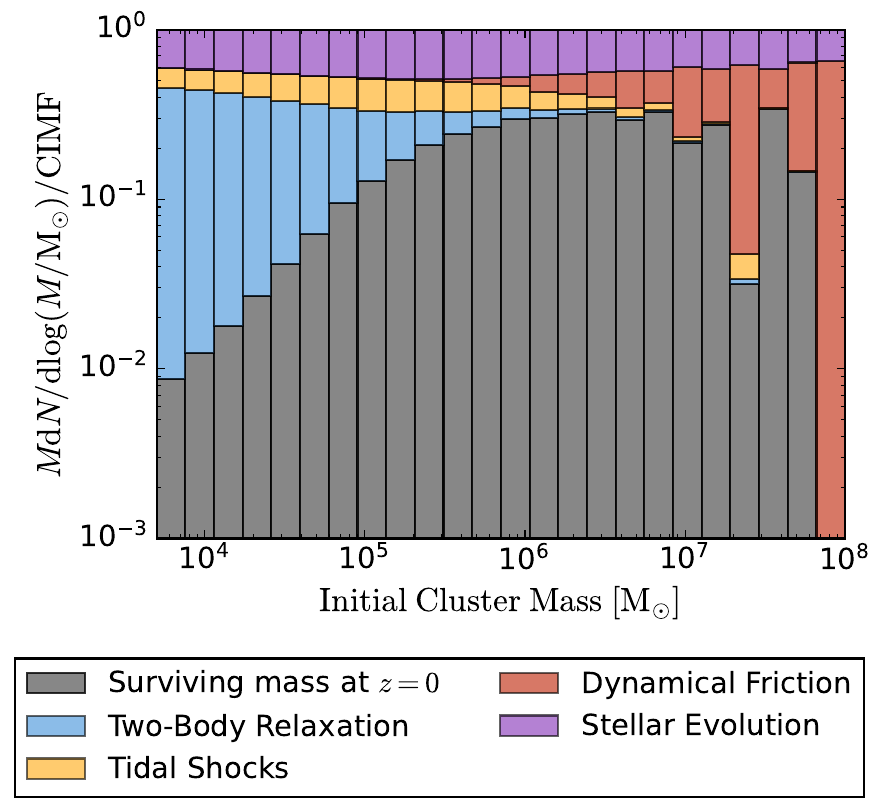}}
    \caption{Relative importance of two-body relaxation, tidal shocks, and dynamical friction disruption processes in the transformation from the CIMF to the evolved SC MF. We show the total surviving mass and mass lost by each process in each bin in initial SC mass, normalised by the CIMF.}
    \label{fig:cimf_to_gcmf}
\end{figure}

\begin{figure*}
    \centering
    \includegraphics[width=18cm]{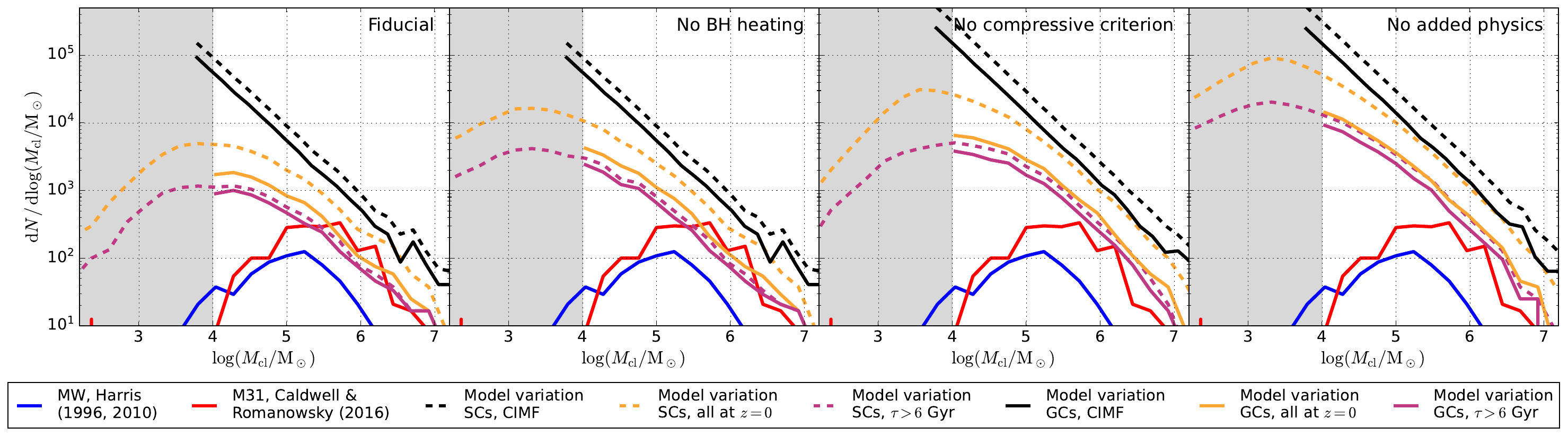}
    \caption{Halo 6 cluster mass functions for the variations of the star cluster model choices (see Table~\ref{tab:modelvariations}). From \textit{left} to \textit{right}, the cluster mass functions for: \texttt{fiducial} model with all the physics described in Sect.~\ref{sec2:globes}, \texttt{no BH heating} variation of the SCs evolution, \texttt{no compressive criterion} variation of the SCs formation, and the \texttt{no added physics} base model. The MFs for all SCs and for the GC selection (see Sect. \ref{sec3.2:gcselection}) are included for all the model variations, as well as the CIMFs. For comparison, the observed GCMFs from the MW and M31 are included \citep[as obtained from][]{Harris_1996, Harris_2010, Caldwell_2016}. The shaded region in all the panels denotes the SC mass range that is excluded in our GC selection.}
    \label{fig:allmodels_mfs}
\end{figure*}

\begin{figure}
    \centering
    \resizebox{0.92\hsize}{!}{\includegraphics{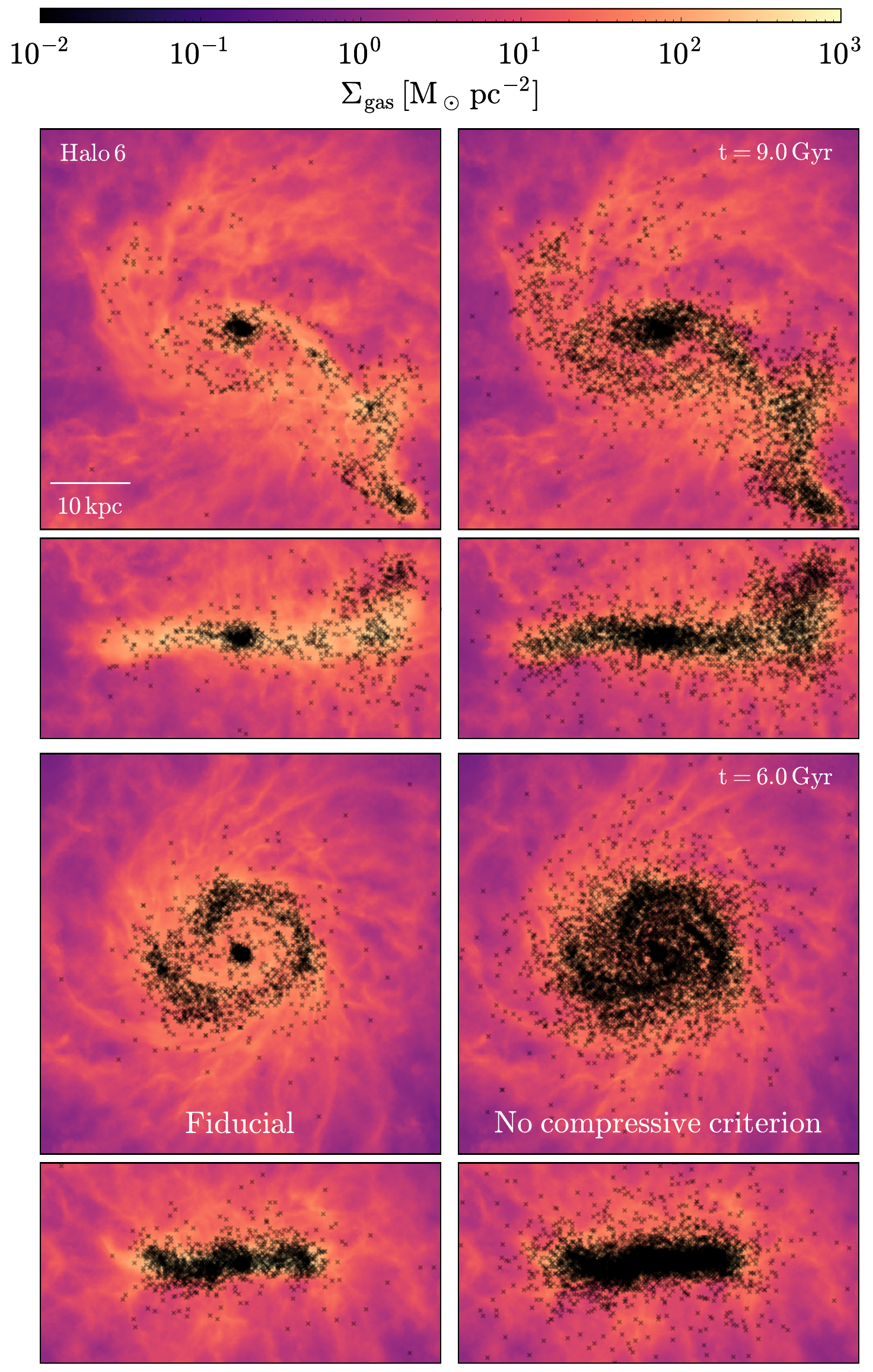}}
    \caption{Face-on and edge-on gas surface density projections of halo 6 at 9~Gyr (\textit{top}) and 6~Gyr (\textit{bottom}) of lookback time for two different model choices for SCs formation. The location of newly formed SCs, i.e. $\tau \lesssim 200 \, \textrm{Myr}$, is highlighted with black crosses. \textit{Left:} \texttt{Fiducial} model with SCs forming only in tidally compressive gas. \textit{Right:} \texttt{no compressive criterion} where SCs formation is not constrained to these environments only.}
    \label{fig:scs_birthloc_constrain}
\end{figure}

As a first step, we explore the effect of modifying the cluster formation and evolution model. It is instructive to first see the relative importance of each of the disruption mechanisms when transforming the CIMF into the evolved SC MF in our model. We show this in Fig.~\ref{fig:cimf_to_gcmf}, where we plot the fraction of cluster mass lost via different mechanisms for each initial cluster mass bin. The relative importance of each mechanism is mass dependent: two-body relaxation is the main disruption mechanism for low initial SC masses ($m_0<10^5\,\mathrm{M}_\odot$), while dynamical friction is the major disruption mechanism at the high mass end ($m_0\gtrsim10^6\,\mathrm{M}_\odot$). Tidal shocks and two-body relaxation play a comparable role in the $10^5<m_0/\mathrm{M}_\odot<10^6$ mass range. This validation mainly demonstrates that, in our model, the main disruption processes are two-body relaxation and dynamical friction. Tidal shocks have a similar, less important, disruption contribution at most scales.

Next, we compared the evolved MFs of the simulated SCs for the first four model variations of halo 6, as shown in Fig.~\ref{fig:allmodels_mfs}. For comparison, the GCMFs for the MW and M31 are included \citep[as obtained from][]{Harris_1996, Harris_2010, Caldwell_2016}.

We first describe the behaviour of the MFs of our \texttt{fiducial} model (leftmost panel), which includes both the formation constraint to tidally compressive environments and the enhanced mass loss due to compact object heating. Although the MFs have their peak around an order of magnitude lower than the observational GCMFs (at $10^4 \, \mathrm{M}_\odot$, coinciding with our lower mass cut for the GC selection) and they retain an extended tail towards lower masses, they are consistent with a transformation of an initial Schechter CIMF into a lognormal shaped GCMF, as expected from the long-term evolution of SCs \citep[also demonstrated by][]{Elmegreen_2010a, Pfeffer_2018, Reina_Campos_2022}. In addition, the simulated GCMFs are consistent with the observed ones for masses above $10^5\,\mathrm{M}_\odot$, especially for the $\tau > 8$~Gyr age cut, which matches the M31 GCMF. This behaviour highlights the importance of combining both internal and external cluster evolution processes when modelling the long-term evolution of cluster populations.

To assess the role of each ingredient in shaping the MF, we progressively removed elements of the fiducial model. The effect of removing the compact object heating correction on the two-body relaxation mass loss can be seen in the centre-left panel (\texttt{no BH heating} model). Here, the long-term evolution of the cluster population becomes significantly less efficient in removing low-mass clusters. As a consequence, the modelled MFs retain a prominent population of evolved low-mass SCs and the peak of the distribution shifts towards lower masses compared to the \texttt{fiducial} case. This behaviour is similar to the MF reported in E-MOSAICS \cite[figs. 16 and 17 in][]{Pfeffer_2018}. This demonstrates that enhanced mass loss driven by the retained compact object population can play a significant role in accelerating cluster evaporation and shaping the evolved MF \citep[as in][]{Gieles_2023}.

If, instead, the constraint to form SCs only in tidally compressive environments is removed (\texttt{no compressive criterion}, centre-right panel), there is a substantial increase in the number of clusters formed. Consequently, the normalisation of both initial and evolved MFs is higher in comparison to the \texttt{fiducial} model and the evolved MFs exceed the observations at all cluster masses, even for our GC selection at the old age cut. The peak of the old SCs MF is similar to the \texttt{fiducial} case although it is at lower masses in the case of all the SCs at $z=0$. 

Finally, when both ingredients are removed (\texttt{no added physics}, rightmost panel), the model reverts to a simpler prescription for cluster formation and evolution, similar to E-MOSAICS but with the different choice for the initial half-mass radius and its evolution. In this case, the MFs again lie well above the observed GCMFs at all masses and the peak is nearly two orders of magnitude below the observed one.

The excessive number of clusters formed in the last two scenarios suggests that in the absence of additional physical constraints, the model allows for SCs to form in regions not favourable for their bound system nature. In particular, tidally compressive gas environments are known to favor the formation of gravitationally bound clusters \citep{Renaud_2014, Renaud_2015}. We further illustrate the impact of enforcing this condition in Fig.~\ref{fig:scs_birthloc_constrain}, where we plot the gas surface density and the location of the newly formed SCs ($\tau \lesssim 200 \, \textrm{Myr}$) at two evolutionary epochs of halo~6. These, around 9 and 6~Gyr of lookback time, correspond to two episodes of sustained high star formation, with the $t=9$~Gyr being merger induced. We show this for the \texttt{fiducial} model (left column) and the \texttt{no compressive criterion} variation (right column). Although the number of formed clusters for the \texttt{fiducial} model is significantly smaller, the formation constraint ensures that it happens in high density gas regions, e.g. central region and spiral arms. For the \texttt{no compressive criterion} variation, cluster formation is widespread throughout the galaxy, happening also in low surface density regions. Consequently, ensuring a compressive environment for SCs formation reduces their overall formation rate and brings the normalisation of the MFs closer to the observational constraints in Fig.~\ref{fig:allmodels_mfs}. Taken together, these comparisons indicate that both the environmental regulation of cluster formation through the implementation of the tidally compressive criterion and the enhanced dynamical mass loss associated with compact object remnants play an important role in reproducing the observed properties of GC systems in our model.

\begin{figure}
    \centering
    \resizebox{0.95\hsize}{!}{\includegraphics{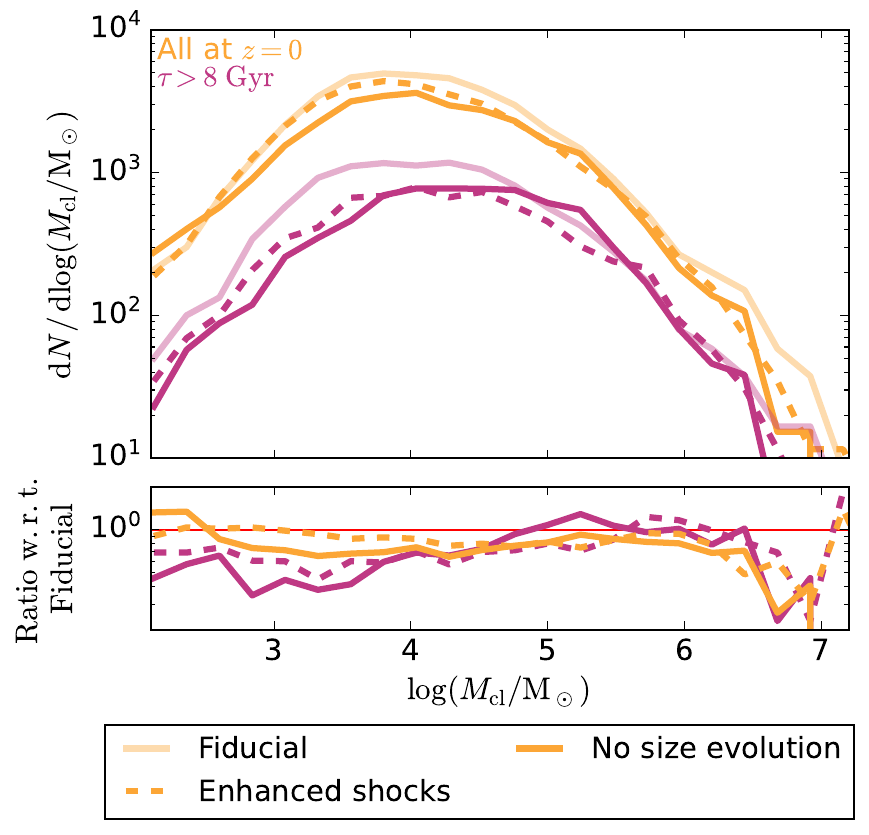}}
    \caption{\textit{Top}: Halo 6 all SCs MFs for the \texttt{enhanced shocks} (boosting tidal shocks by a factor of 10) and for the \texttt{no size evolution} (without evolving the clusters half-mass radius) variations in comparison to the \texttt{fiducial} model (see Table~\ref{tab:modelvariations}). The colours correspond to MFs at different age cuts, as described in the top left corner of the figure. \textit{Bottom:} ratio of the MFs with respect to the correspondent MF from the \texttt{fiducial} model.}
    \label{fig:fiducialvariations}
\end{figure}

We now move to the last two variations (i.e. \texttt{enhanced shocks} and \texttt{no size evolution}) and present their MFs in Fig.~\ref{fig:fiducialvariations} comparing them to those of the \texttt{fiducial} model. With the \texttt{enhanced shocks} variation we aim to see the effect of more disruptive shocks that could be generated from the inclusion of a more sophisticated treatment of the ISM, given that tidal shocks disruption play a minor role in our model. The effect of boosting the tidal shocks mass loss by 10 is evidenced in lowering the normalisation of the MFs in Fig.~\ref{fig:fiducialvariations}. The peak of the MFs, however, remains at the same location as that of the \texttt{fiducial} model and the old SCs MF retains the extended tail towards low masses. We tested a higher boost factor and the behaviour is a further decrease in normalisation but with the same extended tail to low masses (see Appendix~\ref{app:shocks}).

Finally, we check the effect of not including the half-mass radius evolution model, i.e. eq.~\ref{eq:size_evo}, on the SCs disruption and we also compare the produced MFs in Fig.~\ref{fig:fiducialvariations}. Interestingly, disruption becomes more efficient with a higher depletion at the low mass tail of the MFs, the GCMF has a similar normalisation as the  \texttt{enhanced shocks} variation but the peak is shifted to $\approx5\times10^4 \, \mathrm{M}_\odot$. This result is in agreement with \citet{Gieles_2016} and \citet{Reina_Campos_2023} stating that the competition between expansion from two-body relaxation and shrinking from tidal shocks leads to an equilibrium MRR where SCs have a higher resilience to tidal disruption. With no size evolution, disruption is more effective, leading to an old SCs MF more similar to observations, in agreement with the results of \citet{Reina_Campos_2022}. In our case, this is driven by the combined effect of the enhanced mass loss from two-body relaxation and the external environmental disruption.

\subsection{Numerical convergence of the mass functions}
\label{resolution_test}
\begin{figure}
    \centering
    \resizebox{0.95\hsize}{!}{\includegraphics{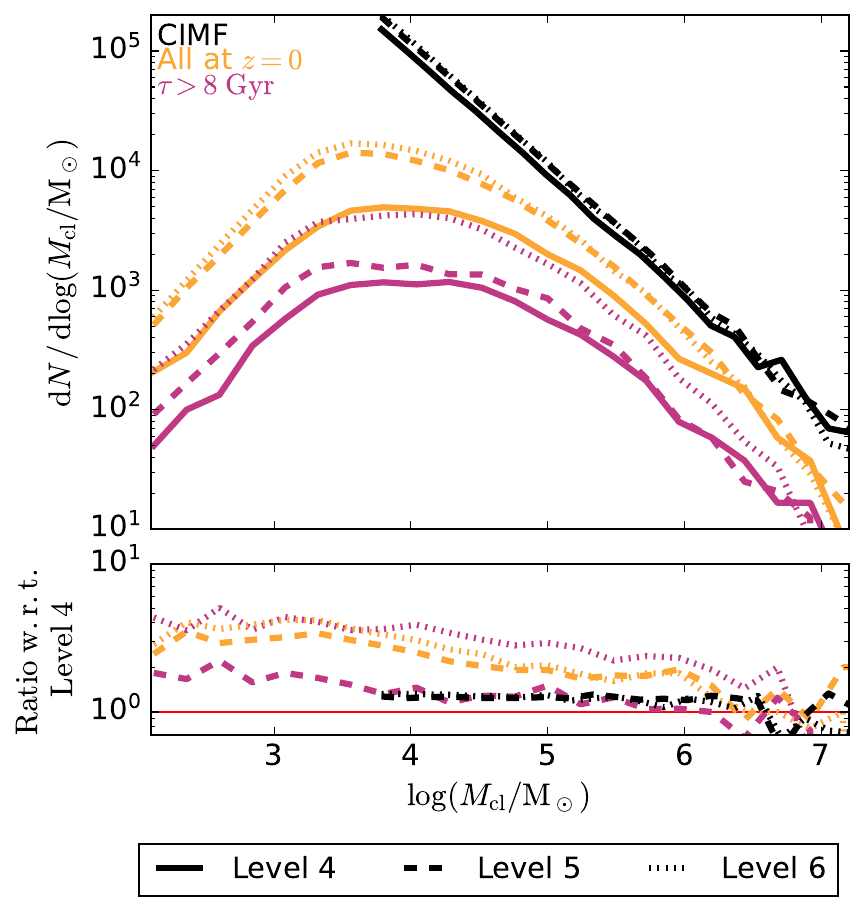}}
    \caption{\textit{Top}: Halo 6 all SCs MFs for the fiducial run at three resolution levels: level~4, level~5, and level~6. The colours correspond to MFs at different age cuts, as described in the top left corner of the figure. \textit{Bottom:} ratio of the MFs with respect to the correspondent MF at level~4.}
    \label{fig:res_test}
\end{figure}

We go on to assess the numerical convergence of our \texttt{fiducial} model by comparing the cluster MFs of halo 6 across three different resolutions: the fiducial resolution level~4, and two lower resolutions, namely, level~5 and level~6, with baryonic mass resolution of $\sim 4 \times 10^5 \, \mathrm{M}_\odot$ and $\sim 4 \times 10^6 \,\mathrm{M}_\odot$, respectively. We present the MFs of all SCs at each resolution in the top panel of Fig.~\ref{fig:res_test}, while in the lower panel we show the ratio of level~5 and level~6 MFs with respect to those of level~4. The CIMFs for the three simulations are nearly identical across resolutions. This is partially an effect of the tidally compressive formation constraint making the cluster formation histories comparable across the three resolutions, even though their star formation histories are markedly different \citep[behaviour across resolutions explained in][]{Grand_2021, Pakmor_2025a}.

The evolved MFs show differences across resolution that cannot be explained by the differences in the CIMF, given that the latter shows negligible change. Instead, they show an interesting systematic trend: as resolution increases, tidal shocks disruption becomes more effective which leads to further depletion of the low mass tail of the MFs. Furthermore, the peak appears to slightly shift to higher masses as resolution increases, again owing to more effective disruption of evolved clusters at higher resolution. Finally, the MFs for the $\tau>8$~Gyr cut in Fig.~\ref{fig:res_test} show a convergent behaviour at the level~4 and level~5 resolutions, with the larger differences being below the $m>10^4\, \mathrm{M}_\odot$ mass cut selection for our GC candidates. This effect is driven by an increased ability to capture tidal shocks (e.g. see Appendix~\ref{app:shocks_capture}) and a deeper galaxy potential which act to enhance SC disruption.

\subsection{Fiducial globular cluster mass functions}
\begin{figure}
    \centering
    \resizebox{0.95\hsize}{!}{\includegraphics{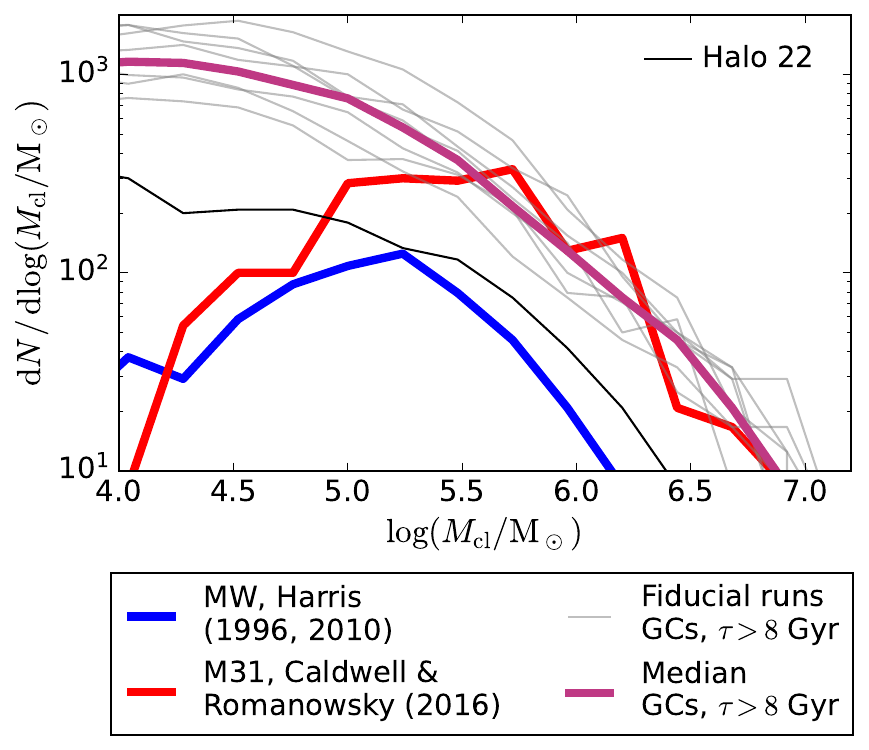}}
    \caption{GCMFs for the \texttt{Fiducial} model of the 9 rerun halos, and their median GCMF, at the $\tau=8$~Gyr age cut. As in Fig.~\ref{fig:allmodels_mfs}, the observed GCMFs for the MW and M31 are included for comparison. The halo with the most similar GCMF to the MW for this age cut is highlighted with black continuous line.}
    \label{fig:fiducial_gcmf}
\end{figure}
Having motivated our \texttt{fiducial} model choice, we now present the GCMFs corresponding to the $\tau > 8$~Gyr age cut of all the re-simulated halos in Fig. \ref{fig:fiducial_gcmf}. We also show the corresponding median GCMF. There is a consistent general behaviour for the GCMFs for their shape and peak location. Their differences primarily arise from the inherent normalisation coming from the differences in stellar mass and star formation histories (SFHs) between the systems (see Sect..~\ref{sec3:age_met}). The median GCMF is compatible with observational constrains for masses above $10^5\,\mathrm{M}_\odot$ and showing more similarity with M31 than with the MW population. This is explained as the median age for the observed GCs in the MW is older, at $\approx12$~Gyr \citep{Kruijssen_2019c}, than that of M31, i.e. $\approx10.7$~Gyr \citep{Usher_2024}. Finally, we single out the GCMF of halo 22 as the one that most resembles that of the MW. In particular, this halo was identified by \citet{Fattahi_2019} as one of the MW-analogue halos with signatures of an early accretion event similar to the Gaia-Enceladus-Sausage (GES) merger for the MW \citep{Belokurov_2018, Helmi_2018}. This connection is particularly relevant as early, massive accretion events are thought to contribute a significant fraction of the GC population and to shape the age and metallicity distributions of GCs \citep[e.g.][]{Kruijssen_2019c, Massari_2019, Newton_2025}.

\subsection{Age distributions}\label{sec3:age_met}
\begin{figure*}
    \sidecaption
    \includegraphics[width=12cm]{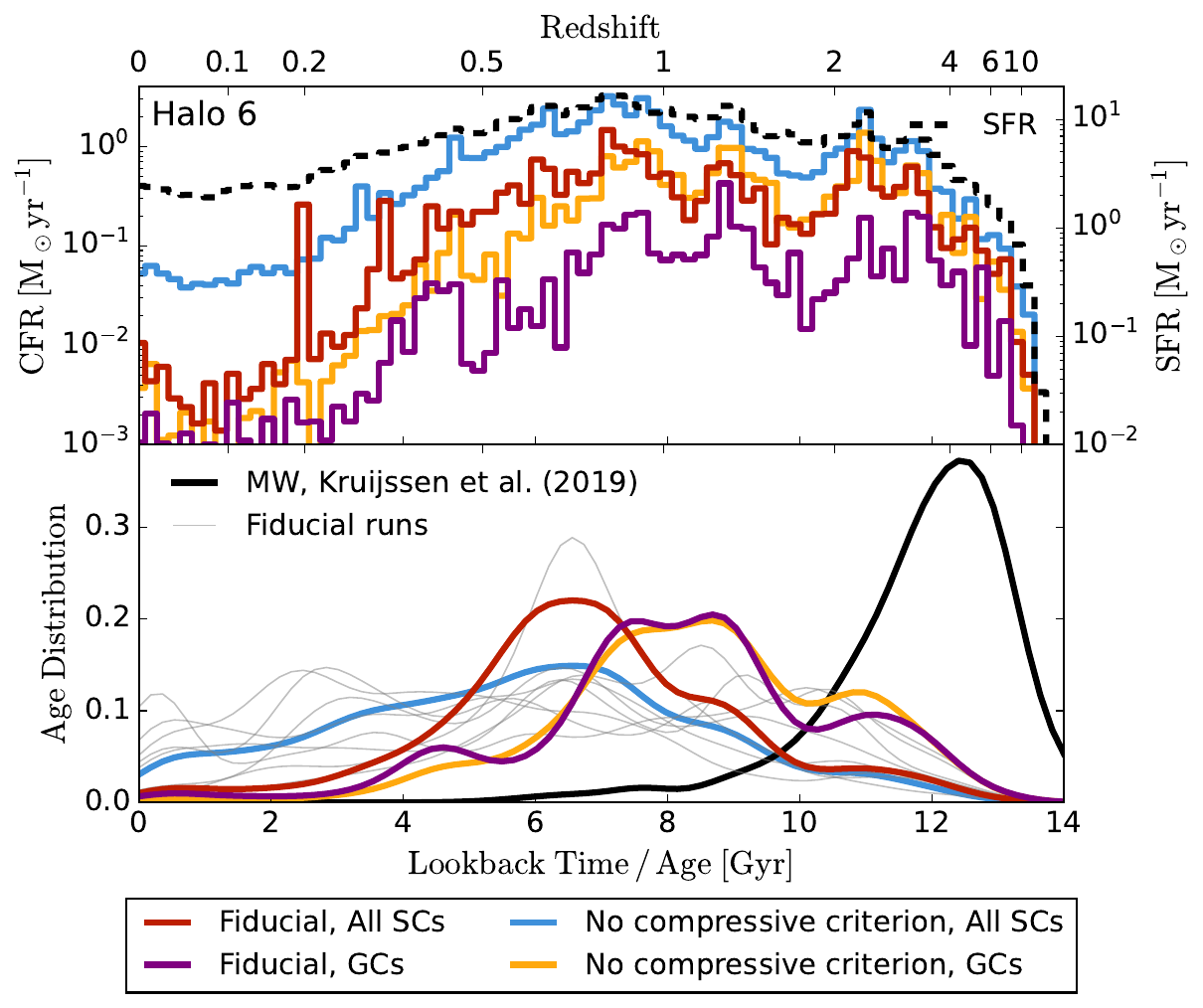}
    \caption{\textit{Top}: All SCs and GC selection cluster formation rate of halo 6 for the \texttt{fiducial} and \texttt{no compressive criterion} variations. The star formation history of the halo is included for comparison. \textit{Bottom}: Age distribution of surviving SCs and GC selection for the same model variations. The MW GCs age distribution is included for comparison, corresponding to the average GCs ages and uncertainties compiled by \citet{Kruijssen_2019c}. Included too are the age distributions for all the rerun halos (thin gray lines), exemplifying the variability of cluster age distributions from halo to halo. The tidal formation constraint heavily impacts the formation rate of SCs while retaining the age distribution of the GC selection.}
    \label{fig:cfr_age}
\end{figure*}

To further characterise the effect of the tidally compressive formation constraint over the formation history of SCs, we plot the cluster formation rate (CFR) and age distribution for all the SCs and for the GCs for the halo 6 in the \texttt{fiducial} and \texttt{no compressive criterion} models in Fig. \ref{fig:cfr_age}. We determined the CFR by binning the initial SC (GC selection) masses by age into 175~Myr intervals. For comparison, we include the star formation rate (SFR) of halo 6, determined by binning the initial stellar masses by age in the same age intervals as for the CFR.
The age distribution corresponds to the sum of normal distributions at the individual cluster ages, assuming a uniform age uncertainty of 0.5 Gyr, following \citet{Valenzuela_2025}. We note that the CFR corresponds to all the formed SCs, while the age distribution lines correspond to only those surviving until redshift $z=0$. In the case of GCs, both curves correspond to surviving populations, given our $m>10^4 \, \mathrm{M}_\odot$ selection. The age distribution of the MW GCs is also included for comparison and was constructed with the same methodology using, in this case, the available uncertainties of the estimated ages. The ages, and their uncertainties, come from the \citet{Kruijssen_2019c} compilation catalogue, averaging GC ages and uncertainty estimates from different studies \citep{Forbes_2010, Dotter_2010, Dotter_2011, VandenBerg_2013}.

The compressive formation constraint has a big impact in inhibiting SC formation while maintaining starburst periods and following the overall SFR of the halo. The biggest impact is at low redshift where the SC formation is heavily reduced. This turns into an age distribution favouring older populations, as seen in the lower panel of Fig.~\ref{fig:cfr_age}. The same effect is visualised for the GC candidates CFR, retaining, however, mostly the same age distribution. This demonstrates that the tidally compressive constraint is successful not only in reducing SC formation but also in retaining the formation of potential GC candidates.

We also include the age distributions for all the fiducial runs in the lower panel of Fig.~\ref{fig:cfr_age}. This illustrates how the variety of assembly histories for the different halos drives different age distributions of their GC populations. However, we note a qualitative mismatch between the simulated distributions and that of the MW. As seen in the lower panel, the empirical distribution is dominated by a peaked, ancient population at $t\approx12$~Gyr, and an almost complete truncation towards surviving clusters at younger ages. In contrast, our simulated populations exhibit much broader, continuous age distributions peaking at intermediate ages ($t\approx 6\,\textrm{--}\,9$~Gyr). This can be partially attributed to the ICs of the simulated halos being not similar enough to the MW and its assembly history, or a consequence of the underlying Auriga galaxy formation model not matching the high SFR of the MW \citep{Snaith_2015} at very high redshift ($z>5$), which delays the simulated peak of SC formation. However, the continuous tails towards younger surviving clusters could imply that significant physical ingredients could be missing from the sub-grid modelling of SCs. Specifically, this points to either a potential underestimation of their disruption, or insufficient suppression of their formation in recent star formation periods. This qualitative mismatch serves as an interesting constraint as it underscores that the precise assembly history of the MW provided a highly specific environment for GC formation and survival that remains challenging to replicate with current sub-grid models.

\begin{figure}
    \centering
    \resizebox{0.95\hsize}{!}{\includegraphics{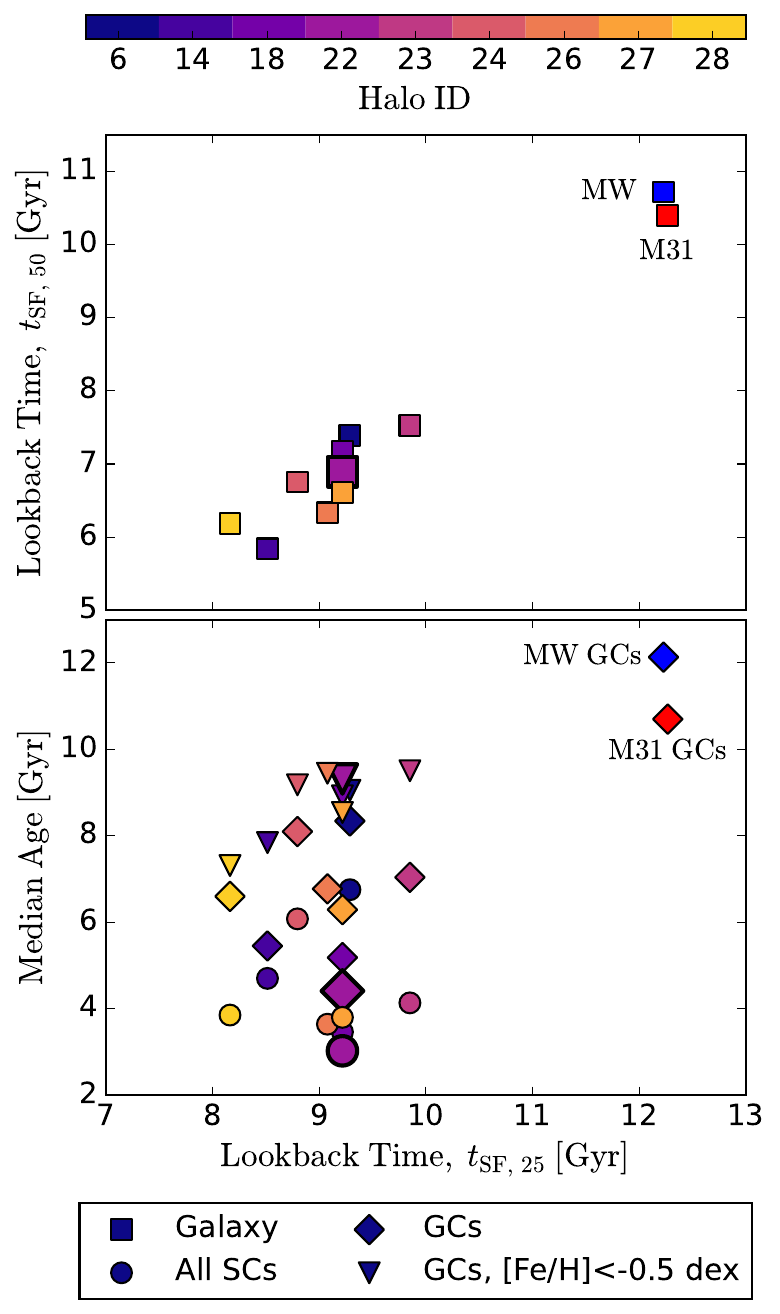}}
    \caption{Individual galaxy stellar mass assembly characteristic timescales and their SC (GC) population median age. \textit{Top}: timescale for each galaxy to form 25 and 50 per cent of its total stellar mass. The timescales for the MW and M31 are included for comparison as calculated from their estimated SFH in \citet{Snaith_2015} and \citet{Williams_2017}, respectively. \textit{Bottom}: median age of their SC/GC system in comparison to the timescale for forming 25 per cent of their total stellar mass. The GC median age for the MW GC system is calculated from \citet{Kruijssen_2019c} while the median age for the M31 GC system is the value quoted in Table 2 of \cite{Usher_2024}. Halo 22, and its SC and GC populations, are highlighted with a larger size marker.}
    \label{fig:tsfr_age}
\end{figure}

We further analyse the diversity of the halos assembly histories and their imprint on the simulated age distributions in Fig.~\ref{fig:tsfr_age} where we show the lookback times at which the host galaxies assembled 25 and 50 per cent of their total stellar mass, $t_\mathrm{SF,25}$ and $t_\mathrm{SF,50}$ (top panel) and the median ages of their SC and GC populations in comparison to $t_\mathrm{SF,25}$ (bottom panel). For comparison, the corresponding timescales and median GC age for the MW and M31 systems are included. The 25 and 50 per cent of star formation timescales where calculated from the estimated SFHs in \cite{Snaith_2015} for the MW and in \cite{Williams_2017} for M31. The GC system median age for the MW is calculated from \citet{Kruijssen_2019c} catalogue while the median age for M31 is the value quoted in Table 2 of \citet{Usher_2024}. Markedly, both the MW and M31 form 25 and 50 per cent of their stellar mass very early on: $t>12$~Gyr of lookback time for the 25 per cent, and $t>10$~Gyr for the 50 percent; with the MW having formed 50 percent of its stellar mass earlier than M31. In contrast, all of the simulated halos have these timescales much later: the halo with earliest star formation, halo 23, forms 25 and 50 per cent of its stellar mass at $t\approx 10$ and $t\approx7.5$~Gyr of lookback time, respectively. The median timescales for the simulated halos are $t\approx 9.22$ and $t\approx 6.75$~Gyr of lookback time. These quantities trace the early assembly history of the galaxies, allowing us to assess how the cluster formation relates to the global buildup of stellar mass.

In the lower panel of Fig.~\ref{fig:tsfr_age}, we show the median ages for all the SCs surviving at $z=0$ in circled markers, for the GC selection given in diamond markers, as well as for the GC selection plus a metallicity cut of $[\mathrm{Fe/H}]<-0.5$~dex, shown as the down-facing triangle markers. With this metallicity cut we aim to target the old population of our GC selection. There is an evident trend across the three selections such that galaxies that assemble their stellar mass earlier tend to host older SC and GC populations, albeit with some scatter. This scatter is driven by the continued formation of clusters at later times and is diminished when targeting the older populations with the metallicity cut. The MW and M31 GC systems, with median ages of $\approx12$~Gyr and $\approx10.7$~Gyr, respectively, confirm this trend. Therefore, this shows that a high SFR at the earliest stages of galaxy assembly is needed in order to obtain GC age distributions comparable to that of the MW or M31. We highlight again halo 22 here with bigger markers: it is one of the halos with an old GC population (after the metallicity cut) with median age close to the observed median age of MW. This is despite having a lower rate of star formation to the one of the MW, specially in the first 2--3~Gyr of evolution. Furthermore, the GC median age following this cut approximately coincides with the GES analogue merger of the halo. We show an extended comparison of the individual halos SFHs with their SC age distributions with respect to the MW in Appendix~\ref{app:age_distribution}. Overall, these results support the interpretation that GCs predominantly trace the early SFH of their host galaxies.

\subsection{Metallicity distributions}
\begin{figure}
    \centering
    \resizebox{0.95\hsize}{!}{\includegraphics{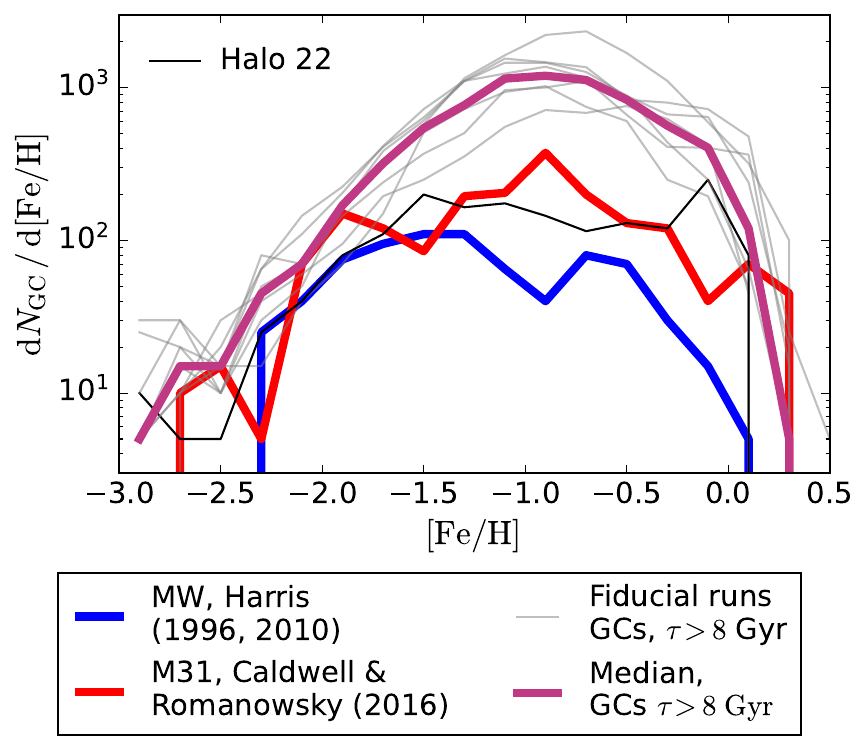}}
    \caption{Metallicity distribution functions of the old ($\tau>8$~Gyr) population of GCs of the fiducial runs, and their median distribution, compared to the MW and M31 MDFs \citep[extracted from][]{Harris_1996, Harris_2010,Caldwell_2016}. The halo with the most similar MDF to the MW for this age cut is highlighted with black continuous line.}
    \label{fig:metallicity}
\end{figure}

Tied to the age distribution of the GCs is their metallicity distribution. In Fig.~\ref{fig:metallicity}, we show the individual metallicity distribution functions (MDFs) of the fiducial runs at the $\tau > 8$~Gyr age cut, as well as the median simulated distribution. The GCs metallicities,  [Fe/H], correspond to the metallicity of their parent star particle\footnote{We apply a correction offset of $-0.5$~dex following \citet{Grand_2021} showing that Auriga galaxies are too metal-rich relative to the observed stellar-mass-to-stellar-metallicity relation \citep[e.g.][]{Kirby_2013}}. The MDFs from the MW and M31 GC populations are included in the figure for comparison \citep[as extracted from][]{Harris_1996, Harris_2010, Caldwell_2016}. The simulated MDFs have a general trend of being richer in metallicity compared to the MW. However, comparing to the M31 MDF, we can see that the simulated MDFs cover the same metallicity range, albeit with a higher number of GCs in the $-1.75$ to $0.3$ [Fe/H] range.

Similarly to the lower panel of Fig.~\ref{fig:cfr_age} (see also Fig.~\ref{fig:halo_age_dist}), the diversity of MDFs in Fig.~\ref{fig:metallicity} is a result of the different assembly histories of the simulated halos. Therefore, the median MDFs illustrate that the simulated clusters are more metal-rich because they are generally formed at later times compared to the MW and M31. Below [Fe/H]~$\approx -1.75$, all the simulated MDFs and their medians agree well with the MW and M31 MDFs and none of them show an important population below the `metallicity floor' of $[\mathrm{Fe/H}]=-2.5$ \citep{Beasley_2019, Kruijssen_2019d}. We note again that halo 22 gives the MDF that is most similar to that of the MW. Furthermore, it shows signatures of bimodality, albeit with a dip at a slightly higher metallicity of $[\mathrm{Fe/H}]\approx-0.7$. Overall, the simulated GC age and metallicity diversity highlight that GCs remain sensitive probes of galactic assembly.

\subsection{Spatial distribution}
\begin{figure}
    \centering
    \resizebox{0.95\hsize}{!}{\includegraphics{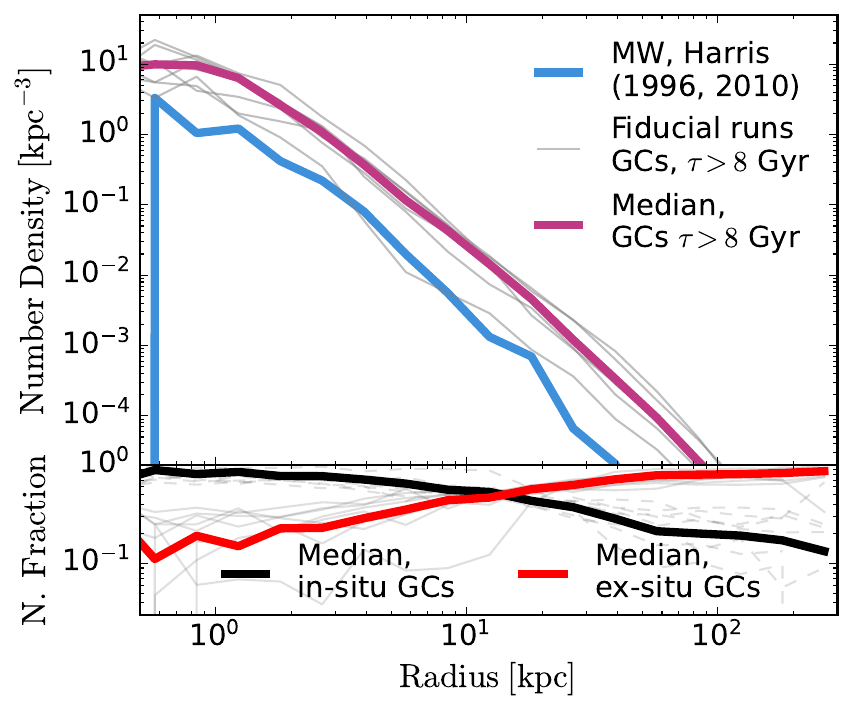}}
    \caption{\textit{Top:} Number density radial profile of the old ($\tau > 8$~Gyr) simulated GCs of the \texttt{fiducial} runs, in gray, in comparison to the distribution of GCs in the MW. \textit{Bottom:} Fraction of in situ (dashed gray lines) and ex situ (continuous gray lines) GC candidates at the same age cut as a function of the galactocentric radius.}
    \label{fig:in_ex_situ}
\end{figure}

In this section, we consider the spatial distribution properties of our simulated populations. As illustrated in Fig.~\ref{fig:spatial_age}, SCs consistent with younger populations are generally distributed along the disc component of the galaxies while older populations populate the dispersed spheroidal component (i.e. a bulge or halo). While the SCs distribution in Fig.~\ref{fig:spatial_age} is from halo 6 only, all the simulations summarised in Table \ref{tab:fiducialruns} have a similar behaviour with a median age difference for the SCs in both components of $\approx 2$~Gyr on average (see also Fig.~\ref{fig:tsfr_age}).

Focusing now in the spatial distribution of our GC selection only, we show in Fig.~\ref{fig:in_ex_situ} the number density radial profile of all the halos with the fiducial model in comparison to the profile of the MW GCs \citep[based on the distance estimates in][]{Harris_1996, Harris_2010}. The simulated GCs follow a similar profile across the galaxy with the difference in the normalisation reflecting the different stellar masses for the different halos. The median profile for the $\tau > 8$~Gyr age cut is a factor of a few higher than the MW profile with a consistent slope. Additionally, in the bottom panel of Fig.~\ref{fig:in_ex_situ} we include the distribution of in situ and ex situ GCs in the simulated populations quantified by their number fraction as a function of radial distance from the galactic centre. In-situ GCs are primarily in the inner regions, while ex situ GCs are located in the outer halo regions. However, there are both ex situ GCs at small radii, such as $R_\textrm{GC} < 10$~kpc, and in situ GCs at high radii, such as $R_\textrm{GC} > 30$~kpc \citep[behaviour also identified in e.g.][]{Pfeffer_2020, Reina_Campos_2022a, Chen_2024b}.

\subsection{GC system-mass-to-halo-mass relation}
\begin{figure}
    \centering
    \resizebox{0.95\hsize}{!}{\includegraphics{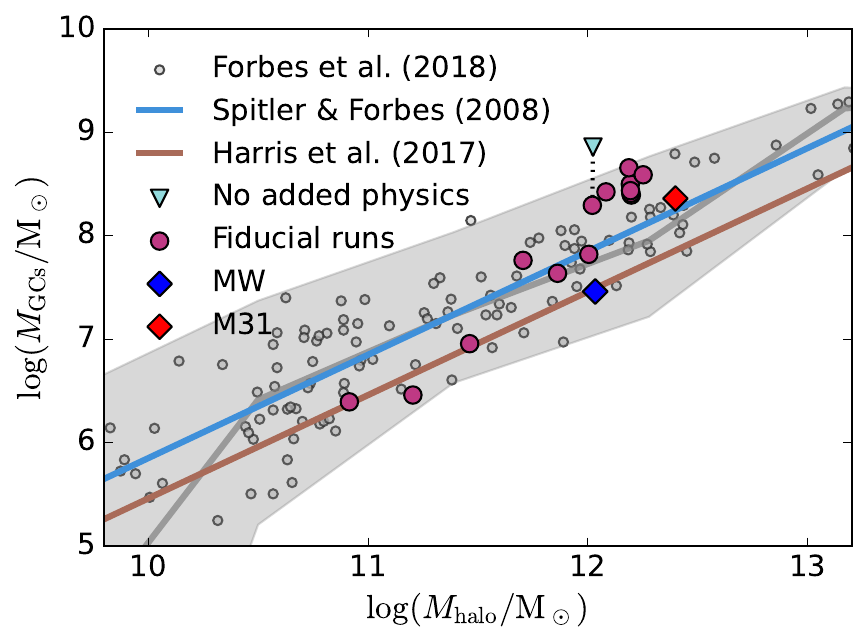}}
    \caption{GC system-mass-to-halo-mass relation for the simulated halos, including a compilation of galaxies across different masses and morphologies presented in \citet{Forbes_2018}, and the MW and M31 systems from the values quoted in Table~\ref{tab:fiducialruns}. The gray continuous line is the median and the shaded region is the $2\sigma$ of the distribution of the galaxies in \citet{Forbes_2018}. Also included are the empirical relations derived by \citet{Spitler_2009, Harris_2017}. For the \texttt{fiducial} model, the simulated GC systems masses, according to the GC candidate selection and old ages ($\tau > 8$~Gyr), generally lie inside the observed scatter of the empirical relation. The  \texttt{no added physics} variation of halo 6, on the other hand, gives a total mass in old GCs that sits above the observational scatter and almost an order of magnitude higher than the \texttt{fiducial} model.}
    \label{fig:gc_halo}
\end{figure}

Finally, we examined the GC system-mass-to-halo-mass relation first introduced in \citet{Spitler_2009} for our \texttt{fiducial} simulations. To probe a wider mass range than $1\text{-}2\times 10^{12} \, \mathrm{M}_\odot$ for the MW and M31, we expand our simulation suite by adding five lower mass halos. These are halos L1 and L10 in the $5\times10^{11} \,\mathrm{M}_\odot <M_{200}(z=0)<10^{12} \, \mathrm{M}_\odot$ mass range \citep[Table 3, \texttt{LowMassMWs} in ][]{Grand_2024} and halos 4, 7, and 10 in the $5\times10^{10} \, \mathrm{M}_\odot <M_{200}(z=0)<5\times10^{11} \, \mathrm{M}_\odot$ mass range \citep[Table 4, \texttt{Halos\_1e11msol} in][]{Grand_2024}. We show the total mass in GCs for the extended \texttt{fiducial} runs in the GC system-mass-to-halo-mass plane in Fig.~\ref{fig:gc_halo}. The simulated halos cover a total dynamical mass range of $8\times10^{10}$ to $2\times10^{12} \,\mathrm{M}_\odot$ and the total GC system mass is calculated from the total mass of our GC selection with ages $\tau > 8$~Gyr. Also in Fig.~\ref{fig:gc_halo}, we show two different fits of the empirical relations presented in \citet{Spitler_2009} and \citet{Harris_2017}, along with the compilation of systems across different morphologies and masses presented in \citet{Forbes_2018}. The gray continuous line and the shaded region correspond to the median and $2\sigma$-region of the distribution of these systems, respectively. We also include the position of the MW and M31 for the values quoted in Table~\ref{tab:fiducialruns}. Lastly, the \texttt{no added physics} model of the halo 6 is also included. The GC system mass from this \texttt{no added physics} model sits well above the $2\sigma$ region of the expected total GCs mass for its halo mass, while the total GC system mass for the \texttt{fiducial} run is inside this scatter. This gives further support for constraining GC formation to fully compressive gas regions for a more realistic modelling of SC formation, and the inclusion of compact object heating. Looking at all the simulated GC populations, they all lie inside the observed scatter of the empirical relation. The trend, however, seems to follow a steeper relation across the simulated mass range. This suggests that a wider study with more simulated halos is needed to constrain the drivers of the slope of the simulated relation. Overall, this picture demonstrates that the fiducial model is able to produce reasonable total GC system masses across a range of halo masses, given the physically motivated candidate selection.

%%%%%%%%%%%%%%%%%%%%%%%%%%%%%%%%%%%%%%%%%%%%%%%%%%%%%%%%%%%%%%
\section{Discussion}\label{sec:discussion}

\begin{figure}
    \centering
    \resizebox{0.95\hsize}{!}{\includegraphics{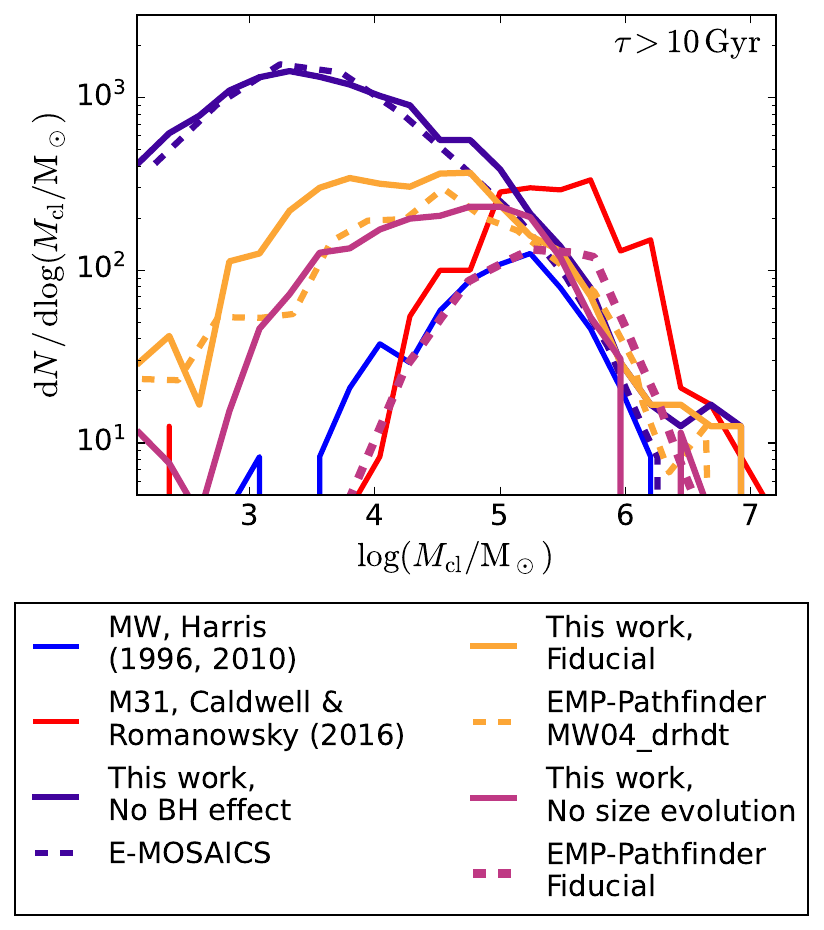}}
    \caption{Comparison of this work simulated MFs with respect to those of E-MOSAICS \citep{Pfeffer_2018, Kruijssen_2019b} and EMP-Pathfinder \citep{Reina_Campos_2022, Reina_Campos_2023} projects. All the simulated MFs correspond to old SCs at the $\tau>10$~Gyr age cut. Form our work, we only plot the models that compare well to the mentioned projects. For E-MOSAICS, the distribution corresponds to the median MF of 25 zoom-in MW-mass simulations plotted in Fig.~8 in \citet{Reina_Campos_2022}. From the same figure, we extracted the fiducial MF of EMP-Pathfinder (with no size evolution) while the MW04\_drhdt model, that includes size evolution, was extracted from Fig.~4 in \citet{Reina_Campos_2023}.}
    \label{fig:sims_compare}
\end{figure}

Our simulated GC populations can be directly compared to those of the aforementioned E-MOSAICS \citep{Pfeffer_2018, Kruijssen_2019b} and EMP-Pathfinder \citep{Reina_Campos_2022, Reina_Campos_2023} projects, owing to similarities of the subgrid models. 
Despite differences in both the underlying galaxy formation and the subgrid cluster models, these simulations manage to reproduce the high mass end of the observed GCMF. By matching their published MFs against our suite of simulations (comparison performed in Fig.~\ref{fig:sims_compare}), we discuss here how different assumptions of star formation and SCs mass loss affect the simulated population of SCs. We first note that, even though the three projects implement the same CFE model \citep{Kruijssen_2012}, we needed to further constrain SCs formation to match the high mass end of the GCMF. This is a consequence of the different galaxy formation models as follows:

EAGLE \citep{Crain_2015, Schaye_2015}, the base model for E-MOSAICS, implements a star formation prescription that is a function of the gas pressure \citep{Schaye2008} and a metallicity dependent density threshold \citep{Schaye2004}. In contrast, the star formation prescription in Auriga is solely a function of the gas density \citep{Springel_2003}. While the characteristic pressure of star-forming gas fluctuates by several orders of magnitude in both models due to mergers, gas accretion rate, and feedback \citep[e.g. see Fig. 5 in][]{Pfeffer_2018}, we have verified that the characteristic pressure of star forming gas in Auriga is persistently higher. Particularly, this pressure corresponds to the effective equation of state of the two phase ISM model. Given that the CFE model is a monotonic function of the pressure, the calculated CFE in Auriga remains high, i.e. $\Gamma\gtrsim 0.1$, at all times. Additionally, the Auriga galaxy formation model produces galaxies more massive in stars for the same halo mass, in comparison with EAGLE, leading to a higher mass budget for SC formation.

For EMP-Pathfinder, the galaxy formation model also gives a continuous SC formation with a high CFE across the galaxy assembly history \citep[see Fig. 12 in][]{Reina_Campos_2022}. Nevertheless, the mass budget for SCs formation is lower than in Auriga from their galaxies being less massive in stars for similar halo masses \citep[see Table 2 in][]{Reina_Campos_2022}. Particularly, the explicit multi-phase ISM model allowing gas to cool down to $T\sim100$~K before reaching the density threshold for star formation could be playing the role of environmental regulation for star clusters formation. Additionally, the more efficient mass loss from tidal shocks contributes to their simulated MFs to match the MW GCMF across all masses.

We now move to compare the low mass end of the MFs across our simulated populations and those of E-MOSAICS and EMP-Pathfinder.  In Fig.~\ref{fig:sims_compare} we show the MFs of our variations comparing well with the MFs reported in E-MOSAICS and EMP-Pathfinder. All of the MFs correspond to SCs at the $\tau>10$~Gyr age cut. We note that the different age cut plotted here, instead of the $\tau>8$~Gyr used in the rest of the paper, is to match the published MFs from these projects. We can see that our \texttt{no BH effect} model matches the MF from E-MOSAICS, demonstrating a consistent behaviour from adopting similar assumptions. The comparison with the MFs from EMP-Pathfinder is not as straightforward: while our \texttt{fiducial} model compares well with the `MW04\_drhdt' MF reported in \citet{Reina_Campos_2023} (for the same initial size prescription, albeit with more clusters in the $10^3<m/\mathrm{M}_\odot<10^4$ mass range), our \texttt{no size evolution} variation sits well above the 'fiducial' MF from \citet{Reina_Campos_2022} (that does not include size evolution) for masses $m<10^5 \,\mathrm{M}_\odot$. Although both models show the same behaviour, the effect of not including size evolution is milder in our model.

Although the quantitative alignment between the simulation suites shows variations due to the distinct sub-grid physics, the comparisons demonstrate that different theoretical approaches can yield similarly evolved MFs. Particularly, this highlights a degree of degeneracy in the mechanisms shaping GC populations. Moreover, both simulation suites confirm that efficient mass-loss prescriptions are needed to sufficiently disrupt the evolved low-mass SCs and recover MFs consistent with observations. In our case, this is achieved through the implementation of the compact object heating correction in the two body relaxation prescription. Furthermore, both models demonstrate that incorporating the size evolution of SCs acts as an agent suppressing the effectiveness of these disruption mechanisms. More studies are needed to better characterise this effect and improve the subgrid modelling of the size evolution of clusters.

Finally, as shown by \citet{Reina_Campos_2022}, tidal shocks-driven disruption of SCs is sensitive to the treatment of the ISM in the galaxy formation model. Our \texttt{enhanced shocks} variation aimed to mimic the effect of the inclusion of a multiphase ISM on the SCs disruption and indeed showed a more disruptive scenario but with a different behaviour than in EMP-Pathfinder. In particular, the peak of the MFs is not shifted to higher masses and the MF keeps a populated low mass tail (see Figs.~\ref{fig:fiducialvariations} and~\ref{fig:cce_test}). Our test, however, is not conclusive and running AuriGLOBES with a different ISM model, such as SMUGGLE \citep{Marinacci_2019}, could help to characterise better the effect of the cold ISM structure in our SC population. Additionally, we also note that following the behaviour of the MFs in Fig.~\ref{fig:res_test}, resolution has an effect on the ability to capture tidal shock disruption. In this sense, the underestimation of tidal shocks disruption seen in \citet{Pfeffer_2018} for E-MOSAICS could partially be attributed to the resolution of the simulation suite.

To summarise, our work complements earlier findings on environmental cluster disruption mechanisms and simulated SC populations across cosmic history. All of these similar subgrid models lay the foundation to constrain GC formation and evolution in a cosmological context. Nevertheless, it is evident that further works are required to characterise the different processes shaping the GC population throughout cosmic history. Additionally, further extensions to this foundation could explore self-consistent formation pathways of different structures such as the stellar stream population arising from GC disruption.

\section{Conclusions}\label{sec:conclusions}
We introduce AuriGLOBES, a physically motivated subgrid model of SC formation and evolution inside the Auriga galaxy formation model. With our model we aim to further constrain our understanding of GC populations and how they have been shaped through cosmic time. To this end we implemented two novel approaches in the subgrid modelling of SCs: to constrain their formation to tidally compressive environments only, and to include a correction factor to account for the dynamical heating from their compact object remnant population. Our main conclusions are:
\begin{itemize}
    \item Our \texttt{fiducial} model produces GC populations that are in agreement with the empirical GC system-mass-to-halo-mass relation across a halo mass range of $10^{11} \,\rm M_{\odot}\textrm{--}10^{12} \,\rm M_{\odot}$, along with a set of plausible properties in terms of their spatial, metallicity, and age distributions. 
    \item An important aspect of our model is the environmental regulation of GC formation via tidally compressive tides. This is in addition to the gas properties, such as density and pressure, that are taken into account in the particular CFE calculation.
    \item Reproducing a GCMF that is consistent with observational constraints requires an effective prescription of tidal shocks disruption, two-body relaxation, and dynamical friction. In particular, we have shown that two-body relaxation prescriptions should also account for the heating of compact object remnants.
    \item The effectiveness of tidal shocks is affected not only by the inclusion of a cold phase in the ISM model but also on the resolution of the simulation.
    \item In agreement with the EMP-Pathfinder simulations \citep{Reina_Campos_2022, Reina_Campos_2023}, the size evolution of SCs suppresses the effectiveness of the disruption mechanisms.
    \item The diversity seen among the age, metallicity, and spatial distributions of our simulated GC populations reflect the variation of assembly and accretion histories of the parent halos, reinforcing the role of GCs as fossil records of galactic growth. In particular, there is a clear correlation between the GC formation with early star formation in the galaxies.
    \item Among the simulated halos, the old GC population of halo 22 is the one with closer resemblance to that of the MW in terms of the mass, metallicity, and age distributions. In particular, this halo has been previously identified to include a GES analogue merger that could play a role shaping its GC population. We defer a dedicated analysis of the imprint of the halos assembly histories on their GC populations to a future work.
\end{itemize}

Finally, we want to highlight that AuriGLOBES represents a reliable framework for SC and GC population studies. Its relatively cheap overhead on top of the Auriga model makes it appealing for its application to larger cosmological volumes or statistical sample of galaxies. In addition, its AREPO-based modularity and its on-the-fly functionality open exciting avenues for further characterise the GC population and streams generated from their disruption. Specifically, our future plans include generating the stellar streams with a spawning routine similar to the SUPERSTARS method in Auriga \citep{Pakmor_2025b} as well as running trials with other models of the ISM with our AuriGLOBES framework.

%%%%%%%%%%%%%%%%%%%%%%%%%%%%%%%%%%%%%%%%%%%%%%%%%%%%%%%%%%%%%%
\begin{acknowledgements}
We thank the referee for the comments and suggestions that helped to improve this manuscript.
This work has been supported by the Spanish Ministry of Science and Innovation and Universities (MCIN/AEI/10.13039/501100011033/ - FEDER) through the research grants PID2021-122603NB-C22 and the Spanish Ministry of Science, Innovation and Universities (MICIU) PID2024-156100NB-C22.
This study made use of Prospero high-performance computing facility at Liverpool John Moores  University.
The authors acknowledge financial support from the Royal Society under grant RGS/R2/242421.
The authors thankfully acknowledge the technical expertise and assistance provided by the Spanish Supercomputing Network (Red Espa\~nola de Supercomputaci\'on), as well as the computer resources used: the LaPalma Supercomputer, located at the Instituto de Astrofísica de Canarias.
The authors wish to acknowledge the contribution of the IAC High-Performance Computing support team and hardware facilities to the results of this research.
RJJG acknowledges support from the STFC Ernest Rutherford Fellowship (ST/W003643/1).
MRC acknowledges funding from the Global Talent Junior Fellowship Programme at the Instituto Galego de Física de Altas Enerxías (IGFAE), supported by grant CEX2023-001318-M (María de Maeztu Unit of Excellence and Agencia Española de Investigación / 10.13039/501100011033).
This work has received financial support from the Xunta de Galicia (CIGUS Network of Research Centres) and the European Union through the Galicia Feder 2021-2027 Program.
MRC gratefully acknowledges the Canadian Institute for Theoretical Astrophysics (CITA) National Fellowship for partial support; this work was supported by the Natural Sciences and Engineering Research Council of Canada (NSERC) [funding reference number 568580].
\end{acknowledgements}

%%%%%%%%%%%%%%%%%%%%%%%%%%%%%%%%%%%%%%%%%%%%%%%%%%%%%%%%%%%%%%
\bibliographystyle{aa}
\bibliography{Bibliography}

@ARTICLE{Genel13,
   author = {{Genel}, S. and {Vogelsberger}, M. and {Nelson}, D. and {Sijacki}, D. and 
	{Springel}, V. and {Hernquist}, L.},
    title = "{Following the flow: tracer particles in astrophysical fluid simulations}",
  journal = {\mnras},
archivePrefix = "arXiv",
   eprint = {1305.2195},
 primaryClass = "astro-ph.IM",
     year = 2013,
    month = oct,
   volume = 435,
    pages = {1426-1442},
      doi = {10.1093/mnras/stt1383},
   adsurl = {http://adsabs.harvard.edu/abs/2013MNRAS.435.1426G}
}

@ARTICLE{Pakmor_2025a,
       author = {{Pakmor}, R{\"u}diger and {Bieri}, Rebekka and {Fragkoudi}, Francesca and {G{\'o}mez}, Facundo A. and {Grand}, Robert J.~J. and {Simpson}, Christine M. and {Talbot}, Rosie Y. and {van de Voort}, Freeke and {Werhahn}, Maria},
        title = "{Quantifying the intrinsic variability due to randomness of the Auriga galaxy formation model}",
      journal = {\mnras},
         year = 2025,
        month = oct,
       volume = {543},
       number = {2},
        pages = {1761-1774},
          doi = {10.1093/mnras/staf1542},
archivePrefix = {arXiv},
       eprint = {2507.13440},
 primaryClass = {astro-ph.GA},
       adsurl = {https://ui.adsabs.harvard.edu/abs/2025MNRAS.543.1761P}
}

@ARTICLE{Adamo_2020,
       author = {{Adamo}, A. and {Hollyhead}, K. and {Messa}, M. and {Ryon}, J.~E. and {Bajaj}, V. and {Runnholm}, A. and {Aalto}, S. and {Calzetti}, D. and {Gallagher}, J.~S. and {Hayes}, M.~J. and {Kruijssen}, J.~M.~D. and {K{\"o}nig}, S. and {Larsen}, S.~S. and {Melinder}, J. and {Sabbi}, E. and {Smith}, L.~J. and {{\"O}stlin}, G.},
        title = "{Star cluster formation in the most extreme environments: insights from the HiPEEC survey}",
      journal = {\mnras},
         year = 2020,
        month = dec,
       volume = {499},
       number = {3},
        pages = {3267-3294},
          doi = {10.1093/mnras/staa2380},
archivePrefix = {arXiv},
       eprint = {2008.12794},
 primaryClass = {astro-ph.GA},
       adsurl = {https://ui.adsabs.harvard.edu/abs/2020MNRAS.499.3267A}
}

@ARTICLE{Planck_2013,
       author = {{Planck Collaboration} and {Ade}, P.~A.~R. and {Aghanim}, N. and {Armitage-Caplan}, C. and {Arnaud}, M. and {Ashdown}, M. and {Atrio-Barandela}, F. and {Aumont}, J. and {Baccigalupi}, C. and {Banday}, A.~J. and {Barreiro}, R.~B. and {Bartlett}, J.~G. and {Battaner}, E. and {Benabed}, K. and {Beno{\^\i}t}, A. and {Benoit-L{\'e}vy}, A. and {Bernard}, J. -P. and {Bersanelli}, M. and {Bielewicz}, P. and {Bobin}, J. and {Bock}, J.~J. and {Bonaldi}, A. and {Bond}, J.~R. and {Borrill}, J. and {Bouchet}, F.~R. and {Bridges}, M. and {Bucher}, M. and {Burigana}, C. and {Butler}, R.~C. and {Calabrese}, E. and {Cappellini}, B. and {Cardoso}, J. -F. and {Catalano}, A. and {Challinor}, A. and {Chamballu}, A. and {Chary}, R. -R. and {Chen}, X. and {Chiang}, H.~C. and {Chiang}, L. -Y. and {Christensen}, P.~R. and {Church}, S. and {Clements}, D.~L. and {Colombi}, S. and {Colombo}, L.~P.~L. and {Couchot}, F. and {Coulais}, A. and {Crill}, B.~P. and {Curto}, A. and {Cuttaia}, F. and {Danese}, L. and {Davies}, R.~D. and {Davis}, R.~J. and {de Bernardis}, P. and {de Rosa}, A. and {de Zotti}, G. and {Delabrouille}, J. and {Delouis}, J. -M. and {D{\'e}sert}, F. -X. and {Dickinson}, C. and {Diego}, J.~M. and {Dolag}, K. and {Dole}, H. and {Donzelli}, S. and {Dor{\'e}}, O. and {Douspis}, M. and {Dunkley}, J. and {Dupac}, X. and {Efstathiou}, G. and {Elsner}, F. and {En{\ss}lin}, T.~A. and {Eriksen}, H.~K. and {Finelli}, F. and {Forni}, O. and {Frailis}, M. and {Fraisse}, A.~A. and {Franceschi}, E. and {Gaier}, T.~C. and {Galeotta}, S. and {Galli}, S. and {Ganga}, K. and {Giard}, M. and {Giardino}, G. and {Giraud-H{\'e}raud}, Y. and {Gjerl{\o}w}, E. and {Gonz{\'a}lez-Nuevo}, J. and {G{\'o}rski}, K.~M. and {Gratton}, S. and {Gregorio}, A. and {Gruppuso}, A. and {Gudmundsson}, J.~E. and {Haissinski}, J. and {Hamann}, J. and {Hansen}, F.~K. and {Hanson}, D. and {Harrison}, D. and {Henrot-Versill{\'e}}, S. and {Hern{\'a}ndez-Monteagudo}, C. and {Herranz}, D. and {Hildebrandt}, S.~R. and {Hivon}, E. and {Hobson}, M. and {Holmes}, W.~A. and {Hornstrup}, A. and {Hou}, Z. and {Hovest}, W. and {Huffenberger}, K.~M. and {Jaffe}, A.~H. and {Jaffe}, T.~R. and {Jewell}, J. and {Jones}, W.~C. and {Juvela}, M. and {Keih{\"a}nen}, E. and {Keskitalo}, R. and {Kisner}, T.~S. and {Kneissl}, R. and {Knoche}, J. and {Knox}, L. and {Kunz}, M. and {Kurki-Suonio}, H. and {Lagache}, G. and {L{\"a}hteenm{\"a}ki}, A. and {Lamarre}, J. -M. and {Lasenby}, A. and {Lattanzi}, M. and {Laureijs}, R.~J. and {Lawrence}, C.~R. and {Leach}, S. and {Leahy}, J.~P. and {Leonardi}, R. and {Le{\'o}n-Tavares}, J. and {Lesgourgues}, J. and {Lewis}, A. and {Liguori}, M. and {Lilje}, P.~B. and {Linden-V{\o}rnle}, M. and {L{\'o}pez-Caniego}, M. and {Lubin}, P.~M. and {Mac{\'\i}as-P{\'e}rez}, J.~F. and {Maffei}, B. and {Maino}, D. and {Mandolesi}, N. and {Maris}, M. and {Marshall}, D.~J. and {Martin}, P.~G. and {Mart{\'\i}nez-Gonz{\'a}lez}, E. and {Masi}, S. and {Massardi}, M. and {Matarrese}, S. and {Matthai}, F. and {Mazzotta}, P. and {Meinhold}, P.~R. and {Melchiorri}, A. and {Melin}, J. -B. and {Mendes}, L. and {Menegoni}, E. and {Mennella}, A. and {Migliaccio}, M. and {Millea}, M. and {Mitra}, S. and {Miville-Desch{\^e}nes}, M. -A. and {Moneti}, A. and {Montier}, L. and {Morgante}, G. and {Mortlock}, D. and {Moss}, A. and {Munshi}, D. and {Murphy}, J.~A. and {Naselsky}, P. and {Nati}, F. and {Natoli}, P. and {Netterfield}, C.~B. and {N{\o}rgaard-Nielsen}, H.~U. and {Noviello}, F. and {Novikov}, D. and {Novikov}, I. and {O'Dwyer}, I.~J. and {Osborne}, S. and {Oxborrow}, C.~A. and {Paci}, F. and {Pagano}, L. and {Pajot}, F. and {Paladini}, R. and {Paoletti}, D. and {Partridge}, B. and {Pasian}, F. and {Patanchon}, G. and {Pearson}, D. and {Pearson}, T.~J. and {Peiris}, H.~V. and {Perdereau}, O. and {Perotto}, L. and {Perrotta}, F. and {Pettorino}, V. and {Piacentini}, F. and {Piat}, M. and {Pierpaoli}, E. and {Pietrobon}, D. and {Plaszczynski}, S. and {Platania}, P. and {Pointecouteau}, E.},
        title = "{Planck 2013 results. XVI. Cosmological parameters}",
      journal = {\aap},
         year = 2014,
        month = nov,
       volume = {571},
          eid = {A16},
        pages = {A16},
          doi = {10.1051/0004-6361/201321591},
archivePrefix = {arXiv},
       eprint = {1303.5076},
 primaryClass = {astro-ph.CO},
       adsurl = {https://ui.adsabs.harvard.edu/abs/2014A&A...571A..16P}
}

@ARTICLE{Alexander_2012,
       author = {{Alexander}, Poul E.~R. and {Gieles}, Mark},
        title = "{A prescription and fast code for the long-term evolution of star clusters}",
      journal = {\mnras},
         year = 2012,
        month = jun,
       volume = {422},
       number = {4},
        pages = {3415-3432},
          doi = {10.1111/j.1365-2966.2012.20867.x},
archivePrefix = {arXiv},
       eprint = {1203.4744},
 primaryClass = {astro-ph.GA},
       adsurl = {https://ui.adsabs.harvard.edu/abs/2012MNRAS.422.3415A}
}

@ARTICLE{Barnes_1986,
       author = {{Barnes}, Josh and {Hut}, Piet},
        title = "{A hierarchical O(N log N) force-calculation algorithm}",
      journal = {\nat},
         year = 1986,
        month = dec,
       volume = {324},
       number = {6096},
        pages = {446-449},
          doi = {10.1038/324446a0},
       adsurl = {https://ui.adsabs.harvard.edu/abs/1986Natur.324..446B}
}

@ARTICLE{Bastian_2008,
       author = {{Bastian}, N.},
        title = "{On the star formation rate - brightest cluster relation: estimating the peak star formation rate in post-merger galaxies}",
      journal = {\mnras},
         year = 2008,
        month = oct,
       volume = {390},
       number = {2},
        pages = {759-768},
          doi = {10.1111/j.1365-2966.2008.13775.x},
archivePrefix = {arXiv},
       eprint = {0807.4687},
 primaryClass = {astro-ph},
       adsurl = {https://ui.adsabs.harvard.edu/abs/2008MNRAS.390..759B}
}

@ARTICLE{Baumgardt_2003,
       author = {{Baumgardt}, Holger and {Makino}, Junichiro},
        title = "{Dynamical evolution of star clusters in tidal fields}",
      journal = {\mnras},
         year = 2003,
        month = mar,
       volume = {340},
       number = {1},
        pages = {227-246},
          doi = {10.1046/j.1365-8711.2003.06286.x},
archivePrefix = {arXiv},
       eprint = {astro-ph/0211471},
 primaryClass = {astro-ph},
       adsurl = {https://ui.adsabs.harvard.edu/abs/2003MNRAS.340..227B}
}

@ARTICLE{Brodie_2006,
       author = {{Brodie}, Jean P. and {Strader}, Jay},
        title = "{Extragalactic Globular Clusters and Galaxy Formation}",
      journal = {\araa},
         year = 2006,
        month = sep,
       volume = {44},
       number = {1},
        pages = {193-267},
          doi = {10.1146/annurev.astro.44.051905.092441},
archivePrefix = {arXiv},
       eprint = {astro-ph/0602601},
 primaryClass = {astro-ph},
       adsurl = {https://ui.adsabs.harvard.edu/abs/2006ARA&A..44..193B}
}

@ARTICLE{Brown_2021,
       author = {{Brown}, Gillen and {Gnedin}, Oleg Y.},
        title = "{Radii of young star clusters in nearby galaxies}",
      journal = {\mnras},
         year = 2021,
        month = dec,
       volume = {508},
       number = {4},
        pages = {5935-5953},
          doi = {10.1093/mnras/stab2907},
archivePrefix = {arXiv},
       eprint = {2106.12420},
 primaryClass = {astro-ph.GA},
       adsurl = {https://ui.adsabs.harvard.edu/abs/2021MNRAS.508.5935B}
}

@ARTICLE{Chabrier_2003,
       author = {{Chabrier}, Gilles},
        title = "{Galactic Stellar and Substellar Initial Mass Function}",
      journal = {\pasp},
         year = 2003,
        month = jul,
       volume = {115},
       number = {809},
        pages = {763-795},
          doi = {10.1086/376392},
archivePrefix = {arXiv},
       eprint = {astro-ph/0304382},
 primaryClass = {astro-ph},
       adsurl = {https://ui.adsabs.harvard.edu/abs/2003PASP..115..763C}
}

@ARTICLE{Chen_2024,
       author = {{Chen}, Yingtian and {Gnedin}, Oleg Y.},
        title = "{Catalogue of model star clusters in the Milky Way and M31 galaxies}",
      journal = {\mnras},
         year = 2024,
        month = jan,
       volume = {527},
       number = {2},
        pages = {3692-3708},
          doi = {10.1093/mnras/stad3345},
archivePrefix = {arXiv},
       eprint = {2309.13374},
 primaryClass = {astro-ph.GA},
       adsurl = {https://ui.adsabs.harvard.edu/abs/2024MNRAS.527.3692C}
}

@ARTICLE{Crain_2015,
       author = {{Crain}, Robert A. and {Schaye}, Joop and {Bower}, Richard G. and {Furlong}, Michelle and {Schaller}, Matthieu and {Theuns}, Tom and {Dalla Vecchia}, Claudio and {Frenk}, Carlos S. and {McCarthy}, Ian G. and {Helly}, John C. and {Jenkins}, Adrian and {Rosas-Guevara}, Yetli M. and {White}, Simon D.~M. and {Trayford}, James W.},
        title = "{The EAGLE simulations of galaxy formation: calibration of subgrid physics and model variations}",
      journal = {\mnras},
         year = 2015,
        month = jun,
       volume = {450},
       number = {2},
        pages = {1937-1961},
          doi = {10.1093/mnras/stv725},
archivePrefix = {arXiv},
       eprint = {1501.01311},
 primaryClass = {astro-ph.GA},
       adsurl = {https://ui.adsabs.harvard.edu/abs/2015MNRAS.450.1937C}
}

@ARTICLE{Creasey_2019,
       author = {{Creasey}, Peter and {Sales}, Laura V. and {Peng}, Eric W. and {Sameie}, Omid},
        title = "{Globular clusters formed within dark haloes I: present-day abundance, distribution, and kinematics}",
      journal = {\mnras},
         year = 2019,
        month = jan,
       volume = {482},
       number = {1},
        pages = {219-230},
          doi = {10.1093/mnras/sty2701},
archivePrefix = {arXiv},
       eprint = {1806.11118},
 primaryClass = {astro-ph.GA},
       adsurl = {https://ui.adsabs.harvard.edu/abs/2019MNRAS.482..219C}
}

@ARTICLE{De_Lucia_2024,
       author = {{De Lucia}, Gabriella and {Kruijssen}, J.~M. Diederik and {Trujillo-Gomez}, Sebastian and {Hirschmann}, Michaela and {Xie}, Lizhi},
        title = "{On the origin of globular clusters in a hierarchical universe}",
      journal = {\mnras},
         year = 2024,
        month = may,
       volume = {530},
       number = {3},
        pages = {2760-2777},
          doi = {10.1093/mnras/stae1006},
archivePrefix = {arXiv},
       eprint = {2307.02530},
 primaryClass = {astro-ph.GA},
       adsurl = {https://ui.adsabs.harvard.edu/abs/2024MNRAS.530.2760D}
}

@ARTICLE{Duerr_1982,
       author = {{Duerr}, R. and {Imhoff}, C.~L. and {Lada}, C.~J.},
        title = "{Star formation in the lam ORI region. I. The distribution of young objects.}",
      journal = {\apj},
         year = 1982,
        month = oct,
       volume = {261},
        pages = {135-150},
          doi = {10.1086/160325},
       adsurl = {https://ui.adsabs.harvard.edu/abs/1982ApJ...261..135D}
}

@ARTICLE{Ekstrom_2012,
       author = {{Ekstr{\"o}m}, S. and {Georgy}, C. and {Eggenberger}, P. and {Meynet}, G. and {Mowlavi}, N. and {Wyttenbach}, A. and {Granada}, A. and {Decressin}, T. and {Hirschi}, R. and {Frischknecht}, U. and {Charbonnel}, C. and {Maeder}, A.},
        title = "{Grids of stellar models with rotation. I. Models from 0.8 to 120 M$_{{\ensuremath{\odot}}}$ at solar metallicity (Z = 0.014)}",
      journal = {\aap},
         year = 2012,
        month = jan,
       volume = {537},
          eid = {A146},
        pages = {A146},
          doi = {10.1051/0004-6361/201117751},
archivePrefix = {arXiv},
       eprint = {1110.5049},
 primaryClass = {astro-ph.SR},
       adsurl = {https://ui.adsabs.harvard.edu/abs/2012A&A...537A.146E}
}

@ARTICLE{Elmegreen_2002,
       author = {{Elmegreen}, Bruce G.},
        title = "{Star Formation from Galaxies to Globules}",
      journal = {\apj},
         year = 2002,
        month = sep,
       volume = {577},
       number = {1},
        pages = {206-220},
          doi = {10.1086/342177},
archivePrefix = {arXiv},
       eprint = {astro-ph/0207114},
 primaryClass = {astro-ph},
       adsurl = {https://ui.adsabs.harvard.edu/abs/2002ApJ...577..206E}
}

@ARTICLE{Elmegreen_2010a,
       author = {{Elmegreen}, Bruce G. and {Hunter}, Deidre A.},
        title = "{On the Disruption of Star Clusters in a Hierarchical Interstellar Medium}",
      journal = {\apj},
         year = 2010,
        month = mar,
       volume = {712},
       number = {1},
        pages = {604-623},
          doi = {10.1088/0004-637X/712/1/604},
archivePrefix = {arXiv},
       eprint = {1002.2823},
 primaryClass = {astro-ph.GA},
       adsurl = {https://ui.adsabs.harvard.edu/abs/2010ApJ...712..604E}
}

@ARTICLE{Fall_2001,
       author = {{Fall}, S. Michael and {Zhang}, Qing},
        title = "{Dynamical Evolution of the Mass Function of Globular Star Clusters}",
      journal = {\apj},
         year = 2001,
        month = nov,
       volume = {561},
       number = {2},
        pages = {751-765},
          doi = {10.1086/323358},
archivePrefix = {arXiv},
       eprint = {astro-ph/0107298},
 primaryClass = {astro-ph},
       adsurl = {https://ui.adsabs.harvard.edu/abs/2001ApJ...561..751F}
}

@ARTICLE{Faucher_2009,
       author = {{Faucher-Gigu{\`e}re}, Claude-Andr{\'e} and {Lidz}, Adam and {Zaldarriaga}, Matias and {Hernquist}, Lars},
        title = "{A New Calculation of the Ionizing Background Spectrum and the Effects of He II Reionization}",
      journal = {\apj},
         year = 2009,
        month = oct,
       volume = {703},
       number = {2},
        pages = {1416-1443},
          doi = {10.1088/0004-637X/703/2/1416},
archivePrefix = {arXiv},
       eprint = {0901.4554},
 primaryClass = {astro-ph.CO},
       adsurl = {https://ui.adsabs.harvard.edu/abs/2009ApJ...703.1416F}
}

@ARTICLE{Forbes_2010,
       author = {{Forbes}, Duncan A. and {Bridges}, Terry},
        title = "{Accreted versus in situ Milky Way globular clusters}",
      journal = {\mnras},
         year = 2010,
        month = may,
       volume = {404},
       number = {3},
        pages = {1203-1214},
          doi = {10.1111/j.1365-2966.2010.16373.x},
archivePrefix = {arXiv},
       eprint = {1001.4289},
 primaryClass = {astro-ph.GA},
       adsurl = {https://ui.adsabs.harvard.edu/abs/2010MNRAS.404.1203F}
}

@ARTICLE{Gieles_2014,
       author = {{Gieles}, Mark and {Alexander}, Poul E.~R. and {Lamers}, Henny J.~G.~L.~M. and {Baumgardt}, Holger},
        title = "{A prescription and fast code for the long-term evolution of star clusters - II. Unbalanced and core evolution}",
      journal = {\mnras},
         year = 2014,
        month = jan,
       volume = {437},
       number = {1},
        pages = {916-929},
          doi = {10.1093/mnras/stt1980},
archivePrefix = {arXiv},
       eprint = {1310.3631},
 primaryClass = {astro-ph.GA},
       adsurl = {https://ui.adsabs.harvard.edu/abs/2014MNRAS.437..916G}
}

@ARTICLE{Gieles_2023,
       author = {{Gieles}, Mark and {Gnedin}, Oleg Y.},
        title = "{The mass-loss rates of star clusters with stellar-mass black holes: implications for the globular cluster mass function}",
      journal = {\mnras},
         year = 2023,
        month = jul,
       volume = {522},
       number = {4},
        pages = {5340-5357},
          doi = {10.1093/mnras/stad1287},
archivePrefix = {arXiv},
       eprint = {2303.03791},
 primaryClass = {astro-ph.GA},
       adsurl = {https://ui.adsabs.harvard.edu/abs/2023MNRAS.522.5340G}
}

@ARTICLE{Gieles_2011,
       author = {{Gieles}, Mark and {Heggie}, Douglas C. and {Zhao}, Hongsheng},
        title = "{The life cycle of star clusters in a tidal field}",
      journal = {\mnras},
         year = 2011,
        month = jun,
       volume = {413},
       number = {4},
        pages = {2509-2524},
          doi = {10.1111/j.1365-2966.2011.18320.x},
archivePrefix = {arXiv},
       eprint = {1101.1821},
 primaryClass = {astro-ph.GA},
       adsurl = {https://ui.adsabs.harvard.edu/abs/2011MNRAS.413.2509G}
}

@ARTICLE{Gieles_2006,
       author = {{Gieles}, M. and {Portegies Zwart}, S.~F. and {Baumgardt}, H. and {Athanassoula}, E. and {Lamers}, H.~J.~G.~L.~M. and {Sipior}, M. and {Leenaarts}, J.},
        title = "{Star cluster disruption by giant molecular clouds}",
      journal = {\mnras},
         year = 2006,
        month = sep,
       volume = {371},
       number = {2},
        pages = {793-804},
          doi = {10.1111/j.1365-2966.2006.10711.x},
archivePrefix = {arXiv},
       eprint = {astro-ph/0606451},
 primaryClass = {astro-ph},
       adsurl = {https://ui.adsabs.harvard.edu/abs/2006MNRAS.371..793G}
}

@ARTICLE{Gieles_2016,
       author = {{Gieles}, Mark and {Renaud}, Florent},
        title = "{If it does not kill them, it makes them stronger: collisional evolution of star clusters with tidal shocks}",
      journal = {\mnras},
         year = 2016,
        month = nov,
       volume = {463},
       number = {1},
        pages = {L103-L107},
          doi = {10.1093/mnrasl/slw163},
archivePrefix = {arXiv},
       eprint = {1605.05940},
 primaryClass = {astro-ph.GA},
       adsurl = {https://ui.adsabs.harvard.edu/abs/2016MNRAS.463L.103G}
}

@ARTICLE{Giersz_1994,
       author = {{Giersz}, M. and {Heggie}, D.~C.},
        title = "{Statistics of N-Body Simulations - Part One - Equal Masses Before Core Collapse}",
      journal = {\mnras},
         year = 1994,
        month = may,
       volume = {268},
        pages = {257},
          doi = {10.1093/mnras/268.1.257},
archivePrefix = {arXiv},
       eprint = {astro-ph/9305008},
 primaryClass = {astro-ph},
       adsurl = {https://ui.adsabs.harvard.edu/abs/1994MNRAS.268..257G}
}

@ARTICLE{Gnedin_1999a,
       author = {{Gnedin}, Oleg Y. and {Hernquist}, Lars and {Ostriker}, Jeremiah P.},
        title = "{Tidal Shocking by Extended Mass Distributions}",
      journal = {\apj},
         year = 1999,
        month = mar,
       volume = {514},
       number = {1},
        pages = {109-118},
          doi = {10.1086/306910},
       adsurl = {https://ui.adsabs.harvard.edu/abs/1999ApJ...514..109G}
}

@ARTICLE{Gnedin_1999b,
       author = {{Gnedin}, Oleg Y. and {Lee}, Hyung Mok and {Ostriker}, Jeremiah P.},
        title = "{Effects of Tidal Shocks on the Evolution of Globular Clusters}",
      journal = {\apj},
         year = 1999,
        month = sep,
       volume = {522},
       number = {2},
        pages = {935-949},
          doi = {10.1086/307659},
archivePrefix = {arXiv},
       eprint = {astro-ph/9806245},
 primaryClass = {astro-ph},
       adsurl = {https://ui.adsabs.harvard.edu/abs/1999ApJ...522..935G}
}

@ARTICLE{Gnedin_1997,
       author = {{Gnedin}, Oleg Y. and {Ostriker}, Jeremiah P.},
        title = "{Destruction of the Galactic Globular Cluster System}",
      journal = {\apj},
         year = 1997,
        month = jan,
       volume = {474},
       number = {1},
        pages = {223-255},
          doi = {10.1086/303441},
archivePrefix = {arXiv},
       eprint = {astro-ph/9603042},
 primaryClass = {astro-ph},
       adsurl = {https://ui.adsabs.harvard.edu/abs/1997ApJ...474..223G}
}

@ARTICLE{Grand_2017,
       author = {{Grand}, Robert J.~J. and {G{\'o}mez}, Facundo A. and {Marinacci}, Federico and {Pakmor}, R{\"u}diger and {Springel}, Volker and {Campbell}, David J.~R. and {Frenk}, Carlos S. and {Jenkins}, Adrian and {White}, Simon D.~M.},
        title = "{The Auriga Project: the properties and formation mechanisms of disc galaxies across cosmic time}",
      journal = {\mnras},
         year = 2017,
        month = may,
       volume = {467},
       number = {1},
        pages = {179-207},
          doi = {10.1093/mnras/stx071},
archivePrefix = {arXiv},
       eprint = {1610.01159},
 primaryClass = {astro-ph.GA},
       adsurl = {https://ui.adsabs.harvard.edu/abs/2017MNRAS.467..179G}
}

@ARTICLE{Grand_2024,
       author = {{Grand}, Robert J.~J. and {Fragkoudi}, Francesca and {G{\'o}mez}, Facundo A. and {Jenkins}, Adrian and {Marinacci}, Federico and {Pakmor}, R{\"u}diger and {Springel}, Volker},
        title = "{Overview and public data release of the augmented Auriga Project: cosmological simulations of dwarf and Milky Way-mass galaxies}",
      journal = {\mnras},
         year = 2024,
        month = aug,
       volume = {532},
       number = {2},
        pages = {1814-1831},
          doi = {10.1093/mnras/stae1598},
archivePrefix = {arXiv},
       eprint = {2401.08750},
 primaryClass = {astro-ph.GA},
       adsurl = {https://ui.adsabs.harvard.edu/abs/2024MNRAS.532.1814G}
}

@ARTICLE{Grisdale_2019,
       author = {{Grisdale}, Kearn and {Agertz}, Oscar and {Renaud}, Florent and {Romeo}, Alessandro B. and {Devriendt}, Julien and {Slyz}, Adrianne},
        title = "{On the observed diversity of star formation efficiencies in Giant Molecular Clouds}",
      journal = {\mnras},
         year = 2019,
        month = jul,
       volume = {486},
       number = {4},
        pages = {5482-5491},
          doi = {10.1093/mnras/stz1201},
archivePrefix = {arXiv},
       eprint = {1902.00518},
 primaryClass = {astro-ph.GA},
       adsurl = {https://ui.adsabs.harvard.edu/abs/2019MNRAS.486.5482G}
}

@ARTICLE{Grudic_2020,
       author = {{Grudi{\'c}}, Michael Y. and {Hopkins}, Philip F.},
        title = "{A general-purpose time-step criterion for simulations with gravity}",
      journal = {\mnras},
         year = 2020,
        month = jul,
       volume = {495},
       number = {4},
        pages = {4306-4313},
          doi = {10.1093/mnras/staa1453},
archivePrefix = {arXiv},
       eprint = {1910.06349},
 primaryClass = {astro-ph.IM},
       adsurl = {https://ui.adsabs.harvard.edu/abs/2020MNRAS.495.4306G}
}

@ARTICLE{Halbesma_2020,
       author = {{Halbesma}, Timo L.~R. and {Grand}, Robert J.~J. and {G{\'o}mez}, Facundo A. and {Marinacci}, Federico and {Pakmor}, R{\"u}diger and {Trick}, Wilma H. and {Busch}, Philipp and {White}, Simon D.~M.},
        title = "{The globular cluster system of the Auriga simulations}",
      journal = {\mnras},
         year = 2020,
        month = jul,
       volume = {496},
       number = {1},
        pages = {638-648},
          doi = {10.1093/mnras/staa1380},
archivePrefix = {arXiv},
       eprint = {1909.02630},
 primaryClass = {astro-ph.GA},
       adsurl = {https://ui.adsabs.harvard.edu/abs/2020MNRAS.496..638H}
}

@ARTICLE{Harris_1991,
       author = {{Harris}, William E.},
        title = "{Globular cluster systems in galaxies beyond the Local Group.}",
      journal = {\araa},
         year = 1991,
        month = jan,
       volume = {29},
        pages = {543-579},
          doi = {10.1146/annurev.aa.29.090191.002551},
       adsurl = {https://ui.adsabs.harvard.edu/abs/1991ARA&A..29..543H}
}

@ARTICLE{Henon_1961,
       author = {{H{\'e}non}, M.},
        title = "{Sur l'{\'e}volution dynamique des amas globulaires}",
      journal = {Annales d'Astrophysique},
         year = 1961,
        month = feb,
       volume = {24},
        pages = {369},
       adsurl = {https://ui.adsabs.harvard.edu/abs/1961AnAp...24..369H}
}

@ARTICLE{Henon_1965,
       author = {{H{\'e}non}, M.},
        title = "{Exploration num{\'e}rique du probl{\`e}me restreint. II. Masses {\'e}gales, stabilit{\'e} des orbites p{\'e}riodiques}",
      journal = {Annales d'Astrophysique},
         year = 1965,
        month = feb,
       volume = {28},
        pages = {992},
       adsurl = {https://ui.adsabs.harvard.edu/abs/1965AnAp...28..992H}
}

@ARTICLE{Ibata_1994,
       author = {{Ibata}, R.~A. and {Gilmore}, G. and {Irwin}, M.~J.},
        title = "{A dwarf satellite galaxy in Sagittarius}",
      journal = {\nat},
         year = 1994,
        month = jul,
       volume = {370},
       number = {6486},
        pages = {194-196},
          doi = {10.1038/370194a0},
       adsurl = {https://ui.adsabs.harvard.edu/abs/1994Natur.370..194I}
}

@ARTICLE{Jenkins_2013,
       author = {{Jenkins}, Adrian},
        title = "{A new way of setting the phases for cosmological multiscale Gaussian initial conditions}",
      journal = {\mnras},
         year = 2013,
        month = sep,
       volume = {434},
       number = {3},
        pages = {2094-2120},
          doi = {10.1093/mnras/stt1154},
archivePrefix = {arXiv},
       eprint = {1306.5968},
 primaryClass = {astro-ph.CO},
       adsurl = {https://ui.adsabs.harvard.edu/abs/2013MNRAS.434.2094J}
}

@ARTICLE{Johnson_2015,
       author = {{Johnson}, K.~E. and {Leroy}, A.~K. and {Indebetouw}, R. and {Brogan}, C.~L. and {Whitmore}, B.~C. and {Hibbard}, J. and {Sheth}, K. and {Evans}, A.~S.},
        title = "{The Physical Conditions in a Pre-super Star Cluster Molecular Cloud in the Antennae Galaxies}",
      journal = {\apj},
         year = 2015,
        month = jun,
       volume = {806},
       number = {1},
          eid = {35},
        pages = {35},
          doi = {10.1088/0004-637X/806/1/35},
archivePrefix = {arXiv},
       eprint = {1503.06477},
 primaryClass = {astro-ph.GA},
       adsurl = {https://ui.adsabs.harvard.edu/abs/2015ApJ...806...35J}
}

@ARTICLE{Johnson_2017,
       author = {{Johnson}, L. Clifton and {Seth}, Anil C. and {Dalcanton}, Julianne J. and {Beerman}, Lori C. and {Fouesneau}, Morgan and {Weisz}, Daniel R. and {Bell}, Timothy A. and {Dolphin}, Andrew E. and {Sandstrom}, Karin and {Williams}, Benjamin F.},
        title = "{Panchromatic Hubble Andromeda Treasury. XVIII. The High-mass Truncation of the Star Cluster Mass Function}",
      journal = {\apj},
         year = 2017,
        month = apr,
       volume = {839},
       number = {2},
          eid = {78},
        pages = {78},
          doi = {10.3847/1538-4357/aa6a1f},
archivePrefix = {arXiv},
       eprint = {1703.10312},
 primaryClass = {astro-ph.GA},
       adsurl = {https://ui.adsabs.harvard.edu/abs/2017ApJ...839...78J}
}

@ARTICLE{Jordan_2007,
       author = {{Jord{\'a}n}, Andr{\'e}s and {McLaughlin}, Dean E. and {C{\^o}t{\'e}}, Patrick and {Ferrarese}, Laura and {Peng}, Eric W. and {Mei}, Simona and {Villegas}, Daniela and {Merritt}, David and {Tonry}, John L. and {West}, Michael J.},
        title = "{The ACS Virgo Cluster Survey. XII. The Luminosity Function of Globular Clusters in Early-Type Galaxies}",
      journal = {\apjs},
         year = 2007,
        month = jul,
       volume = {171},
       number = {1},
        pages = {101-145},
          doi = {10.1086/516840},
archivePrefix = {arXiv},
       eprint = {astro-ph/0702496},
 primaryClass = {astro-ph},
       adsurl = {https://ui.adsabs.harvard.edu/abs/2007ApJS..171..101J}
}

@ARTICLE{Kim_2018,
       author = {{Kim}, Ji-hoon and {Ma}, Xiangcheng and {Grudi{\'c}}, Michael Y. and {Hopkins}, Philip F. and {Hayward}, Christopher C. and {Wetzel}, Andrew and {Faucher-Gigu{\`e}re}, Claude-Andr{\'e} and {Kere{\v{s}}}, Du{\v{s}}an and {Garrison-Kimmel}, Shea and {Murray}, Norman},
        title = "{Formation of globular cluster candidates in merging proto-galaxies at high redshift: a view from the FIRE cosmological simulations}",
      journal = {\mnras},
         year = 2018,
        month = mar,
       volume = {474},
       number = {3},
        pages = {4232-4244},
          doi = {10.1093/mnras/stx2994},
archivePrefix = {arXiv},
       eprint = {1704.02988},
 primaryClass = {astro-ph.GA},
       adsurl = {https://ui.adsabs.harvard.edu/abs/2018MNRAS.474.4232K}
}

@ARTICLE{King_1962,
       author = {{King}, Ivan},
        title = "{The structure of star clusters. I. an empirical density law}",
      journal = {\aj},
         year = 1962,
        month = oct,
       volume = {67},
        pages = {471},
          doi = {10.1086/108756},
       adsurl = {https://ui.adsabs.harvard.edu/abs/1962AJ.....67..471K}
}

@ARTICLE{Kruijssen_2012,
       author = {{Kruijssen}, J.~M. Diederik},
        title = "{On the fraction of star formation occurring in bound stellar clusters}",
      journal = {\mnras},
         year = 2012,
        month = nov,
       volume = {426},
       number = {4},
        pages = {3008-3040},
          doi = {10.1111/j.1365-2966.2012.21923.x},
archivePrefix = {arXiv},
       eprint = {1208.2963},
 primaryClass = {astro-ph.CO},
       adsurl = {https://ui.adsabs.harvard.edu/abs/2012MNRAS.426.3008K}
}

@ARTICLE{Kruijssen_2015,
       author = {{Kruijssen}, J.~M. Diederik},
        title = "{Globular clusters as the relics of regular star formation in `normal' high-redshift galaxies}",
      journal = {\mnras},
         year = 2015,
        month = dec,
       volume = {454},
       number = {2},
        pages = {1658-1686},
          doi = {10.1093/mnras/stv2026},
archivePrefix = {arXiv},
       eprint = {1509.02163},
 primaryClass = {astro-ph.GA},
       adsurl = {https://ui.adsabs.harvard.edu/abs/2015MNRAS.454.1658K}
}

@ARTICLE{Kruijssen_2009,
       author = {{Kruijssen}, J.~M.~D.},
        title = "{The evolution of the stellar mass function in star clusters}",
      journal = {\aap},
         year = 2009,
        month = dec,
       volume = {507},
       number = {3},
        pages = {1409-1423},
          doi = {10.1051/0004-6361/200913325},
archivePrefix = {arXiv},
       eprint = {0910.4579},
 primaryClass = {astro-ph.GA},
       adsurl = {https://ui.adsabs.harvard.edu/abs/2009A&A...507.1409K}
}

@ARTICLE{Kruijssen_2014,
       author = {{Kruijssen}, J.~M. Diederik},
        title = "{Globular cluster formation in the context of galaxy formation and evolution}",
      journal = {Classical and Quantum Gravity},
         year = 2014,
        month = dec,
       volume = {31},
       number = {24},
          eid = {244006},
        pages = {244006},
          doi = {10.1088/0264-9381/31/24/244006},
archivePrefix = {arXiv},
       eprint = {1407.2953},
 primaryClass = {astro-ph.GA},
       adsurl = {https://ui.adsabs.harvard.edu/abs/2014CQGra..31x4006K}
}

@ARTICLE{Kruijssen_2008,
       author = {{Kruijssen}, J.~M.~D. and {Lamers}, H.~J.~G.~L.~M.},
        title = "{The photometric evolution of star clusters and the preferential loss of low-mass bodies - with an application to globular clusters}",
      journal = {\aap},
         year = 2008,
        month = oct,
       volume = {490},
       number = {1},
        pages = {151-171},
          doi = {10.1051/0004-6361:200810167},
archivePrefix = {arXiv},
       eprint = {0809.0307},
 primaryClass = {astro-ph},
       adsurl = {https://ui.adsabs.harvard.edu/abs/2008A&A...490..151K}
}

@ARTICLE{Kruijssen_2011,
       author = {{Kruijssen}, J.~M. Diederik and {Pelupessy}, F. Inti and {Lamers}, Henny J.~G.~L.~M. and {Portegies Zwart}, Simon F. and {Icke}, Vincent},
        title = "{Modelling the formation and evolution of star cluster populations in galaxy simulations}",
      journal = {\mnras},
         year = 2011,
        month = jun,
       volume = {414},
       number = {2},
        pages = {1339-1364},
          doi = {10.1111/j.1365-2966.2011.18467.x},
archivePrefix = {arXiv},
       eprint = {1102.1013},
 primaryClass = {astro-ph.CO},
       adsurl = {https://ui.adsabs.harvard.edu/abs/2011MNRAS.414.1339K}
}

@ARTICLE{Kruijssen_2019b,
       author = {{Kruijssen}, J.~M. Diederik and {Pfeffer}, Joel L. and {Crain}, Robert A. and {Bastian}, Nate},
        title = "{The E-MOSAICS project: tracing galaxy formation and assembly with the age-metallicity distribution of globular clusters}",
      journal = {\mnras},
         year = 2019,
        month = jul,
       volume = {486},
       number = {3},
        pages = {3134-3179},
          doi = {10.1093/mnras/stz968},
archivePrefix = {arXiv},
       eprint = {1904.04261},
 primaryClass = {astro-ph.GA},
       adsurl = {https://ui.adsabs.harvard.edu/abs/2019MNRAS.486.3134K}
}

@ARTICLE{Kruijssen_2019a,
       author = {{Kruijssen}, J.~M. Diederik and {Schruba}, Andreas and {Chevance}, M{\'e}lanie and {Longmore}, Steven N. and {Hygate}, Alexander P.~S. and {Haydon}, Daniel T. and {McLeod}, Anna F. and {Dalcanton}, Julianne J. and {Tacconi}, Linda J. and {van Dishoeck}, Ewine F.},
        title = "{Fast and inefficient star formation due to short-lived molecular clouds and rapid feedback}",
      journal = {\nat},
         year = 2019,
        month = may,
       volume = {569},
       number = {7757},
        pages = {519-522},
          doi = {10.1038/s41586-019-1194-3},
archivePrefix = {arXiv},
       eprint = {1905.08801},
 primaryClass = {astro-ph.GA},
       adsurl = {https://ui.adsabs.harvard.edu/abs/2019Natur.569..519K}
}

@ARTICLE{Krumholz_2005,
       author = {{Krumholz}, Mark R. and {McKee}, Christopher F.},
        title = "{A General Theory of Turbulence-regulated Star Formation, from Spirals to Ultraluminous Infrared Galaxies}",
      journal = {\apj},
         year = 2005,
        month = sep,
       volume = {630},
       number = {1},
        pages = {250-268},
          doi = {10.1086/431734},
archivePrefix = {arXiv},
       eprint = {astro-ph/0505177},
 primaryClass = {astro-ph},
       adsurl = {https://ui.adsabs.harvard.edu/abs/2005ApJ...630..250K}
}

@ARTICLE{Kundic_1995,
       author = {{Kundic}, Tomislav and {Ostriker}, Jeremiah P.},
        title = "{Tidal-Shock Relaxation: A Reexamination of Tidal Shocks in Stellar Systems}",
      journal = {\apj},
         year = 1995,
        month = jan,
       volume = {438},
        pages = {702},
          doi = {10.1086/175114},
       adsurl = {https://ui.adsabs.harvard.edu/abs/1995ApJ...438..702K}
}

@ARTICLE{Lacey_1993,
       author = {{Lacey}, Cedric and {Cole}, Shaun},
        title = "{Merger rates in hierarchical models of galaxy formation}",
      journal = {\mnras},
         year = 1993,
        month = jun,
       volume = {262},
       number = {3},
        pages = {627-649},
          doi = {10.1093/mnras/262.3.627},
       adsurl = {https://ui.adsabs.harvard.edu/abs/1993MNRAS.262..627L}
}

@ARTICLE{Lahen_2020,
       author = {{Lah{\'e}n}, Natalia and {Naab}, Thorsten and {Johansson}, Peter H. and {Elmegreen}, Bruce and {Hu}, Chia-Yu and {Walch}, Stefanie and {Steinwandel}, Ulrich P. and {Moster}, Benjamin P.},
        title = "{The GRIFFIN Project{\textemdash}Formation of Star Clusters with Individual Massive Stars in a Simulated Dwarf Galaxy Starburst}",
      journal = {\apj},
         year = 2020,
        month = mar,
       volume = {891},
       number = {1},
          eid = {2},
        pages = {2},
          doi = {10.3847/1538-4357/ab7190},
archivePrefix = {arXiv},
       eprint = {1911.05093},
 primaryClass = {astro-ph.GA},
       adsurl = {https://ui.adsabs.harvard.edu/abs/2020ApJ...891....2L}
}

@ARTICLE{Li_2017,
       author = {{Li}, Hui and {Gnedin}, Oleg Y. and {Gnedin}, Nickolay Y. and {Meng}, Xi and {Semenov}, Vadim A. and {Kravtsov}, Andrey V.},
        title = "{Star Cluster Formation in Cosmological Simulations. I. Properties of Young Clusters}",
      journal = {\apj},
         year = 2017,
        month = jan,
       volume = {834},
       number = {1},
          eid = {69},
        pages = {69},
          doi = {10.3847/1538-4357/834/1/69},
archivePrefix = {arXiv},
       eprint = {1608.03244},
 primaryClass = {astro-ph.GA},
       adsurl = {https://ui.adsabs.harvard.edu/abs/2017ApJ...834...69L}
}

@ARTICLE{Li_2018,
       author = {{Li}, Hui and {Gnedin}, Oleg Y. and {Gnedin}, Nickolay Y.},
        title = "{Star Cluster Formation in Cosmological Simulations. II. Effects of Star Formation Efficiency and Stellar Feedback}",
      journal = {\apj},
         year = 2018,
        month = jul,
       volume = {861},
       number = {2},
          eid = {107},
        pages = {107},
          doi = {10.3847/1538-4357/aac9b8},
archivePrefix = {arXiv},
       eprint = {1712.01219},
 primaryClass = {astro-ph.GA},
       adsurl = {https://ui.adsabs.harvard.edu/abs/2018ApJ...861..107L}
}

@ARTICLE{Li_2019,
       author = {{Li}, Hui and {Gnedin}, Oleg Y.},
        title = "{Star cluster formation in cosmological simulations - III. Dynamical and chemical evolution}",
      journal = {\mnras},
         year = 2019,
        month = jul,
       volume = {486},
       number = {3},
        pages = {4030-4043},
          doi = {10.1093/mnras/stz1114},
archivePrefix = {arXiv},
       eprint = {1810.11036},
 primaryClass = {astro-ph.GA},
       adsurl = {https://ui.adsabs.harvard.edu/abs/2019MNRAS.486.4030L}
}

@ARTICLE{Li_2022,
       author = {{Li}, Hui and {Vogelsberger}, Mark and {Bryan}, Greg L. and {Marinacci}, Federico and {Sales}, Laura V. and {Torrey}, Paul},
        title = "{Formation and evolution of young massive clusters in galaxy mergers: the SMUGGLE view}",
      journal = {\mnras},
         year = 2022,
        month = jul,
       volume = {514},
       number = {1},
        pages = {265-279},
          doi = {10.1093/mnras/stac1136},
archivePrefix = {arXiv},
       eprint = {2109.10356},
 primaryClass = {astro-ph.GA},
       adsurl = {https://ui.adsabs.harvard.edu/abs/2022MNRAS.514..265L}
}

@ARTICLE{Ma_2020,
       author = {{Ma}, Xiangcheng and {Grudi{\'c}}, Michael Y. and {Quataert}, Eliot and {Hopkins}, Philip F. and {Faucher-Gigu{\`e}re}, Claude-Andr{\'e} and {Boylan-Kolchin}, Michael and {Wetzel}, Andrew and {Kim}, Ji-hoon and {Murray}, Norman and {Kere{\v{s}}}, Du{\v{s}}an},
        title = "{Self-consistent proto-globular cluster formation in cosmological simulations of high-redshift galaxies}",
      journal = {\mnras},
         year = 2020,
        month = apr,
       volume = {493},
       number = {3},
        pages = {4315-4332},
          doi = {10.1093/mnras/staa527},
archivePrefix = {arXiv},
       eprint = {1906.11261},
 primaryClass = {astro-ph.GA},
       adsurl = {https://ui.adsabs.harvard.edu/abs/2020MNRAS.493.4315M}
}

@ARTICLE{Mandelker_2014,
       author = {{Mandelker}, Nir and {Dekel}, Avishai and {Ceverino}, Daniel and {Tweed}, Dylan and {Moody}, Christopher E. and {Primack}, Joel},
        title = "{The population of giant clumps in simulated high-z galaxies: in situ and ex situ migration and survival}",
      journal = {\mnras},
         year = 2014,
        month = oct,
       volume = {443},
       number = {4},
        pages = {3675-3702},
          doi = {10.1093/mnras/stu1340},
archivePrefix = {arXiv},
       eprint = {1311.0013},
 primaryClass = {astro-ph.CO},
       adsurl = {https://ui.adsabs.harvard.edu/abs/2014MNRAS.443.3675M}
}

@ARTICLE{Marinacci_2014,
       author = {{Marinacci}, Federico and {Pakmor}, R{\"u}diger and {Springel}, Volker},
        title = "{The formation of disc galaxies in high-resolution moving-mesh cosmological simulations}",
      journal = {\mnras},
         year = 2014,
        month = jan,
       volume = {437},
       number = {2},
        pages = {1750-1775},
          doi = {10.1093/mnras/stt2003},
archivePrefix = {arXiv},
       eprint = {1305.5360},
 primaryClass = {astro-ph.CO},
       adsurl = {https://ui.adsabs.harvard.edu/abs/2014MNRAS.437.1750M}
}

@ARTICLE{Murali_1997a,
       author = {{Murali}, Chigurupati and {Weinberg}, Martin D.},
        title = "{Evolution of the Galactic globular cluster system}",
      journal = {\mnras},
         year = 1997,
        month = nov,
       volume = {291},
       number = {4},
        pages = {717-731},
          doi = {10.1093/mnras/291.4.717},
       adsurl = {https://ui.adsabs.harvard.edu/abs/1997MNRAS.291..717M}
}

@ARTICLE{Murali_1997b,
       author = {{Murali}, Chigurupati and {Weinberg}, Martin D.},
        title = "{The effect of the Galactic spheroid on globular cluster evolution}",
      journal = {\mnras},
         year = 1997,
        month = jul,
       volume = {288},
       number = {3},
        pages = {749-766},
          doi = {10.1093/mnras/288.3.749},
archivePrefix = {arXiv},
       eprint = {astro-ph/9604049},
 primaryClass = {astro-ph},
       adsurl = {https://ui.adsabs.harvard.edu/abs/1997MNRAS.288..749M}
}

@ARTICLE{Murali_1997c,
       author = {{Murali}, Chigurupati and {Weinberg}, Martin D.},
        title = "{Globular cluster evolution in M87 and fundamental plane ellipticals}",
      journal = {\mnras},
         year = 1997,
        month = jul,
       volume = {288},
       number = {3},
        pages = {767-776},
          doi = {10.1093/mnras/288.3.767},
archivePrefix = {arXiv},
       eprint = {astro-ph/9602058},
 primaryClass = {astro-ph},
       adsurl = {https://ui.adsabs.harvard.edu/abs/1997MNRAS.288..767M}
}

@ARTICLE{Murray_2011,
       author = {{Murray}, Norman},
        title = "{Star Formation Efficiencies and Lifetimes of Giant Molecular Clouds in the Milky Way}",
      journal = {\apj},
         year = 2011,
        month = mar,
       volume = {729},
       number = {2},
          eid = {133},
        pages = {133},
          doi = {10.1088/0004-637X/729/2/133},
archivePrefix = {arXiv},
       eprint = {1007.3270},
 primaryClass = {astro-ph.GA},
       adsurl = {https://ui.adsabs.harvard.edu/abs/2011ApJ...729..133M}
}

@ARTICLE{Ostriker_1972,
       author = {{Ostriker}, Jeremiah P. and {Spitzer}, Jr., Lyman and {Chevalier}, Roger A.},
        title = "{On the Evolution of Globular Clusters}",
      journal = {\apjl},
         year = 1972,
        month = sep,
       volume = {176},
        pages = {L51},
          doi = {10.1086/181018},
       adsurl = {https://ui.adsabs.harvard.edu/abs/1972ApJ...176L..51O}
}

@ARTICLE{Pakmor_2017,
       author = {{Pakmor}, R{\"u}diger and {G{\'o}mez}, Facundo A. and {Grand}, Robert J.~J. and {Marinacci}, Federico and {Simpson}, Christine M. and {Springel}, Volker and {Campbell}, David J.~R. and {Frenk}, Carlos S. and {Guillet}, Thomas and {Pfrommer}, Christoph and {White}, Simon D.~M.},
        title = "{Magnetic field formation in the Milky Way like disc galaxies of the Auriga project}",
      journal = {\mnras},
         year = 2017,
        month = aug,
       volume = {469},
       number = {3},
        pages = {3185-3199},
          doi = {10.1093/mnras/stx1074},
archivePrefix = {arXiv},
       eprint = {1701.07028},
 primaryClass = {astro-ph.GA},
       adsurl = {https://ui.adsabs.harvard.edu/abs/2017MNRAS.469.3185P}
}

@ARTICLE{Pascale_2025,
       author = {{Pascale}, R. and {Calura}, F. and {Vesperini}, E. and {Rosdahl}, J. and {Nipoti}, C. and {Giunchi}, E. and {Lacchin}, E. and {Lupi}, A. and {Messa}, M. and {Meneghetti}, M. and {Ragagnin}, A. and {Vanzella}, E. and {Zanella}, A.},
        title = "{SIEGE: IV. Compact star clusters in cosmological simulations with a high star formation efficiency and subparsec resolution}",
      journal = {\aap},
         year = 2025,
        month = jun,
       volume = {699},
          eid = {A31},
        pages = {A31},
          doi = {10.1051/0004-6361/202453252},
archivePrefix = {arXiv},
       eprint = {2505.06346},
 primaryClass = {astro-ph.GA},
       adsurl = {https://ui.adsabs.harvard.edu/abs/2025A&A...699A..31P}
}

@ARTICLE{Pfeffer_2018,
       author = {{Pfeffer}, Joel and {Kruijssen}, J.~M. Diederik and {Crain}, Robert A. and {Bastian}, Nate},
        title = "{The E-MOSAICS project: simulating the formation and co-evolution of galaxies and their star cluster populations}",
      journal = {\mnras},
         year = 2018,
        month = apr,
       volume = {475},
       number = {4},
        pages = {4309-4346},
          doi = {10.1093/mnras/stx3124},
archivePrefix = {arXiv},
       eprint = {1712.00019},
 primaryClass = {astro-ph.GA},
       adsurl = {https://ui.adsabs.harvard.edu/abs/2018MNRAS.475.4309P}
}

@ARTICLE{Phipps_2020,
       author = {{Phipps}, Frederika and {Khochfar}, Sadegh and {Varri}, Anna Lisa and {Dalla Vecchia}, Claudio},
        title = "{The First Billion Years project: Finding infant globular clusters at z = 6}",
      journal = {\aap},
         year = 2020,
        month = sep,
       volume = {641},
          eid = {A132},
        pages = {A132},
          doi = {10.1051/0004-6361/202037884},
archivePrefix = {arXiv},
       eprint = {1910.09924},
 primaryClass = {astro-ph.GA},
       adsurl = {https://ui.adsabs.harvard.edu/abs/2020A&A...641A.132P}
}

@ARTICLE{Caldwell_2016,
       author = {{Caldwell}, Nelson and {Romanowsky}, Aaron J.},
        title = "{Star Clusters in M31. VII. Global Kinematics and Metallicity Subpopulations of the Globular Clusters}",
      journal = {\apj},
         year = 2016,
        month = jun,
       volume = {824},
       number = {1},
          eid = {42},
        pages = {42},
          doi = {10.3847/0004-637X/824/1/42},
archivePrefix = {arXiv},
       eprint = {1603.06947},
 primaryClass = {astro-ph.GA},
       adsurl = {https://ui.adsabs.harvard.edu/abs/2016ApJ...824...42C}
}

@ARTICLE{Prieto_2008,
       author = {{Prieto}, Jos{\'e} L. and {Gnedin}, Oleg Y.},
        title = "{Dynamical Evolution of Globular Clusters in Hierarchical Cosmology}",
      journal = {\apj},
         year = 2008,
        month = dec,
       volume = {689},
       number = {2},
        pages = {919-935},
          doi = {10.1086/591777},
archivePrefix = {arXiv},
       eprint = {astro-ph/0608069},
 primaryClass = {astro-ph},
       adsurl = {https://ui.adsabs.harvard.edu/abs/2008ApJ...689..919P}
}

@ARTICLE{Reina_Campos_2022,
       author = {{Reina-Campos}, Marta and {Keller}, Benjamin W. and {Kruijssen}, J.~M. Diederik and {Gensior}, Jindra and {Trujillo-Gomez}, Sebastian and {Jeffreson}, Sarah M.~R. and {Pfeffer}, Joel L. and {Sills}, Alison},
        title = "{Introducing EMP-Pathfinder: modelling the simultaneous formation and evolution of stellar clusters in their host galaxies}",
      journal = {\mnras},
         year = 2022,
        month = dec,
       volume = {517},
       number = {3},
        pages = {3144-3180},
          doi = {10.1093/mnras/stac1934},
archivePrefix = {arXiv},
       eprint = {2202.06961},
 primaryClass = {astro-ph.GA},
       adsurl = {https://ui.adsabs.harvard.edu/abs/2022MNRAS.517.3144R}
}

@ARTICLE{Reina_Campos_2017,
       author = {{Reina-Campos}, Marta and {Kruijssen}, J.~M. Diederik},
        title = "{A unified model for the maximum mass scales of molecular clouds, stellar clusters and high-redshift clumps}",
      journal = {\mnras},
         year = 2017,
        month = aug,
       volume = {469},
       number = {2},
        pages = {1282-1298},
          doi = {10.1093/mnras/stx790},
archivePrefix = {arXiv},
       eprint = {1704.00732},
 primaryClass = {astro-ph.GA},
       adsurl = {https://ui.adsabs.harvard.edu/abs/2017MNRAS.469.1282R}
}

@ARTICLE{Reina_Campos_2023,
       author = {{Reina-Campos}, Marta and {Sills}, Alison and {Bichon}, Godefroy},
        title = "{Initial sizes of star clusters: implications for cluster dissolution during galaxy evolution}",
      journal = {\mnras},
         year = 2023,
        month = sep,
       volume = {524},
       number = {1},
        pages = {968-980},
          doi = {10.1093/mnras/stad1879},
archivePrefix = {arXiv},
       eprint = {2306.17701},
 primaryClass = {astro-ph.GA},
       adsurl = {https://ui.adsabs.harvard.edu/abs/2023MNRAS.524..968R}
}

@ARTICLE{Renaud_2017,
       author = {{Renaud}, Florent and {Agertz}, Oscar and {Gieles}, Mark},
        title = "{The origin of the Milky Way globular clusters}",
      journal = {\mnras},
         year = 2017,
        month = mar,
       volume = {465},
       number = {3},
        pages = {3622-3636},
          doi = {10.1093/mnras/stw2969},
archivePrefix = {arXiv},
       eprint = {1610.03101},
 primaryClass = {astro-ph.GA},
       adsurl = {https://ui.adsabs.harvard.edu/abs/2017MNRAS.465.3622R}
}

@ARTICLE{Renaud_2009,
       author = {{Renaud}, F. and {Boily}, C.~M. and {Naab}, T. and {Theis}, Ch.},
        title = "{Fully Compressive Tides in Galaxy Mergers}",
      journal = {\apj},
         year = 2009,
        month = nov,
       volume = {706},
       number = {1},
        pages = {67-82},
          doi = {10.1088/0004-637X/706/1/67},
archivePrefix = {arXiv},
       eprint = {0910.0196},
 primaryClass = {astro-ph.CO},
       adsurl = {https://ui.adsabs.harvard.edu/abs/2009ApJ...706...67R}
}

@ARTICLE{Renaud_2015,
       author = {{Renaud}, Florent and {Bournaud}, Fr{\'e}d{\'e}ric and {Duc}, Pierre-Alain},
        title = "{A parsec-resolution simulation of the Antennae galaxies: formation of star clusters during the merger}",
      journal = {\mnras},
         year = 2015,
        month = jan,
       volume = {446},
       number = {2},
        pages = {2038-2054},
          doi = {10.1093/mnras/stu2208},
archivePrefix = {arXiv},
       eprint = {1410.5754},
 primaryClass = {astro-ph.GA},
       adsurl = {https://ui.adsabs.harvard.edu/abs/2015MNRAS.446.2038R}
}

@ARTICLE{Renaud_2014,
       author = {{Renaud}, F. and {Bournaud}, F. and {Kraljic}, K. and {Duc}, P. -A.},
        title = "{Starbursts triggered by intergalactic tides andinterstellar compressive turbulence.}",
      journal = {\mnras},
         year = 2014,
        month = jul,
       volume = {442},
        pages = {L33-L37},
          doi = {10.1093/mnrasl/slu050},
archivePrefix = {arXiv},
       eprint = {1403.7316},
 primaryClass = {astro-ph.GA},
       adsurl = {https://ui.adsabs.harvard.edu/abs/2014MNRAS.442L..33R}
}

@ARTICLE{Rodriguez_2023,
       author = {{Rodriguez}, Carl L. and {Hafen}, Zachary and {Grudi{\'c}}, Michael Y. and {Lamberts}, Astrid and {Sharma}, Kuldeep and {Faucher-Gigu{\`e}re}, Claude-Andr{\'e} and {Wetzel}, Andrew},
        title = "{Great balls of FIRE II: The evolution and destruction of star clusters across cosmic time in a Milky Way-mass galaxy}",
      journal = {\mnras},
         year = 2023,
        month = may,
       volume = {521},
       number = {1},
        pages = {124-147},
          doi = {10.1093/mnras/stad578},
archivePrefix = {arXiv},
       eprint = {2203.16547},
 primaryClass = {astro-ph.GA},
       adsurl = {https://ui.adsabs.harvard.edu/abs/2023MNRAS.521..124R}
}

@ARTICLE{Schaal_2015,
       author = {{Schaal}, Kevin and {Springel}, Volker},
        title = "{Shock finding on a moving mesh - I. Shock statistics in non-radiative cosmological simulations}",
      journal = {\mnras},
         year = 2015,
        month = feb,
       volume = {446},
       number = {4},
        pages = {3992-4007},
          doi = {10.1093/mnras/stu2386},
archivePrefix = {arXiv},
       eprint = {1407.4117},
 primaryClass = {astro-ph.CO},
       adsurl = {https://ui.adsabs.harvard.edu/abs/2015MNRAS.446.3992S}
}

@ARTICLE{Schaal_2016,
       author = {{Schaal}, Kevin and {Springel}, Volker and {Pakmor}, R{\"u}diger and {Pfrommer}, Christoph and {Nelson}, Dylan and {Vogelsberger}, Mark and {Genel}, Shy and {Pillepich}, Annalisa and {Sijacki}, Debora and {Hernquist}, Lars},
        title = "{Shock finding on a moving-mesh - II. Hydrodynamic shocks in the Illustris universe}",
      journal = {\mnras},
         year = 2016,
        month = oct,
       volume = {461},
       number = {4},
        pages = {4441-4465},
          doi = {10.1093/mnras/stw1587},
archivePrefix = {arXiv},
       eprint = {1604.07401},
 primaryClass = {astro-ph.CO},
       adsurl = {https://ui.adsabs.harvard.edu/abs/2016MNRAS.461.4441S}
}

@ARTICLE{Schaye_2015,
       author = {{Schaye}, Joop and {Crain}, Robert A. and {Bower}, Richard G. and {Furlong}, Michelle and {Schaller}, Matthieu and {Theuns}, Tom and {Dalla Vecchia}, Claudio and {Frenk}, Carlos S. and {McCarthy}, I.~G. and {Helly}, John C. and {Jenkins}, Adrian and {Rosas-Guevara}, Y.~M. and {White}, Simon D.~M. and {Baes}, Maarten and {Booth}, C.~M. and {Camps}, Peter and {Navarro}, Julio F. and {Qu}, Yan and {Rahmati}, Alireza and {Sawala}, Till and {Thomas}, Peter A. and {Trayford}, James},
        title = "{The EAGLE project: simulating the evolution and assembly of galaxies and their environments}",
      journal = {\mnras},
         year = 2015,
        month = jan,
       volume = {446},
       number = {1},
        pages = {521-554},
          doi = {10.1093/mnras/stu2058},
archivePrefix = {arXiv},
       eprint = {1407.7040},
 primaryClass = {astro-ph.GA},
       adsurl = {https://ui.adsabs.harvard.edu/abs/2015MNRAS.446..521S}
}

@ARTICLE{Schechter_1976,
       author = {{Schechter}, P.},
        title = "{An analytic expression for the luminosity function for galaxies.}",
      journal = {\apj},
         year = 1976,
        month = jan,
       volume = {203},
        pages = {297-306},
          doi = {10.1086/154079},
       adsurl = {https://ui.adsabs.harvard.edu/abs/1976ApJ...203..297S}
}

@ARTICLE{Searle_1978,
       author = {{Searle}, L. and {Zinn}, R.},
        title = "{Composition of halo clusters and the formation of the galactic halo.}",
      journal = {\apj},
         year = 1978,
        month = oct,
       volume = {225},
        pages = {357-379},
          doi = {10.1086/156499},
       adsurl = {https://ui.adsabs.harvard.edu/abs/1978ApJ...225..357S}
}

@BOOK{Spitzer_1987,
       author = {{Spitzer}, Lyman},
        title = "{Dynamical evolution of globular clusters}",
         year = 1987,
       adsurl = {https://ui.adsabs.harvard.edu/abs/1987degc.book.....S}
}

@ARTICLE{Springel_2005,
       author = {{Springel}, Volker},
        title = "{The cosmological simulation code GADGET-2}",
      journal = {\mnras},
         year = 2005,
        month = dec,
       volume = {364},
       number = {4},
        pages = {1105-1134},
          doi = {10.1111/j.1365-2966.2005.09655.x},
archivePrefix = {arXiv},
       eprint = {astro-ph/0505010},
 primaryClass = {astro-ph},
       adsurl = {https://ui.adsabs.harvard.edu/abs/2005MNRAS.364.1105S}
}

@ARTICLE{Springel_2010,
       author = {{Springel}, Volker},
        title = "{E pur si muove: Galilean-invariant cosmological hydrodynamical simulations on a moving mesh}",
      journal = {\mnras},
         year = 2010,
        month = jan,
       volume = {401},
       number = {2},
        pages = {791-851},
          doi = {10.1111/j.1365-2966.2009.15715.x},
archivePrefix = {arXiv},
       eprint = {0901.4107},
 primaryClass = {astro-ph.CO},
       adsurl = {https://ui.adsabs.harvard.edu/abs/2010MNRAS.401..791S}
}

@ARTICLE{Springel_2003,
       author = {{Springel}, Volker and {Hernquist}, Lars},
        title = "{Cosmological smoothed particle hydrodynamics simulations: a hybrid multiphase model for star formation}",
      journal = {\mnras},
         year = 2003,
        month = feb,
       volume = {339},
       number = {2},
        pages = {289-311},
          doi = {10.1046/j.1365-8711.2003.06206.x},
archivePrefix = {arXiv},
       eprint = {astro-ph/0206393},
 primaryClass = {astro-ph},
       adsurl = {https://ui.adsabs.harvard.edu/abs/2003MNRAS.339..289S}
}

@ARTICLE{Strader_2005,
       author = {{Strader}, Jay and {Brodie}, Jean P. and {Cenarro}, A.~J. and {Beasley}, Michael A. and {Forbes}, Duncan A.},
        title = "{Extragalactic Globular Clusters: Old Spectroscopic Ages and New Views on Their Formation}",
      journal = {\aj},
         year = 2005,
        month = oct,
       volume = {130},
       number = {4},
        pages = {1315-1323},
          doi = {10.1086/432717},
archivePrefix = {arXiv},
       eprint = {astro-ph/0506289},
 primaryClass = {astro-ph},
       adsurl = {https://ui.adsabs.harvard.edu/abs/2005AJ....130.1315S}
}

@ARTICLE{Toomre_1964,
       author = {{Toomre}, A.},
        title = "{On the gravitational stability of a disk of stars.}",
      journal = {\apj},
         year = 1964,
        month = may,
       volume = {139},
        pages = {1217-1238},
          doi = {10.1086/147861},
       adsurl = {https://ui.adsabs.harvard.edu/abs/1964ApJ...139.1217T}
}

@ARTICLE{Utomo_2018,
       author = {{Utomo}, Dyas and {Sun}, Jiayi and {Leroy}, Adam K. and {Kruijssen}, J.~M. Diederik and {Schinnerer}, Eva and {Schruba}, Andreas and {Bigiel}, Frank and {Blanc}, Guillermo A. and {Chevance}, M{\'e}lanie and {Emsellem}, Eric and {Herrera}, Cinthya and {Hygate}, Alexander P.~S. and {Kreckel}, Kathryn and {Ostriker}, Eve C. and {Pety}, Jerome and {Querejeta}, Miguel and {Rosolowsky}, Erik and {Sandstrom}, Karin M. and {Usero}, Antonio},
        title = "{Star Formation Efficiency per Free-fall Time in nearby Galaxies}",
      journal = {\apjl},
         year = 2018,
        month = jul,
       volume = {861},
       number = {2},
          eid = {L18},
        pages = {L18},
          doi = {10.3847/2041-8213/aacf8f},
archivePrefix = {arXiv},
       eprint = {1806.11121},
 primaryClass = {astro-ph.GA},
       adsurl = {https://ui.adsabs.harvard.edu/abs/2018ApJ...861L..18U}
}

@ARTICLE{Valenzuela_2021,
       author = {{Valenzuela}, Lucas M. and {Moster}, Benjamin P. and {Remus}, Rhea-Silvia and {O'Leary}, Joseph A. and {Burkert}, Andreas},
        title = "{Globular cluster numbers in dark matter haloes in a dual formation scenario: an empirical model within EMERGE}",
      journal = {\mnras},
         year = 2021,
        month = aug,
       volume = {505},
       number = {4},
        pages = {5815-5832},
          doi = {10.1093/mnras/stab1701},
archivePrefix = {arXiv},
       eprint = {2012.09172},
 primaryClass = {astro-ph.GA},
       adsurl = {https://ui.adsabs.harvard.edu/abs/2021MNRAS.505.5815V}
}

@ARTICLE{VandenBerg_2013,
       author = {{VandenBerg}, Don A. and {Brogaard}, K. and {Leaman}, R. and {Casagrande}, L.},
        title = "{The Ages of 55 Globular Clusters as Determined Using an Improved \textbackslashDelta V\^HB\_TO Method along with Color-Magnitude Diagram Constraints, and Their Implications for Broader Issues}",
      journal = {\apj},
         year = 2013,
        month = oct,
       volume = {775},
       number = {2},
          eid = {134},
        pages = {134},
          doi = {10.1088/0004-637X/775/2/134},
archivePrefix = {arXiv},
       eprint = {1308.2257},
 primaryClass = {astro-ph.GA},
       adsurl = {https://ui.adsabs.harvard.edu/abs/2013ApJ...775..134V}
}

@ARTICLE{Vesperini_2001,
       author = {{Vesperini}, E.},
        title = "{Evolution of globular cluster systems in elliptical galaxies - II. Power-law initial mass function}",
      journal = {\mnras},
         year = 2001,
        month = apr,
       volume = {322},
       number = {2},
        pages = {247-256},
          doi = {10.1046/j.1365-8711.2001.04072.x},
archivePrefix = {arXiv},
       eprint = {astro-ph/0010111},
 primaryClass = {astro-ph},
       adsurl = {https://ui.adsabs.harvard.edu/abs/2001MNRAS.322..247V}
}

@ARTICLE{Vogelsberger_2013,
       author = {{Vogelsberger}, Mark and {Genel}, Shy and {Sijacki}, Debora and {Torrey}, Paul and {Springel}, Volker and {Hernquist}, Lars},
        title = "{A model for cosmological simulations of galaxy formation physics}",
      journal = {\mnras},
         year = 2013,
        month = dec,
       volume = {436},
       number = {4},
        pages = {3031-3067},
          doi = {10.1093/mnras/stt1789},
archivePrefix = {arXiv},
       eprint = {1305.2913},
 primaryClass = {astro-ph.CO},
       adsurl = {https://ui.adsabs.harvard.edu/abs/2013MNRAS.436.3031V}
}

@ARTICLE{Wadsley_2008,
       author = {{Wadsley}, J.~W. and {Veeravalli}, G. and {Couchman}, H.~M.~P.},
        title = "{On the treatment of entropy mixing in numerical cosmology}",
      journal = {\mnras},
         year = 2008,
        month = jun,
       volume = {387},
       number = {1},
        pages = {427-438},
          doi = {10.1111/j.1365-2966.2008.13260.x},
       adsurl = {https://ui.adsabs.harvard.edu/abs/2008MNRAS.387..427W}
}

@ARTICLE{Wainer_2022,
       author = {{Wainer}, Tobin M. and {Johnson}, L. Clifton and {Seth}, Anil C. and {Torresvillanueva}, Estephani E. and {Dalcanton}, Julianne J. and {Durbin}, Meredith J. and {Dolphin}, Andrew and {Weisz}, Daniel R. and {Williams}, Benjamin F. and {Phatter Collaboration}},
        title = "{The Panchromatic Hubble Andromeda Treasury: Triangulum Extended Region (PHATTER). III. The Mass Function of Young Stellar Clusters in M33}",
      journal = {\apj},
         year = 2022,
        month = mar,
       volume = {928},
       number = {1},
          eid = {15},
        pages = {15},
          doi = {10.3847/1538-4357/ac51cf},
archivePrefix = {arXiv},
       eprint = {2201.04161},
 primaryClass = {astro-ph.GA},
       adsurl = {https://ui.adsabs.harvard.edu/abs/2022ApJ...928...15W}
}

@ARTICLE{Weinberg_1994a,
       author = {{Weinberg}, Martin D.},
        title = "{Adiobatic Invariants in Stellar Dynamics. I. Basic Concepts}",
      journal = {\aj},
         year = 1994,
        month = oct,
       volume = {108},
        pages = {1398},
          doi = {10.1086/117161},
archivePrefix = {arXiv},
       eprint = {astro-ph/9404015},
 primaryClass = {astro-ph},
       adsurl = {https://ui.adsabs.harvard.edu/abs/1994AJ....108.1398W}
}

@ARTICLE{Weinberg_1994b,
       author = {{Weinberg}, Martin D.},
        title = "{Adiabatic Invariants in Stellar Dynamics. II. Gravitational Shocking}",
      journal = {\aj},
         year = 1994,
        month = oct,
       volume = {108},
        pages = {1403},
          doi = {10.1086/117162},
archivePrefix = {arXiv},
       eprint = {astro-ph/9404016},
 primaryClass = {astro-ph},
       adsurl = {https://ui.adsabs.harvard.edu/abs/1994AJ....108.1403W}
}

@ARTICLE{Weinberg_1994c,
       author = {{Weinberg}, Martin D.},
        title = "{Adiabatic Invariants in Stellar Dynamics . III. Application to Globular Cluster Evolution}",
      journal = {\aj},
         year = 1994,
        month = oct,
       volume = {108},
        pages = {1414},
          doi = {10.1086/117163},
archivePrefix = {arXiv},
       eprint = {astro-ph/9404017},
 primaryClass = {astro-ph},
       adsurl = {https://ui.adsabs.harvard.edu/abs/1994AJ....108.1414W}
}

@ARTICLE{Whitmore_2010,
       author = {{Whitmore}, Bradley C. and {Chandar}, Rupali and {Schweizer}, Fran{\c{c}}ois and {Rothberg}, Barry and {Leitherer}, Claus and {Rieke}, Marcia and {Rieke}, George and {Blair}, W.~P. and {Mengel}, S. and {Alonso-Herrero}, A.},
        title = "{The Antennae Galaxies (NGC 4038/4039) Revisited: Advanced Camera for Surveys and NICMOS Observations of a Prototypical Merger}",
      journal = {\aj},
         year = 2010,
        month = jul,
       volume = {140},
       number = {1},
        pages = {75-109},
          doi = {10.1088/0004-6256/140/1/75},
archivePrefix = {arXiv},
       eprint = {1005.0629},
 primaryClass = {astro-ph.EP},
       adsurl = {https://ui.adsabs.harvard.edu/abs/2010AJ....140...75W}
}

@ARTICLE{Williams_2025,
       author = {{Williams}, Claire E. and {Naoz}, Smadar and {Lake}, William and {Burkhart}, Blakesley and {Marinacci}, Federico and {Vogelsberger}, Mark and {Yoshida}, Naoki and {Menon}, Shyam H. and {Chen}, Avi and {Adamo}, Angela},
        title = "{{\ensuremath{\Lambda}}CDM Star Clusters at Cosmic Dawn: Stellar Densities, Environment, and Equilibrium}",
      journal = {\apj},
         year = 2025,
        month = sep,
       volume = {990},
       number = {2},
          eid = {135},
        pages = {135},
          doi = {10.3847/1538-4357/adf19d},
archivePrefix = {arXiv},
       eprint = {2502.17561},
 primaryClass = {astro-ph.GA},
       adsurl = {https://ui.adsabs.harvard.edu/abs/2025ApJ...990..135W}
}

@ARTICLE{Zhang_1999,
       author = {{Zhang}, Qing and {Fall}, S. Michael},
        title = "{The Mass Function of Young Star Clusters in the ``Antennae'' Galaxies}",
      journal = {\apjl},
         year = 1999,
        month = dec,
       volume = {527},
       number = {2},
        pages = {L81-L84},
          doi = {10.1086/312412},
archivePrefix = {arXiv},
       eprint = {astro-ph/9911229},
 primaryClass = {astro-ph},
       adsurl = {https://ui.adsabs.harvard.edu/abs/1999ApJ...527L..81Z}
}

@ARTICLE{Portegies_Zwart_2010,
       author = {{Portegies Zwart}, Simon F. and {McMillan}, Stephen L.~W. and {Gieles}, Mark},
        title = "{Young Massive Star Clusters}",
      journal = {\araa},
         year = 2010,
        month = sep,
       volume = {48},
        pages = {431-493},
          doi = {10.1146/annurev-astro-081309-130834},
archivePrefix = {arXiv},
       eprint = {1002.1961},
 primaryClass = {astro-ph.GA},
       adsurl = {https://ui.adsabs.harvard.edu/abs/2010ARA&A..48..431P}
}

@ARTICLE{Spitler_2009,
       author = {{Spitler}, L.~R. and {Forbes}, D.~A.},
        title = "{A new method for estimating dark matter halo masses using globular cluster systems}",
      journal = {\mnras},
         year = 2009,
        month = jan,
       volume = {392},
       number = {1},
        pages = {L1-L5},
          doi = {10.1111/j.1745-3933.2008.00567.x},
archivePrefix = {arXiv},
       eprint = {0809.5057},
 primaryClass = {astro-ph},
       adsurl = {https://ui.adsabs.harvard.edu/abs/2009MNRAS.392L...1S}
}

@ARTICLE{Forbes_2018,
       author = {{Forbes}, Duncan A. and {Read}, Justin I. and {Gieles}, Mark and {Collins}, Michelle L.~M.},
        title = "{Extending the globular cluster system-halo mass relation to the lowest galaxy masses}",
      journal = {\mnras},
         year = 2018,
        month = dec,
       volume = {481},
       number = {4},
        pages = {5592-5605},
          doi = {10.1093/mnras/sty2584},
archivePrefix = {arXiv},
       eprint = {1809.07831},
 primaryClass = {astro-ph.GA},
       adsurl = {https://ui.adsabs.harvard.edu/abs/2018MNRAS.481.5592F}
}

@ARTICLE{Harris_2017,
       author = {{Harris}, William E. and {Blakeslee}, John P. and {Harris}, Gretchen L.~H.},
        title = "{Galactic Dark Matter Halos and Globular Cluster Populations. III. Extension to Extreme Environments}",
      journal = {\apj},
         year = 2017,
        month = feb,
       volume = {836},
       number = {1},
          eid = {67},
        pages = {67},
          doi = {10.3847/1538-4357/836/1/67},
archivePrefix = {arXiv},
       eprint = {1701.04845},
 primaryClass = {astro-ph.GA},
       adsurl = {https://ui.adsabs.harvard.edu/abs/2017ApJ...836...67H}
}

@ARTICLE{Dickson_2024,
       author = {{Dickson}, N. and {Smith}, P.~J. and {H{\'e}nault-Brunet}, V. and {Gieles}, M. and {Baumgardt}, H.},
        title = "{Multimass modelling of milky way globular clusters - II. Present-day black hole populations}",
      journal = {\mnras},
         year = 2024,
        month = mar,
       volume = {529},
       number = {1},
        pages = {331-347},
          doi = {10.1093/mnras/stae470},
archivePrefix = {arXiv},
       eprint = {2308.13037},
 primaryClass = {astro-ph.GA},
       adsurl = {https://ui.adsabs.harvard.edu/abs/2024MNRAS.529..331D}
}

@ARTICLE{Kremer_2020,
       author = {{Kremer}, Kyle and {Ye}, Claire S. and {Rui}, Nicholas Z. and {Weatherford}, Newlin C. and {Chatterjee}, Sourav and {Fragione}, Giacomo and {Rodriguez}, Carl L. and {Spera}, Mario and {Rasio}, Frederic A.},
        title = "{Modeling Dense Star Clusters in the Milky Way and Beyond with the CMC Cluster Catalog}",
      journal = {\apjs},
         year = 2020,
        month = apr,
       volume = {247},
       number = {2},
          eid = {48},
        pages = {48},
          doi = {10.3847/1538-4365/ab7919},
archivePrefix = {arXiv},
       eprint = {1911.00018},
 primaryClass = {astro-ph.HE},
       adsurl = {https://ui.adsabs.harvard.edu/abs/2020ApJS..247...48K}
}

@ARTICLE{Weatherford_2020,
       author = {{Weatherford}, Newlin C. and {Chatterjee}, Sourav and {Kremer}, Kyle and {Rasio}, Frederic A.},
        title = "{A Dynamical Survey of Stellar-mass Black Holes in 50 Milky Way Globular Clusters}",
      journal = {\apj},
         year = 2020,
        month = aug,
       volume = {898},
       number = {2},
          eid = {162},
        pages = {162},
          doi = {10.3847/1538-4357/ab9f98},
archivePrefix = {arXiv},
       eprint = {1911.09125},
 primaryClass = {astro-ph.SR},
       adsurl = {https://ui.adsabs.harvard.edu/abs/2020ApJ...898..162W}
}

@ARTICLE{Harris_1996,
       author = {{Harris}, William E.},
        title = "{A Catalog of Parameters for Globular Clusters in the Milky Way}",
      journal = {\aj},
         year = 1996,
        month = oct,
       volume = {112},
        pages = {1487},
          doi = {10.1086/118116},
       adsurl = {https://ui.adsabs.harvard.edu/abs/1996AJ....112.1487H}
}

@ARTICLE{Harris_2010,
       author = {{Harris}, William E.},
        title = "{A New Catalog of Globular Clusters in the Milky Way}",
      journal = {arXiv e-prints},
         year = 2010,
        month = dec,
          eid = {arXiv:1012.3224},
        pages = {arXiv:1012.3224},
          doi = {10.48550/arXiv.1012.3224},
archivePrefix = {arXiv},
       eprint = {1012.3224},
 primaryClass = {astro-ph.GA},
       adsurl = {https://ui.adsabs.harvard.edu/abs/2010arXiv1012.3224H}
}

@ARTICLE{Valenzuela_2025,
       author = {{Valenzuela}, Lucas M. and {Forbes}, Duncan A. and {Remus}, Rhea-Silvia},
        title = "{Globular cluster ages and their relation to high-redshift stellar cluster formation times from different globular cluster models}",
      journal = {\mnras},
         year = 2025,
        month = feb,
       volume = {537},
       number = {1},
        pages = {306-320},
          doi = {10.1093/mnras/staf015},
archivePrefix = {arXiv},
       eprint = {2410.12901},
 primaryClass = {astro-ph.GA},
       adsurl = {https://ui.adsabs.harvard.edu/abs/2025MNRAS.537..306V}
}

@ARTICLE{Kruijssen_2019c,
       author = {{Kruijssen}, J.~M. Diederik and {Pfeffer}, Joel L. and {Reina-Campos}, Marta and {Crain}, Robert A. and {Bastian}, Nate},
        title = "{The formation and assembly history of the Milky Way revealed by its globular cluster population}",
      journal = {\mnras},
         year = 2019,
        month = jul,
       volume = {486},
       number = {3},
        pages = {3180-3202},
          doi = {10.1093/mnras/sty1609},
archivePrefix = {arXiv},
       eprint = {1806.05680},
 primaryClass = {astro-ph.GA},
       adsurl = {https://ui.adsabs.harvard.edu/abs/2019MNRAS.486.3180K}
}

@ARTICLE{Dotter_2010,
       author = {{Dotter}, Aaron and {Sarajedini}, Ata and {Anderson}, Jay and {Aparicio}, Antonio and {Bedin}, Luigi R. and {Chaboyer}, Brian and {Majewski}, Steven and {Mar{\'\i}n-Franch}, A. and {Milone}, Antonino and {Paust}, Nathaniel and {Piotto}, Giampaolo and {Reid}, I. Neill and {Rosenberg}, Alfred and {Siegel}, Michael},
        title = "{The ACS Survey of Galactic Globular Clusters. IX. Horizontal Branch Morphology and the Second Parameter Phenomenon}",
      journal = {\apj},
         year = 2010,
        month = jan,
       volume = {708},
       number = {1},
        pages = {698-716},
          doi = {10.1088/0004-637X/708/1/698},
archivePrefix = {arXiv},
       eprint = {0911.2469},
 primaryClass = {astro-ph.SR},
       adsurl = {https://ui.adsabs.harvard.edu/abs/2010ApJ...708..698D}
}

@article{Dotter_2011,
doi = {10.1088/0004-637X/738/1/74},
url = {https://doi.org/10.1088/0004-637X/738/1/74},
year = {2011},
month = {aug},
publisher = {The American Astronomical Society},
volume = {738},
number = {1},
pages = {74},
author = {Dotter, Aaron and Sarajedini, Ata and Anderson, Jay},
title = {GLOBULAR CLUSTERS IN THE OUTER GALACTIC HALO: NEW HUBBLE SPACE TELESCOPE/ADVANCED CAMERA FOR SURVEYS IMAGING OF SIX GLOBULAR CLUSTERS AND THE GALACTIC GLOBULAR CLUSTER AGE–METALLICITY RELATION*},
journal = {\apj},
}

@ARTICLE{Snaith_2015,
       author = {{Snaith}, O. and {Haywood}, M. and {Di Matteo}, P. and {Lehnert}, M.~D. and {Combes}, F. and {Katz}, D. and {G{\'o}mez}, A.},
        title = "{Reconstructing the star formation history of the Milky Way disc(s) from chemical abundances}",
      journal = {\aap},
         year = 2015,
        month = jun,
       volume = {578},
          eid = {A87},
        pages = {A87},
          doi = {10.1051/0004-6361/201424281},
archivePrefix = {arXiv},
       eprint = {1410.3829},
 primaryClass = {astro-ph.GA},
       adsurl = {https://ui.adsabs.harvard.edu/abs/2015A&A...578A..87S}
}

@ARTICLE{Kruijssen_2012a,
       author = {{Kruijssen}, J.~M. Diederik and {Maschberger}, Thomas and {Moeckel}, Nickolas and {Clarke}, Cathie J. and {Bastian}, Nate and {Bonnell}, Ian A.},
        title = "{The dynamical state of stellar structure in star-forming regions}",
      journal = {\mnras},
         year = 2012,
        month = jan,
       volume = {419},
       number = {1},
        pages = {841-853},
          doi = {10.1111/j.1365-2966.2011.19748.x},
archivePrefix = {arXiv},
       eprint = {1109.0986},
 primaryClass = {astro-ph.GA},
       adsurl = {https://ui.adsabs.harvard.edu/abs/2012MNRAS.419..841K}
}

@ARTICLE{Gutcke_2024,
       author = {{Gutcke}, Thales A.},
        title = "{Low-mass Globular Clusters from Stripped Dark Matter Halos}",
      journal = {\apj},
         year = 2024,
        month = aug,
       volume = {971},
       number = {1},
          eid = {103},
        pages = {103},
          doi = {10.3847/1538-4357/ad5c62},
archivePrefix = {arXiv},
       eprint = {2310.03790},
 primaryClass = {astro-ph.GA},
       adsurl = {https://ui.adsabs.harvard.edu/abs/2024ApJ...971..103G}
}

@ARTICLE{Taylor_2025,
       author = {{Taylor}, Ethan D. and {Read}, Justin I. and {Orkney}, Matthew D.~A. and {Kim}, Stacy Y. and {Pontzen}, Andrew and {Agertz}, Oscar and {Rey}, Martin P. and {Andersson}, Eric P. and {Collins}, Michelle L.~M. and {Yates}, Robert M.},
        title = "{The emergence of globular clusters and globular-cluster-like dwarfs}",
      journal = {\nat},
         year = 2025,
        month = sep,
       volume = {645},
       number = {8080},
        pages = {327-331},
          doi = {10.1038/s41586-025-09494-x},
archivePrefix = {arXiv},
       eprint = {2509.09582},
 primaryClass = {astro-ph.GA},
       adsurl = {https://ui.adsabs.harvard.edu/abs/2025Natur.645..327T}
}

@ARTICLE{DellaCroce_2024,
       author = {{Della Croce}, A. and {Aros}, F.~I. and {Vesperini}, E. and {Dalessandro}, E. and {Lanzoni}, B. and {Ferraro}, F.~R. and {Bhat}, B.},
        title = "{Inference of black-hole mass fraction in Galactic globular clusters: A multi-dimensional approach to break the initial-condition degeneracies}",
      journal = {\aap},
         year = 2024,
        month = oct,
       volume = {690},
          eid = {A179},
        pages = {A179},
          doi = {10.1051/0004-6361/202450954},
archivePrefix = {arXiv},
       eprint = {2409.01400},
 primaryClass = {astro-ph.GA},
       adsurl = {https://ui.adsabs.harvard.edu/abs/2024A&A...690A.179D}
}

@ARTICLE{Harris_1994,
       author = {{Harris}, William E. and {Pudritz}, Ralph E.},
        title = "{Supergiant Molecular Clouds and the Formation of Globular Cluster Systems}",
      journal = {\apj},
         year = 1994,
        month = jul,
       volume = {429},
        pages = {177},
          doi = {10.1086/174310},
       adsurl = {https://ui.adsabs.harvard.edu/abs/1994ApJ...429..177H}
}

@ARTICLE{Elmegreen_1997,
       author = {{Elmegreen}, Bruce G. and {Efremov}, Yuri N.},
        title = "{A Universal Formation Mechanism for Open and Globular Clusters in Turbulent Gas}",
      journal = {\apj},
         year = 1997,
        month = may,
       volume = {480},
       number = {1},
        pages = {235-245},
          doi = {10.1086/303966},
       adsurl = {https://ui.adsabs.harvard.edu/abs/1997ApJ...480..235E}
}

@ARTICLE{Maccarone_2007,
       author = {{Maccarone}, Thomas J. and {Kundu}, Arunav and {Zepf}, Stephen E. and {Rhode}, Katherine L.},
        title = "{A black hole in a globular cluster}",
      journal = {\nat},
         year = 2007,
        month = jan,
       volume = {445},
       number = {7124},
        pages = {183-185},
          doi = {10.1038/nature05434},
archivePrefix = {arXiv},
       eprint = {astro-ph/0701310},
 primaryClass = {astro-ph},
       adsurl = {https://ui.adsabs.harvard.edu/abs/2007Natur.445..183M}
}

@ARTICLE{Giesers_2018,
       author = {{Giesers}, Benjamin and {Dreizler}, Stefan and {Husser}, Tim-Oliver and {Kamann}, Sebastian and {Anglada Escud{\'e}}, Guillem and {Brinchmann}, Jarle and {Carollo}, C. Marcella and {Roth}, Martin M. and {Weilbacher}, Peter M. and {Wisotzki}, Lutz},
        title = "{A detached stellar-mass black hole candidate in the globular cluster NGC 3201}",
      journal = {\mnras},
         year = 2018,
        month = mar,
       volume = {475},
       number = {1},
        pages = {L15-L19},
          doi = {10.1093/mnrasl/slx203},
archivePrefix = {arXiv},
       eprint = {1801.05642},
 primaryClass = {astro-ph.SR},
       adsurl = {https://ui.adsabs.harvard.edu/abs/2018MNRAS.475L..15G}
}

@ARTICLE{Schaye2008,
       author = {{Schaye}, Joop and {Dalla Vecchia}, Claudio},
        title = "{On the relation between the Schmidt and Kennicutt-Schmidt star formation laws and its implications for numerical simulations}",
      journal = {\mnras},
         year = 2008,
        month = jan,
       volume = {383},
       number = {3},
        pages = {1210-1222},
          doi = {10.1111/j.1365-2966.2007.12639.x},
archivePrefix = {arXiv},
       eprint = {0709.0292},
 primaryClass = {astro-ph},
       adsurl = {https://ui.adsabs.harvard.edu/abs/2008MNRAS.383.1210S}
}

@ARTICLE{Schaye2004,
       author = {{Schaye}, Joop},
        title = "{Star Formation Thresholds and Galaxy Edges: Why and Where}",
      journal = {\apj},
         year = 2004,
        month = jul,
       volume = {609},
       number = {2},
        pages = {667-682},
          doi = {10.1086/421232},
archivePrefix = {arXiv},
       eprint = {astro-ph/0205125},
 primaryClass = {astro-ph},
       adsurl = {https://ui.adsabs.harvard.edu/abs/2004ApJ...609..667S}
}

@ARTICLE{Whitmore_1995,
       author = {{Whitmore}, Bradley C. and {Schweizer}, Francois},
        title = "{Hubble Space Telescope Observations of Young Star Clusters in NGC 4038/4039, ``The Antennae'' Galaxies}",
      journal = {\aj},
         year = 1995,
        month = mar,
       volume = {109},
        pages = {960},
          doi = {10.1086/117334},
       adsurl = {https://ui.adsabs.harvard.edu/abs/1995AJ....109..960W}
}

@ARTICLE{Adamo_2017,
       author = {{Adamo}, A. and {Ryon}, J.~E. and {Messa}, M. and {Kim}, H. and {Grasha}, K. and {Cook}, D.~O. and {Calzetti}, D. and {Lee}, J.~C. and {Whitmore}, B.~C. and {Elmegreen}, B.~G. and {Ubeda}, L. and {Smith}, L.~J. and {Bright}, S.~N. and {Runnholm}, A. and {Andrews}, J.~E. and {Fumagalli}, M. and {Gouliermis}, D.~A. and {Kahre}, L. and {Nair}, P. and {Thilker}, D. and {Walterbos}, R. and {Wofford}, A. and {Aloisi}, A. and {Ashworth}, G. and {Brown}, T.~M. and {Chandar}, R. and {Christian}, C. and {Cignoni}, M. and {Clayton}, G.~C. and {Dale}, D.~A. and {de Mink}, S.~E. and {Dobbs}, C. and {Elmegreen}, D.~M. and {Evans}, A.~S. and {Gallagher}, III, J.~S. and {Grebel}, E.~K. and {Herrero}, A. and {Hunter}, D.~A. and {Johnson}, K.~E. and {Kennicutt}, R.~C. and {Krumholz}, M.~R. and {Lennon}, D. and {Levay}, K. and {Martin}, C. and {Nota}, A. and {{\"O}stlin}, G. and {Pellerin}, A. and {Prieto}, J. and {Regan}, M.~W. and {Sabbi}, E. and {Sacchi}, E. and {Schaerer}, D. and {Schiminovich}, D. and {Shabani}, F. and {Tosi}, M. and {Van Dyk}, S.~D. and {Zackrisson}, E.},
        title = "{Legacy ExtraGalactic UV Survey with The Hubble Space Telescope: Stellar Cluster Catalogs and First Insights Into Cluster Formation and Evolution in NGC 628}",
      journal = {\apj},
         year = 2017,
        month = jun,
       volume = {841},
       number = {2},
          eid = {131},
        pages = {131},
          doi = {10.3847/1538-4357/aa7132},
archivePrefix = {arXiv},
       eprint = {1705.01588},
 primaryClass = {astro-ph.GA},
       adsurl = {https://ui.adsabs.harvard.edu/abs/2017ApJ...841..131A}
}

@ARTICLE{Holtzman_1992,
       author = {{Holtzman}, J.~A. and {Faber}, S.~M. and {Shaya}, E.~J. and {Lauer}, T.~R. and {Groth}, J. and {Hunter}, D.~A. and {Baum}, W.~A. and {Ewald}, S.~P. and {Hester}, J.~J. and {Light}, R.~M. and {Lynds}, C.~R. and {O'Neil}, Jr., E.~J. and {Westphal}, J.~A.},
        title = "{Planetary Camera Observations of NGC 1275: Discovery of a Central Population of Compact Massive Blue Star Clusters}",
      journal = {\aj},
         year = 1992,
        month = mar,
       volume = {103},
        pages = {691},
          doi = {10.1086/116094},
       adsurl = {https://ui.adsabs.harvard.edu/abs/1992AJ....103..691H}
}

@ARTICLE{Zepf_1999,
       author = {{Zepf}, Stephen E. and {Ashman}, Keith M. and {English}, Jayanne and {Freeman}, Kenneth C. and {Sharples}, Ray M.},
        title = "{The Formation and Evolution of Candidate Young Globular Clusters in NGC 3256}",
      journal = {\aj},
         year = 1999,
        month = aug,
       volume = {118},
       number = {2},
        pages = {752-764},
          doi = {10.1086/300961},
archivePrefix = {arXiv},
       eprint = {astro-ph/9904247},
 primaryClass = {astro-ph},
       adsurl = {https://ui.adsabs.harvard.edu/abs/1999AJ....118..752Z}
}

@ARTICLE{Barnard_2012,
       author = {{Barnard}, R. and {Garcia}, M. and {Murray}, S.~S.},
        title = "{12 Years of X-Ray Variability in M31 Globular Clusters, Including 8 Black Hole Candidates, as Seen by Chandra}",
      journal = {\apj},
         year = 2012,
        month = sep,
       volume = {757},
       number = {1},
          eid = {40},
        pages = {40},
          doi = {10.1088/0004-637X/757/1/40},
archivePrefix = {arXiv},
       eprint = {1203.2583},
 primaryClass = {astro-ph.HE},
       adsurl = {https://ui.adsabs.harvard.edu/abs/2012ApJ...757...40B}
}

@ARTICLE{Maccarone_2011,
       author = {{Maccarone}, Thomas J. and {Kundu}, Arunav and {Zepf}, Stephen E. and {Rhode}, Katherine L.},
        title = "{A new globular cluster black hole in NGC 4472}",
      journal = {\mnras},
         year = 2011,
        month = jan,
       volume = {410},
       number = {3},
        pages = {1655-1659},
          doi = {10.1111/j.1365-2966.2010.17547.x},
archivePrefix = {arXiv},
       eprint = {1008.2896},
 primaryClass = {astro-ph.HE},
       adsurl = {https://ui.adsabs.harvard.edu/abs/2011MNRAS.410.1655M}
}

@ARTICLE{Saracino_2022,
       author = {{Saracino}, S. and {Kamann}, S. and {Guarcello}, M.~G. and {Usher}, C. and {Bastian}, N. and {Cabrera-Ziri}, I. and {Gieles}, M. and {Dreizler}, S. and {Da Costa}, G.~S. and {Husser}, T.-O. and {H{\'e}nault-Brunet}, V.},
        title = "{A black hole detected in the young massive LMC cluster NGC 1850}",
      journal = {\mnras},
         year = 2022,
        month = apr,
       volume = {511},
       number = {2},
        pages = {2914-2924},
          doi = {10.1093/mnras/stab3159},
archivePrefix = {arXiv},
       eprint = {2111.06506},
 primaryClass = {astro-ph.GA},
       adsurl = {https://ui.adsabs.harvard.edu/abs/2022MNRAS.511.2914S}
}

@ARTICLE{Plummer_1911,
       author = {{Plummer}, H.~C.},
        title = "{On the problem of distribution in globular star clusters}",
      journal = {\mnras},
         year = 1911,
        month = mar,
       volume = {71},
        pages = {460-470},
          doi = {10.1093/mnras/71.5.460},
       adsurl = {https://ui.adsabs.harvard.edu/abs/1911MNRAS..71..460P}
}

@ARTICLE{Alexander_2014,
       author = {{Alexander}, Poul E.~R. and {Gieles}, Mark and {Lamers}, Henny J.~G.~L.~M. and {Baumgardt}, Holger},
        title = "{A prescription and fast code for the long-term evolution of star clusters - III. Unequal masses and stellar evolution}",
      journal = {\mnras},
         year = 2014,
        month = aug,
       volume = {442},
       number = {2},
        pages = {1265-1285},
          doi = {10.1093/mnras/stu899},
archivePrefix = {arXiv},
       eprint = {1405.1086},
 primaryClass = {astro-ph.GA},
       adsurl = {https://ui.adsabs.harvard.edu/abs/2014MNRAS.442.1265A}
}

@ARTICLE{Belokurov_2018,
       author = {{Belokurov}, V. and {Erkal}, D. and {Evans}, N.~W. and {Koposov}, S.~E. and {Deason}, A.~J.},
        title = "{Co-formation of the disc and the stellar halo}",
      journal = {\mnras},
         year = 2018,
        month = jul,
       volume = {478},
       number = {1},
        pages = {611-619},
          doi = {10.1093/mnras/sty982},
archivePrefix = {arXiv},
       eprint = {1802.03414},
 primaryClass = {astro-ph.GA},
       adsurl = {https://ui.adsabs.harvard.edu/abs/2018MNRAS.478..611B}
}

@ARTICLE{Helmi_2018,
       author = {{Helmi}, Amina and {Babusiaux}, Carine and {Koppelman}, Helmer H. and {Massari}, Davide and {Veljanoski}, Jovan and {Brown}, Anthony G.~A.},
        title = "{The merger that led to the formation of the Milky Way's inner stellar halo and thick disk}",
      journal = {\nat},
         year = 2018,
        month = oct,
       volume = {563},
       number = {7729},
        pages = {85-88},
          doi = {10.1038/s41586-018-0625-x},
archivePrefix = {arXiv},
       eprint = {1806.06038},
 primaryClass = {astro-ph.GA},
       adsurl = {https://ui.adsabs.harvard.edu/abs/2018Natur.563...85H}
}

@ARTICLE{Fattahi_2019,
       author = {{Fattahi}, Azadeh and {Belokurov}, Vasily and {Deason}, Alis J. and {Frenk}, Carlos S. and {G{\'o}mez}, Facundo A. and {Grand}, Robert J.~J. and {Marinacci}, Federico and {Pakmor}, R{\"u}diger and {Springel}, Volker},
        title = "{The origin of galactic metal-rich stellar halo components with highly eccentric orbits}",
      journal = {\mnras},
         year = 2019,
        month = apr,
       volume = {484},
       number = {4},
        pages = {4471-4483},
          doi = {10.1093/mnras/stz159},
archivePrefix = {arXiv},
       eprint = {1810.07779},
 primaryClass = {astro-ph.GA},
       adsurl = {https://ui.adsabs.harvard.edu/abs/2019MNRAS.484.4471F}
}

@ARTICLE{Lian_2025,
       author = {{Lian}, Jianhui and {Wang}, Tao and {Feng}, Qikang and {Huang}, Yang and {Guo}, Helong},
        title = "{The Milky Way Is a Less Massive Galaxy{\textemdash}New Estimates of the Milky Way's Local and Global Stellar Masses}",
      journal = {\apjl},
         year = 2025,
        month = sep,
       volume = {990},
       number = {2},
          eid = {L37},
        pages = {L37},
          doi = {10.3847/2041-8213/adfc73},
archivePrefix = {arXiv},
       eprint = {2508.13665},
 primaryClass = {astro-ph.GA},
       adsurl = {https://ui.adsabs.harvard.edu/abs/2025ApJ...990L..37L}
}

@ARTICLE{Licquia_2015,
       author = {{Licquia}, Timothy C. and {Newman}, Jeffrey A.},
        title = "{Improved Estimates of the Milky Way's Stellar Mass and Star Formation Rate from Hierarchical Bayesian Meta-Analysis}",
      journal = {\apj},
         year = 2015,
        month = jun,
       volume = {806},
       number = {1},
          eid = {96},
        pages = {96},
          doi = {10.1088/0004-637X/806/1/96},
archivePrefix = {arXiv},
       eprint = {1407.1078},
 primaryClass = {astro-ph.GA},
       adsurl = {https://ui.adsabs.harvard.edu/abs/2015ApJ...806...96L}
}

@ARTICLE{Dornan_2025,
       author = {{Dornan}, Veronika and {Harris}, William E.},
        title = "{Major Mergers Mean Major Offset: Drivers of Intrinsic Scatter in the M$_{GCS}$─M$_{h}$ Scaling Relation for Massive Elliptical Galaxies}",
      journal = {\apj},
         year = 2025,
        month = jul,
       volume = {988},
       number = {1},
          eid = {70},
        pages = {70},
          doi = {10.3847/1538-4357/ade05e},
archivePrefix = {arXiv},
       eprint = {2505.24154},
 primaryClass = {astro-ph.GA},
       adsurl = {https://ui.adsabs.harvard.edu/abs/2025ApJ...988...70D}
}

@ARTICLE{Forbes_2025,
       author = {{Forbes}, Duncan A. and {Gannon}, Jonah S.},
        title = "{Do ultra-diffuse galaxies follow the globular cluster{\textendash}halo mass relation?}",
      journal = {\mnras},
         year = 2025,
        month = oct,
       volume = {543},
       number = {1},
        pages = {L1-L5},
          doi = {10.1093/mnrasl/slaf084},
archivePrefix = {arXiv},
       eprint = {2507.20687},
 primaryClass = {astro-ph.GA},
       adsurl = {https://ui.adsabs.harvard.edu/abs/2025MNRAS.543L...1F}
}

@ARTICLE{Deng_2024,
       author = {{Deng}, Yunwei and {Li}, Hui and {Liu}, Boyuan and {Kannan}, Rahul and {Smith}, Aaron and {Bryan}, Greg L.},
        title = "{RIGEL: Simulating dwarf galaxies at solar mass resolution with radiative transfer and feedback from individual massive stars}",
      journal = {\aap},
         year = 2024,
        month = nov,
       volume = {691},
          eid = {A231},
        pages = {A231},
          doi = {10.1051/0004-6361/202450699},
archivePrefix = {arXiv},
       eprint = {2405.08869},
 primaryClass = {astro-ph.GA},
       adsurl = {https://ui.adsabs.harvard.edu/abs/2024A&A...691A.231D}
}

@ARTICLE{Claeyssens_2026,
       author = {{Claeyssens}, Ad{\'e}la{\"\i}de and {Adamo}, Angela and {Kokorev}, Vasily and {Furtak}, Lukas and {Richard}, Johan and {Beauchesne}, Benjamin and {Dessauges-Zavadsky}, Miroslava and {Atek}, Hakim and {Chisholm}, John and {Endsley}, Ryan and {Fujimoto}, Seiji and {Korber}, Damien and {Pan}, Richard and {Saldana-Lopez}, Alberto and {Schaerer}, Daniel},
        title = "{A first GLIMPSE into star clusters populations across cosmic time}",
      journal = {arXiv e-prints},
         year = 2026,
        month = jan,
          eid = {arXiv:2601.16281},
        pages = {arXiv:2601.16281},
          doi = {10.48550/arXiv.2601.16281},
archivePrefix = {arXiv},
       eprint = {2601.16281},
 primaryClass = {astro-ph.GA},
       adsurl = {https://ui.adsabs.harvard.edu/abs/2026arXiv260116281C}
}

@ARTICLE{Claeyssens_2025,
       author = {{Claeyssens}, Ad{\'e}la{\"\i}de and {Adamo}, Angela and {Messa}, Matteo and {Dessauges-Zavadsky}, Miroslava and {Richard}, Johan and {Kramarenko}, Ivan and {Matthee}, Jorryt and {Naidu}, Rohan P.},
        title = "{Tracing star formation across cosmic time at tens of parsec-scales in the lensing cluster field Abell 2744}",
      journal = {\mnras},
         year = 2025,
        month = mar,
       volume = {537},
       number = {3},
        pages = {2535-2558},
          doi = {10.1093/mnras/staf058},
archivePrefix = {arXiv},
       eprint = {2410.10974},
 primaryClass = {astro-ph.GA},
       adsurl = {https://ui.adsabs.harvard.edu/abs/2025MNRAS.537.2535C}
}

@ARTICLE{Mowla_2024,
       author = {{Mowla}, Lamiya and {Iyer}, Kartheik and {Asada}, Yoshihisa and {Desprez}, Guillaume and {Tan}, Vivian Yun Yan and {Martis}, Nicholas and {Sarrouh}, Ghassan and {Strait}, Victoria and {Abraham}, Roberto and {Brada{\v{c}}}, Maru{\v{s}}a and {Brammer}, Gabriel and {Muzzin}, Adam and {Pacifici}, Camilla and {Ravindranath}, Swara and {Sawicki}, Marcin and {Willott}, Chris and {Estrada-Carpenter}, Vince and {Jahan}, Nusrath and {Noirot}, Ga{\"e}l and {Matharu}, Jasleen and {Rihtar{\v{s}}i{\v{c}}}, Gregor and {Zabl}, Johannes},
        title = "{Formation of a low-mass galaxy from star clusters in a 600-million-year-old Universe}",
      journal = {\nat},
         year = 2024,
        month = dec,
       volume = {636},
       number = {8042},
        pages = {332-336},
          doi = {10.1038/s41586-024-08293-0},
archivePrefix = {arXiv},
       eprint = {2402.08696},
 primaryClass = {astro-ph.GA},
       adsurl = {https://ui.adsabs.harvard.edu/abs/2024Natur.636..332M}
}

@ARTICLE{Adamo_2024,
       author = {{Adamo}, Angela and {Bradley}, Larry D. and {Vanzella}, Eros and {Claeyssens}, Ad{\'e}la{\"\i}de and {Welch}, Brian and {Diego}, Jose M. and {Mahler}, Guillaume and {Oguri}, Masamune and {Sharon}, Keren and {Abdurro'uf} and {Hsiao}, Tiger Yu-Yang and {Xu}, Xinfeng and {Messa}, Matteo and {Lassen}, Augusto E. and {Zackrisson}, Erik and {Brammer}, Gabriel and {Coe}, Dan and {Kokorev}, Vasily and {Ricotti}, Massimo and {Zitrin}, Adi and {Fujimoto}, Seiji and {Inoue}, Akio K. and {Resseguier}, Tom and {Rigby}, Jane R. and {Jim{\'e}nez-Teja}, Yolanda and {Windhorst}, Rogier A. and {Hashimoto}, Takuya and {Tamura}, Yoichi},
        title = "{Bound star clusters observed in a lensed galaxy 460 Myr after the Big Bang}",
      journal = {\nat},
         year = 2024,
        month = aug,
       volume = {632},
       number = {8025},
        pages = {513-516},
          doi = {10.1038/s41586-024-07703-7},
archivePrefix = {arXiv},
       eprint = {2401.03224},
 primaryClass = {astro-ph.GA},
       adsurl = {https://ui.adsabs.harvard.edu/abs/2024Natur.632..513A}
}

@ARTICLE{Sun_2023,
       author = {{Sun}, Jiayi and {Leroy}, Adam K. and {Ostriker}, Eve C. and {Meidt}, Sharon and {Rosolowsky}, Erik and {Schinnerer}, Eva and {Wilson}, Christine D. and {Utomo}, Dyas and {Belfiore}, Francesco and {Blanc}, Guillermo A. and {Emsellem}, Eric and {Faesi}, Christopher and {Groves}, Brent and {Hughes}, Annie and {Koch}, Eric W. and {Kreckel}, Kathryn and {Liu}, Daizhong and {Pan}, Hsi-An and {Pety}, J{\'e}r{\^o}me and {Querejeta}, Miguel and {Razza}, Alessandro and {Saito}, Toshiki and {Sardone}, Amy and {Usero}, Antonio and {Williams}, Thomas G. and {Bigiel}, Frank and {Bolatto}, Alberto D. and {Chevance}, M{\'e}lanie and {Dale}, Daniel A. and {Gensior}, Jindra and {Glover}, Simon C.~O. and {Grasha}, Kathryn and {Henshaw}, Jonathan D. and {Jim{\'e}nez-Donaire}, Mar{\'\i}a J. and {Klessen}, Ralf S. and {Kruijssen}, J.~M. Diederik and {Murphy}, Eric J. and {Neumann}, Lukas and {Teng}, Yu-Hsuan and {Thilker}, David A.},
        title = "{Star Formation Laws and Efficiencies across 80 Nearby Galaxies}",
      journal = {\apjl},
         year = 2023,
        month = mar,
       volume = {945},
       number = {2},
          eid = {L19},
        pages = {L19},
          doi = {10.3847/2041-8213/acbd9c},
archivePrefix = {arXiv},
       eprint = {2302.12267},
 primaryClass = {astro-ph.GA},
       adsurl = {https://ui.adsabs.harvard.edu/abs/2023ApJ...945L..19S}
}

@ARTICLE{Usher_2024,
       author = {{Usher}, Christopher and {Caldwell}, Nelson and {Cabrera-Ziri}, Ivan},
        title = "{Measuring M31 globular cluster ages and metallicities using both photometry and spectroscopy}",
      journal = {\mnras},
         year = 2024,
        month = mar,
       volume = {528},
       number = {4},
        pages = {6010-6024},
          doi = {10.1093/mnras/stae282},
archivePrefix = {arXiv},
       eprint = {2401.13918},
 primaryClass = {astro-ph.GA},
       adsurl = {https://ui.adsabs.harvard.edu/abs/2024MNRAS.528.6010U}
}

@ARTICLE{Williams_2017,
       author = {{Williams}, Benjamin F. and {Dolphin}, Andrew E. and {Dalcanton}, Julianne J. and {Weisz}, Daniel R. and {Bell}, Eric F. and {Lewis}, Alexia R. and {Rosenfield}, Philip and {Choi}, Yumi and {Skillman}, Evan and {Monachesi}, Antonela},
        title = "{PHAT. XIX. The Ancient Star Formation History of the M31 Disk}",
      journal = {\apj},
         year = 2017,
        month = sep,
       volume = {846},
       number = {2},
          eid = {145},
        pages = {145},
          doi = {10.3847/1538-4357/aa862a},
       adsurl = {https://ui.adsabs.harvard.edu/abs/2017ApJ...846..145W}
}

@ARTICLE{Grand_2021,
       author = {{Grand}, Robert J.~J. and {Marinacci}, Federico and {Pakmor}, R{\"u}diger and {Simpson}, Christine M. and {Kelly}, Ashley J. and {G{\'o}mez}, Facundo A. and {Jenkins}, Adrian and {Springel}, Volker and {Frenk}, Carlos S. and {White}, Simon D.~M.},
        title = "{Determining the full satellite population of a Milky Way-mass halo in a highly resolved cosmological hydrodynamic simulation}",
      journal = {\mnras},
         year = 2021,
        month = nov,
       volume = {507},
       number = {4},
        pages = {4953-4967},
          doi = {10.1093/mnras/stab2492},
archivePrefix = {arXiv},
       eprint = {2105.04560},
 primaryClass = {astro-ph.GA},
       adsurl = {https://ui.adsabs.harvard.edu/abs/2021MNRAS.507.4953G}
}

@ARTICLE{Pakmor_2025b,
       author = {{Pakmor}, R{\"u}diger and {Fragkoudi}, Francesca and {Grand}, Robert J.~J. and {Simpson}, Christine M. and {G{\'o}mez}, Facundo A. and {van de Voort}, Freeke and {Bieri}, Rebekka and {Trick}, Wilma and {Werhahn}, Maria and {Talbot}, Rosie Y.},
        title = "{Auriga Superstars: Improving the resolution and fidelity of stellar dynamics in cosmological galaxy simulations}",
      journal = {\mnras},
         year = 2025,
        month = nov,
       volume = {543},
       number = {4},
        pages = {4355-4368},
          doi = {10.1093/mnras/staf1755},
archivePrefix = {arXiv},
       eprint = {2507.22104},
 primaryClass = {astro-ph.GA},
       adsurl = {https://ui.adsabs.harvard.edu/abs/2025MNRAS.543.4355P}
}

@ARTICLE{Kruijssen_2019d,
       author = {{Kruijssen}, J.~M. Diederik},
        title = "{The minimum metallicity of globular clusters and its physical origin - implications for the galaxy mass-metallicity relation and observations of proto-globular clusters at high redshift}",
      journal = {\mnras},
         year = 2019,
        month = jun,
       volume = {486},
       number = {1},
        pages = {L20-L25},
          doi = {10.1093/mnrasl/slz052},
archivePrefix = {arXiv},
       eprint = {1904.09987},
 primaryClass = {astro-ph.GA},
       adsurl = {https://ui.adsabs.harvard.edu/abs/2019MNRAS.486L..20K}
}

@ARTICLE{Beasley_2019,
       author = {{Beasley}, Michael A. and {Leaman}, Ryan and {Gallart}, Carme and {Larsen}, S{\o}ren S. and {Battaglia}, Giuseppina and {Monelli}, Matteo and {Pedreros}, Mario H.},
        title = "{An old, metal-poor globular cluster in Sextans A and the metallicity floor of globular cluster systems}",
      journal = {\mnras},
         year = 2019,
        month = aug,
       volume = {487},
       number = {2},
        pages = {1986-1993},
          doi = {10.1093/mnras/stz1349},
archivePrefix = {arXiv},
       eprint = {1904.01084},
 primaryClass = {astro-ph.GA},
       adsurl = {https://ui.adsabs.harvard.edu/abs/2019MNRAS.487.1986B}
}

@ARTICLE{Pfeffer_2020,
       author = {{Pfeffer}, Joel L. and {Trujillo-Gomez}, Sebastian and {Kruijssen}, J.~M.~D. and {Crain}, Robert A. and {Hughes}, Meghan E. and {Reina-Campos}, Marta and {Bastian}, Nate},
        title = "{Predicting accreted satellite galaxy masses and accretion redshifts based on globular cluster orbits in the E-MOSAICS simulations}",
      journal = {\mnras},
         year = 2020,
        month = dec,
       volume = {499},
       number = {4},
        pages = {4863-4875},
          doi = {10.1093/mnras/staa3109},
archivePrefix = {arXiv},
       eprint = {2003.00076},
 primaryClass = {astro-ph.GA},
       adsurl = {https://ui.adsabs.harvard.edu/abs/2020MNRAS.499.4863P}
}

@ARTICLE{Reina_Campos_2022a,
       author = {{Reina-Campos}, Marta and {Trujillo-Gomez}, Sebastian and {Deason}, Alis J. and {Kruijssen}, J.~M. Diederik and {Pfeffer}, Joel L. and {Crain}, Robert A. and {Bastian}, Nate and {Hughes}, Meghan E.},
        title = "{Radial distributions of globular clusters trace their host dark matter halo: insights from the E-MOSAICS simulations}",
      journal = {\mnras},
         year = 2022,
        month = jul,
       volume = {513},
       number = {3},
        pages = {3925-3945},
          doi = {10.1093/mnras/stac1126},
archivePrefix = {arXiv},
       eprint = {2106.07652},
 primaryClass = {astro-ph.GA},
       adsurl = {https://ui.adsabs.harvard.edu/abs/2022MNRAS.513.3925R}
}

@ARTICLE{Chen_2024b,
       author = {{Chen}, Yingtian and {Gnedin}, Oleg Y.},
        title = "{Galaxy assembly revealed by globular clusters}",
      journal = {The Open Journal of Astrophysics},
         year = 2024,
        month = mar,
       volume = {7},
          eid = {23},
        pages = {23},
          doi = {10.33232/001c.116169},
archivePrefix = {arXiv},
       eprint = {2401.17420},
 primaryClass = {astro-ph.GA},
       adsurl = {https://ui.adsabs.harvard.edu/abs/2024OJAp....7E..23C}
}

@ARTICLE{Cantun_2020,
       author = {{Cautun}, Marius and {Ben{\'\i}tez-Llambay}, Alejandro and {Deason}, Alis J. and {Frenk}, Carlos S. and {Fattahi}, Azadeh and {G{\'o}mez}, Facundo A. and {Grand}, Robert J.~J. and {Oman}, Kyle A. and {Navarro}, Julio F. and {Simpson}, Christine M.},
        title = "{The milky way total mass profile as inferred from Gaia DR2}",
      journal = {\mnras},
         year = 2020,
        month = may,
       volume = {494},
       number = {3},
        pages = {4291-4313},
          doi = {10.1093/mnras/staa1017},
archivePrefix = {arXiv},
       eprint = {1911.04557},
 primaryClass = {astro-ph.GA},
       adsurl = {https://ui.adsabs.harvard.edu/abs/2020MNRAS.494.4291C}
}

@INPROCEEDINGS{Sick_2015,
       author = {{Sick}, Jonathan and {Courteau}, Stephane and {Cuillandre}, Jean-Charles and {Dalcanton}, Julianne and {de Jong}, Roelof and {McDonald}, Michael and {Simard}, Dana and {Tully}, R. Brent},
        title = "{The Stellar Mass of M31 as inferred by the Andromeda Optical \& Infrared Disk Survey}",
    booktitle = {Galaxy Masses as Constraints of Formation Models},
         year = 2015,
       editor = {{Cappellari}, Michele and {Courteau}, St{\'e}phane},
       series = {IAU Symposium},
       volume = {311},
        month = apr,
        pages = {82-85},
          doi = {10.1017/S1743921315003440},
archivePrefix = {arXiv},
       eprint = {1410.0017},
 primaryClass = {astro-ph.GA},
       adsurl = {https://ui.adsabs.harvard.edu/abs/2015IAUS..311...82S}
}

@ARTICLE{Villanueva_Domingo_2023,
       author = {{Villanueva-Domingo}, Pablo and {Villaescusa-Navarro}, Francisco and {Genel}, Shy and {Angl{\'e}s-Alc{\'a}zar}, Daniel and {Hernquist}, Lars and {Marinacci}, Federico and {Spergel}, David N. and {Vogelsberger}, Mark and {Narayanan}, Desika},
        title = "{Weighing the Milky Way and Andromeda galaxies with artificial intelligence}",
      journal = {\prd},
         year = 2023,
        month = may,
       volume = {107},
       number = {10},
          eid = {103003},
        pages = {103003},
          doi = {10.1103/PhysRevD.107.103003},
archivePrefix = {arXiv},
       eprint = {2111.14874},
 primaryClass = {astro-ph.GA},
       adsurl = {https://ui.adsabs.harvard.edu/abs/2023PhRvD.107j3003V}
}

@ARTICLE{Marinacci_2019,
       author = {{Marinacci}, Federico and {Sales}, Laura V. and {Vogelsberger}, Mark and {Torrey}, Paul and {Springel}, Volker},
        title = "{Simulating the interstellar medium and stellar feedback on a moving mesh: implementation and isolated galaxies}",
      journal = {\mnras},
         year = 2019,
        month = nov,
       volume = {489},
       number = {3},
        pages = {4233-4260},
          doi = {10.1093/mnras/stz2391},
archivePrefix = {arXiv},
       eprint = {1905.08806},
 primaryClass = {astro-ph.GA},
       adsurl = {https://ui.adsabs.harvard.edu/abs/2019MNRAS.489.4233M}
}

@ARTICLE{Kirby_2013,
       author = {{Kirby}, Evan N. and {Cohen}, Judith G. and {Guhathakurta}, Puragra and {Cheng}, Lucy and {Bullock}, James S. and {Gallazzi}, Anna},
        title = "{The Universal Stellar Mass-Stellar Metallicity Relation for Dwarf Galaxies}",
      journal = {\apj},
         year = 2013,
        month = dec,
       volume = {779},
       number = {2},
          eid = {102},
        pages = {102},
          doi = {10.1088/0004-637X/779/2/102},
archivePrefix = {arXiv},
       eprint = {1310.0814},
 primaryClass = {astro-ph.GA},
       adsurl = {https://ui.adsabs.harvard.edu/abs/2013ApJ...779..102K}
}

@ARTICLE{Massari_2019,
       author = {{Massari}, D. and {Koppelman}, H.~H. and {Helmi}, A.},
        title = "{Origin of the system of globular clusters in the Milky Way}",
      journal = {\aap},
         year = 2019,
        month = oct,
       volume = {630},
          eid = {L4},
        pages = {L4},
          doi = {10.1051/0004-6361/201936135},
archivePrefix = {arXiv},
       eprint = {1906.08271},
 primaryClass = {astro-ph.GA},
       adsurl = {https://ui.adsabs.harvard.edu/abs/2019A&A...630L...4M}
}

@ARTICLE{Andersson_2024,
       author = {{Andersson}, Eric P. and {Mac Low}, Mordecai-Mark and {Agertz}, Oscar and {Renaud}, Florent and {Li}, Hui},
        title = "{Pre-supernova feedback sets the star cluster mass function to a power law and reduces the cluster formation efficiency}",
      journal = {\aap},
         year = 2024,
        month = jan,
       volume = {681},
          eid = {A28},
        pages = {A28},
          doi = {10.1051/0004-6361/202347792},
archivePrefix = {arXiv},
       eprint = {2308.12363},
 primaryClass = {astro-ph.GA},
       adsurl = {https://ui.adsabs.harvard.edu/abs/2024A&A...681A..28A}
}

@ARTICLE{Keller_2019,
       author = {{Keller}, B.~W. and {Wadsley}, J.~W. and {Wang}, L. and {Kruijssen}, J.~M.~D.},
        title = "{Chaos and variance in galaxy formation}",
      journal = {\mnras},
         year = 2019,
        month = jan,
       volume = {482},
       number = {2},
        pages = {2244-2261},
          doi = {10.1093/mnras/sty2859},
archivePrefix = {arXiv},
       eprint = {1803.05445},
 primaryClass = {astro-ph.GA},
       adsurl = {https://ui.adsabs.harvard.edu/abs/2019MNRAS.482.2244K}
}

@ARTICLE{Newton_2025,
       author = {{Newton}, Oliver and {Davies}, Jonathan J. and {Pfeffer}, Joel and {Crain}, Robert A. and {Kruijssen}, J.~M. Diederik and {Pontzen}, Andrew and {Bastian}, Nate},
        title = "{The formation and disruption of globular cluster populations in simulations of present-day L* galaxies with controlled assembly histories}",
      journal = {\mnras},
         year = 2025,
        month = sep,
       volume = {542},
       number = {2},
        pages = {591-607},
          doi = {10.1093/mnras/staf1226},
archivePrefix = {arXiv},
       eprint = {2409.04516},
 primaryClass = {astro-ph.GA},
       adsurl = {https://ui.adsabs.harvard.edu/abs/2025MNRAS.542..591N}
}

%%%%%%%%%%%%%%%%%%%%%%%%%%%%%%%%%%%%%%%%%%%%%%%%%%%%%%%%%%%%%%%
\begin{appendix}

\section{Additional variation model runs}
For the appendix, some other model variations have been run at the same or lower resolution for halo 6. These are summarised in Table \ref{tab:appendixvariations}.

\begin{table}[h]
\centering
\small
\caption{Additional variations for halo 6. }
\label{tab:appendixvariations}
\begin{tabular}{ccc}
\hline \hline
Model & Variation & Variation remarks \\
\hline
\multicolumn{3}{c}{Resolution level~4 ($M_\mathrm{bary}\sim5\times 10^4 \mathrm{M}_\odot$, $\epsilon_\mathrm{phys}\sim375$~pc)} \\
x50in100myr & Enhanced shocks & x50 during 100~Myr \\
x50in200myr & Enhanced shocks & x50 during 200~Myr \\
x50fullevo & Enhanced shocks & x50 full evolution \\
\hline
\multicolumn{3}{c}{Resolution level~5 ($M_\mathrm{bary}\sim4\times 10^5 \mathrm{M}_\odot$, $\epsilon_\mathrm{phys}\sim750$~pc)} \\
Full tensor & With long range forces & $T_{ij} = T_{ij,\mathrm{Tree}}+T_{ij,\mathrm{PM}}$ \\
rhevo\_f1/2 & $f=1/2$ in eq.~\ref{eq:size_evo} & $(dr_\mathrm{h})_\mathrm{sh}=0 ; \; (d\rho_\mathrm{h})_\mathrm{sh}<0$ \\
rhevo\_f1 & $f=1$ in eq.~\ref{eq:size_evo} & $(dr_\mathrm{h})_\mathrm{sh}<0 ; \; (d\rho_\mathrm{h})_\mathrm{sh}>0$ \\
\hline

\end{tabular}

\raggedright
\textit{Notes.} The columns are (i) the model variation identifier, and
(ii) and (iii) the model variation description.
\end{table}
\FloatBarrier

\section{Tidal tensor}
\subsection{Tidal tensor calculation}
\label{app:tidal_tensor}
\begin{figure}
    \centering
    \resizebox{0.95\hsize}{!}{\includegraphics{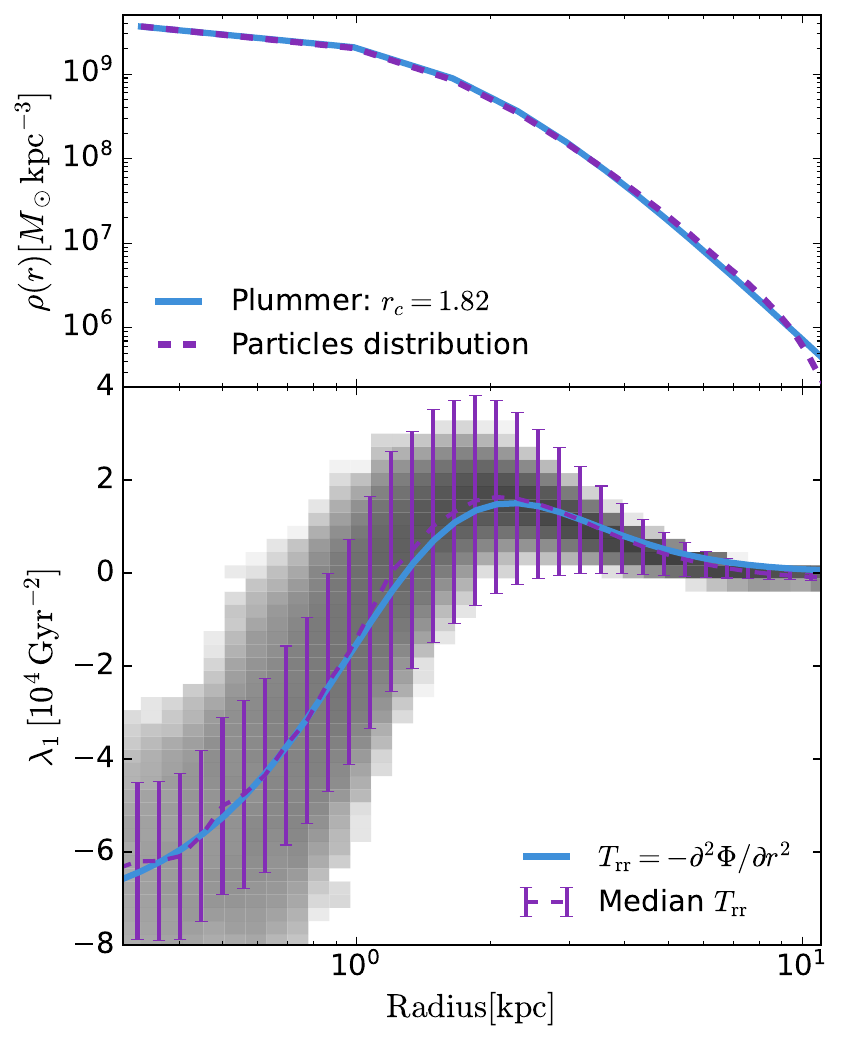}}
    \caption{Tidal tensor from a simulated Plummer sphere. \textit{Top}: Profile fit of the simulated sphere. \textit{Bottom}: Distribution of the highest tidal tensor eigenvalue, as calculated in the simulation compared to the analytical eigenvalue.}
    \label{fig:plummer_test}
\end{figure}

As the formation and evolution of the SCs in this model are heavily reliant on the tidal tensor, ensuring that its calculus is correct becomes fundamental. We reproduce the test included in both \citet{Pfeffer_2018} and \citet{Reina_Campos_2022} for obtaining the tidal tensor on a \citet{Plummer_1911} sphere in Fig. \ref{fig:plummer_test}. The same sphere of total mass $M=10^{11} \mathrm{M}_\odot$ and $N=10^6$ particles used by \citet{Reina_Campos_2022} was run. Their initial distribution is described by the Plummer potential,
\begin{equation}
\Phi_\mathrm{PL}(r)=-\frac{GM}{\sqrt{r^2 - r_c^2 }}.
\end{equation}
The ensemble of particles is run for $\sim 1$~Gyr and, after relaxation, evolves to a characteristic radius of $r_c \simeq 1.82$~kpc. In Fig. \ref{fig:plummer_test}, the distribution of the numerical first eigenvalue, as obtained from the particles distribution, is compared to the analytical, expected, eigenvalue which is
\begin{equation}
T_\mathrm{rr}=-\frac{\partial^2 \Phi_\mathrm{PL}}{\partial r^2} = -GM \frac{r_c^2 - 2r^2}{\left( r^2 + r_c^2\right)^{5/2}}.
\end{equation}
It is evident that the median eigenvalue follows the expected analytical solution within the standard deviation of the particles.

\subsection{Long range contributions}
\label{app:long_range_tensor}
\begin{figure}
    \centering
    \resizebox{0.95\hsize}{!}{\includegraphics{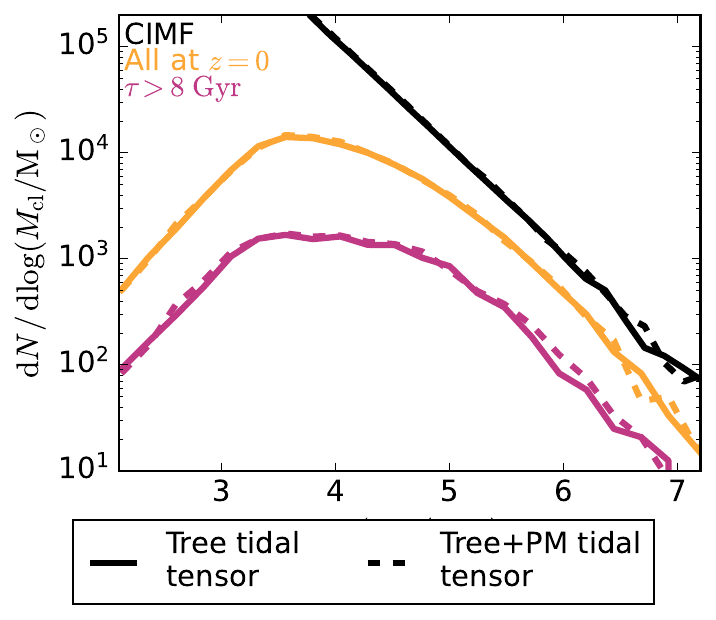}}
    \caption{Mass functions for the simulated SC populations when including long range contributions in the calculus of the tidal tensor.}
    \label{fig:fulltensortest}
\end{figure}

\begin{figure*}
    \sidecaption
    \includegraphics[width=12cm]{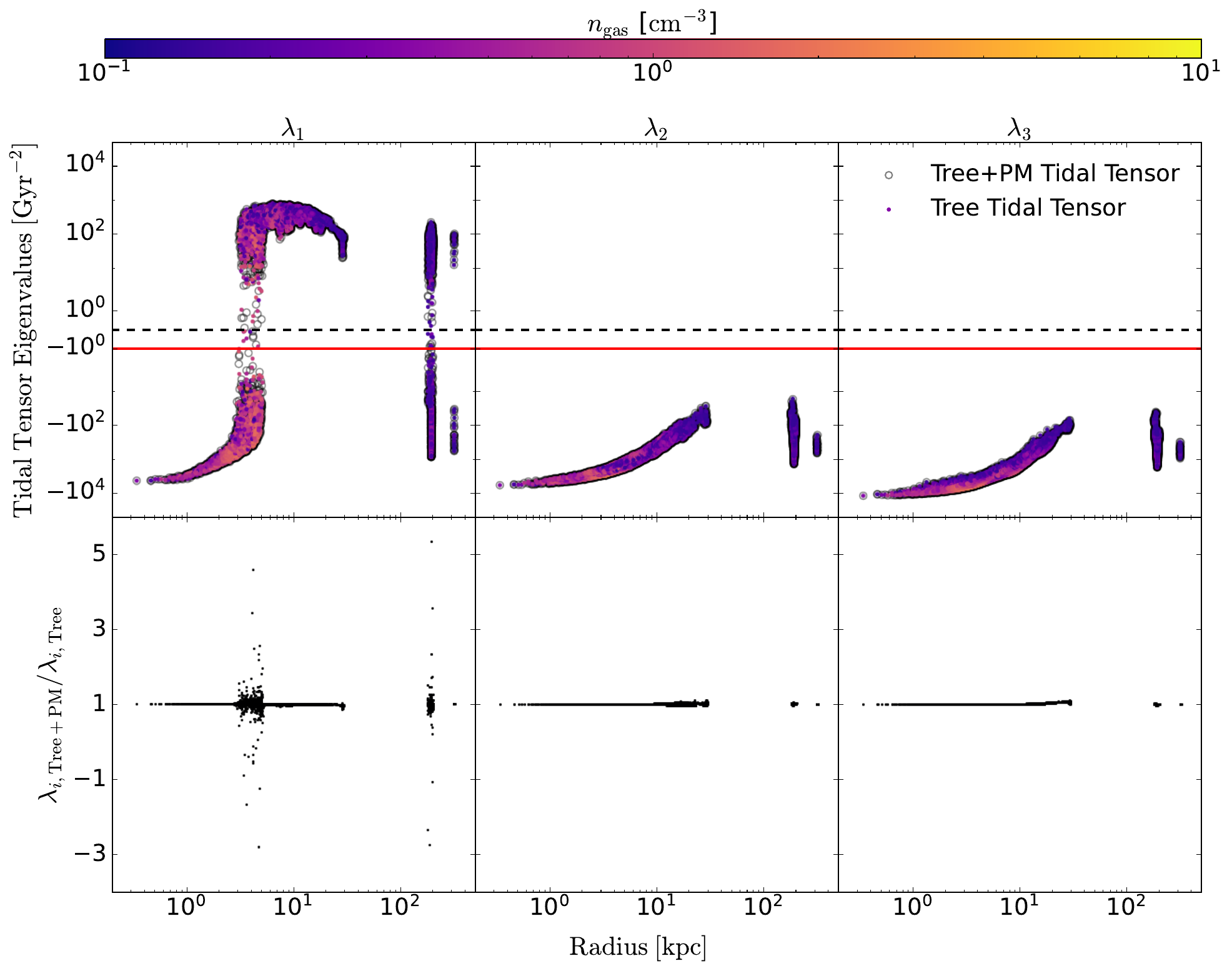}
    \caption{Star forming gas tidal tensor eigenvalues distribution at $z = 0$, for the halo 6. \textit{Top}: Distribution for the \texttt{fiducial} variation at lower resolution with/without including the tidal tensor from long-range gravity contributions. \textit{Bottom}: Ratio of the eigenvalues for the two different calculations of the tidal tensor.}
    \label{fig:fulltensoreig}
\end{figure*}

Additionally, in Sect. \ref{sec:simdescription} was mentioned that the long range force contributions to the tidal tensor are not taken into account as they are minimal and should not influence whether the medium is compressive or not. Furthermore, the tidal shocks mass loss evolution should not be sensitive to these contributions. Here we put this to test by running two simulations of the fiducial model of AuriGLOBES at resolution level~5 for the halo 6. The first run without including the long-range force contributions to the tidal tensor (\texttt{fiducial} model), and the second one (\texttt{full tensor}) including them by means of differentiating the Poisson equation in Fourier space (by "pulling down" $k's$), i.e.
\begin{equation*}
    \Phi(\mathbf{k}) \propto \frac{\delta(\mathbf{k})}{k^2} \Longrightarrow \widetilde{T}_{ij}(\mathbf{k}) \propto \frac{k_i k_j}{k^2}\delta(\mathbf{k}),
\end{equation*}
where $\delta(\mathbf{k})$ is the density contrast field.

The simulated GCMFs for both simulations are shown in Fig. \ref{fig:fulltensortest}. It can be seen that, down to the $\tau>8$~Gyr age cut, they are nearly identical with their differences relying on the differences of the CIMF. This mainly shows that the effect on the mass loss evolution is in fact minimal. At formation, however, there is a minimal effect on the amount of clusters being formed. This can be verified in Fig. \ref{fig:fulltensoreig}, by comparing the tidal tensor eigenvalues of star forming gas at redshift $z=0$ for both simulations. Although the distribution is nearly identical, close to the zero crossing, and to our numerical threshold to define a tidally compressive region, there are differences on some cells lying above or below this line. Therefore, SCs that were formed in some environments at one variation, may not be formed in the other. Nevertheless, this test shows that neglecting the long-range force contributions to the tidal tensor is justified given that including it increases the computational time by $\sim 70$~\% and the effect in the simulated populations is minimal.

\subsection{Estimating $\Omega$ and $\kappa$}
\label{app:omega_kappa}
\begin{figure}
    \centering
    \resizebox{0.95\hsize}{!}{\includegraphics{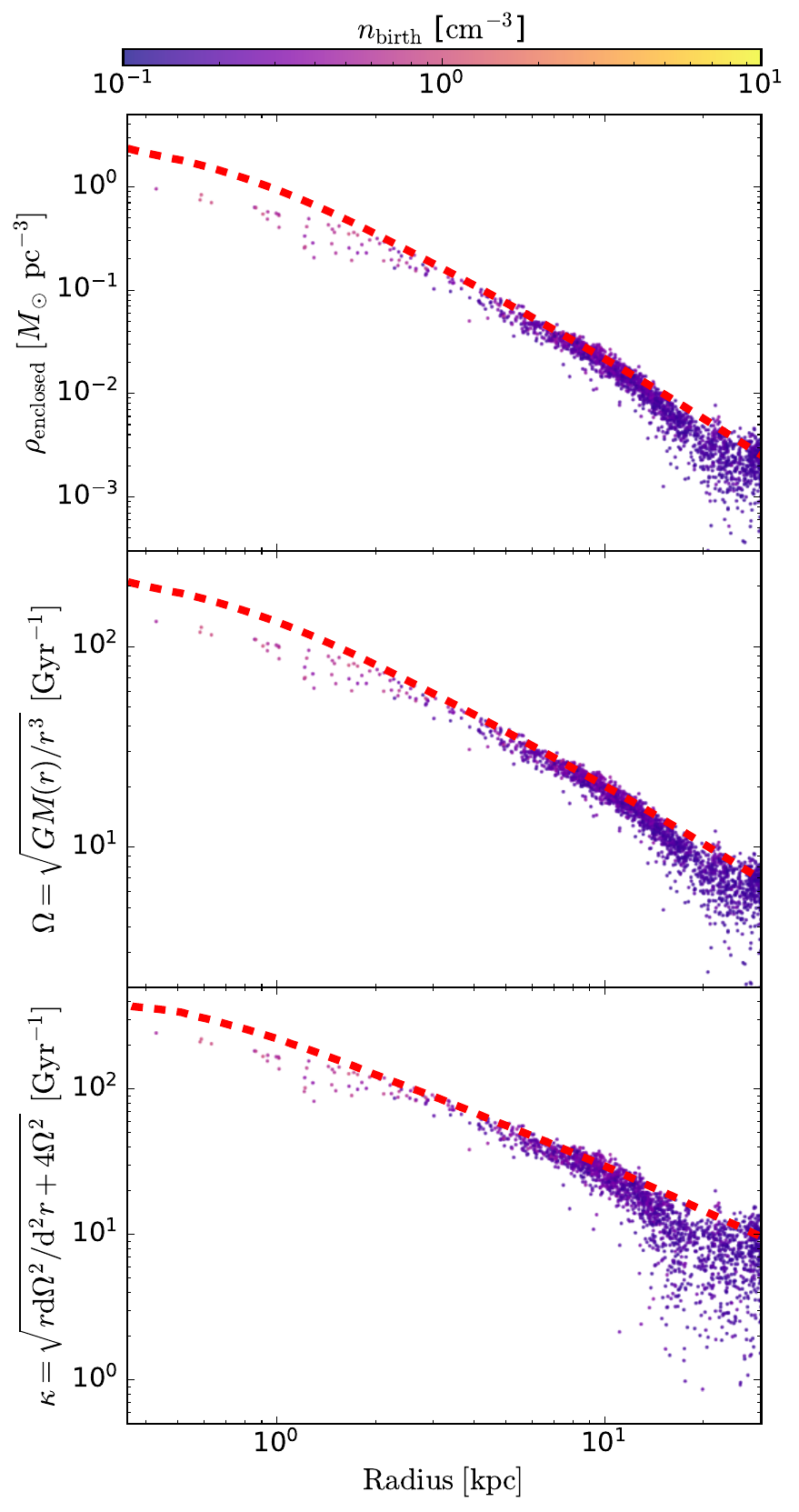}}
    \caption{Enclosed mean density (\textit{top}), circular frequency (\textit{middle}), and epicyclic frequency (\textit{bottom}) distributions, as estimated from the tidal tensor eigenvalues. The distribution corresponds to stars born during the last 50~Myr of galaxy evolution for the halo 6. The profile, as calculated at post-processing, of these quantities is included in red dashed line in each panel for comparison.}
    \label{fig:frequenciestest}
\end{figure}

The circular frequency, $\Omega$, and the epicyclic frequency, $\kappa$, are quantities estimated locally to calculate properties both at the formation (equations \ref{eq:toomre_mass} and \ref{eq:fb_time}), and evolution (equation \ref{eq:tidal_radius}), of the SCs. Here we demonstrate that the estimate of these quantities from the tidal tensor agrees well with the values calculated at post-processing. The circular and epicyclic frequencies can be derived from the tidal tensor eigenvalues via
\begin{equation}
    \Omega^2 = \frac{1}{3}\left|\sum_i \lambda_i\right|,
\end{equation}
and
\begin{equation}
    \kappa^2 = \left| 3\Omega^2 - \lambda_1 \right|.
\end{equation}
With $\lambda_1$ being the first (maximal) eigenvalue of the tidal tensor. These relations are derived from equating the circular frequency in terms of the mean density enclosed within the radius $r$, i.e. $\Omega^2= v_c^2/r^2= GM(r)/r^3 = 4/3 \pi G \bar{\rho}(r)$, to the Poisson's equation and its relation with the tidal tensor eigenvalues, $\nabla^2\Phi = 4\pi G \rho = -\sum_i\lambda_i$. For the full derivation, we refer to Appendix A of \citet{Pfeffer_2018} or to Appendix C in \citet{Reina_Campos_2022}.

We compare in Fig.~\ref{fig:frequenciestest} the distribution of the enclosed mean density, $\bar{\rho}$, the circular frequency, $\Omega$, and epicyclic frequency, $\kappa$, as estimated from the tidal tensor eigenvalues to the quantities calculated at post processing from the simulated halo 6 at $z=0$. The plotted distribution of the estimated values corresponds to stars formed in the last 50~Myr of the galaxy evolution. It is evident from the figure that the estimate is in an overall agreement with the profile of the simulated galaxy. The bigger differences are in the central $\sim 500$~pc of the galaxy, comparable to the softening length of the simulation, and at larger radii $>10$~kpc where the scatter becomes important. Nevertheless, the estimates are reasonable for the purposes of the model.

\FloatBarrier
\section{Tidal shock capture}
\label{app:shocks_capture}
\begin{figure}
    \centering
    \resizebox{0.95\hsize}{!}{\includegraphics{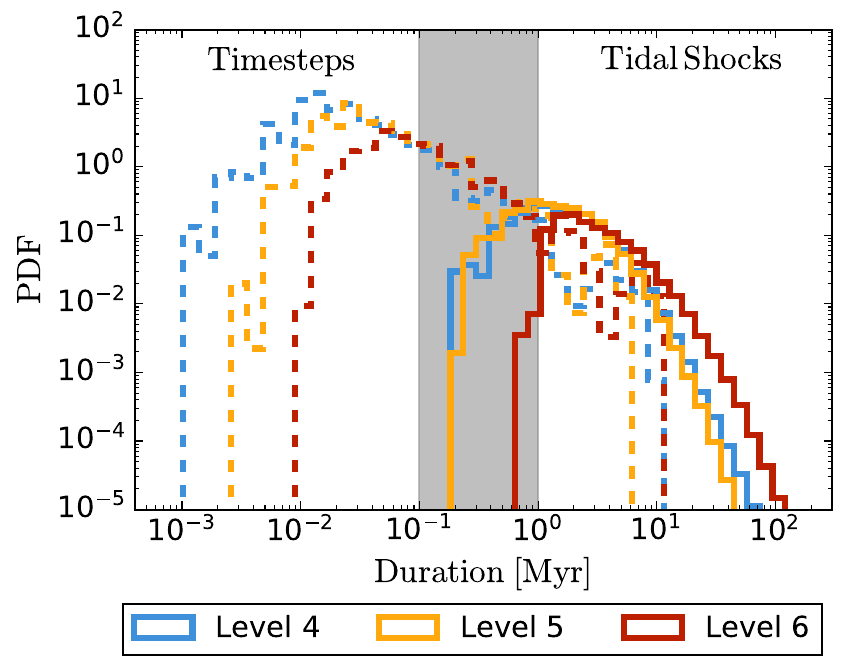}}
    \caption{Distribution of the duration of the full simulation timesteps and of the tidal shocks duration for a subsample of SCs. The grey region denotes the range where tidal shocks capture starts to be not resolved.}
    \label{fig:shocks_capture}
\end{figure}

Given that we do not explicitly limit the duration of the simulation timesteps, it is important to address how the model is able to resolve tidal shocks disruption. Particularly, tidal shocks shorter than the simulation timestep, i.e. $\tau_{ij}<\mathrm{d}t$, may be missed and could impact how well the tidal heating, $I_\mathrm{tid}$, is determined. To quantify how much this could affect our model, we show in Fig.~\ref{fig:shocks_capture} the distribution of the simulation timesteps and of the tidal shocks duration from a sample of 100 SCs per run for which the full dynamical evolution, at the timestep level, is recorded. We show this for the three different resolution levels explored in Sect.~\ref{resolution_test}. There is a sharp cut at the lowest recorded shock duration for each resolution level, and the higher the resolution the shorter this threshold duration. These thresholds correspond to timesteps in the 0.1--1~Myr range, in the long timesteps tail of the distribution. With this, we interpret that tidal heating may not be correctly captured in timesteps equal or larger than this threshold region and is one of the drivers of higher shocks disruption from increasing the resolution. Nevertheless, we note that the majority of the simulation timesteps are shorter than the shortest shock duration and, therefore, we interpret that the majority of the shocks are well resolved.

\FloatBarrier
\section{On the $r_h$ evolution model}
\label{app:rh_evolution}
\begin{figure}
    \centering
    \resizebox{0.95\hsize}{!}{\includegraphics{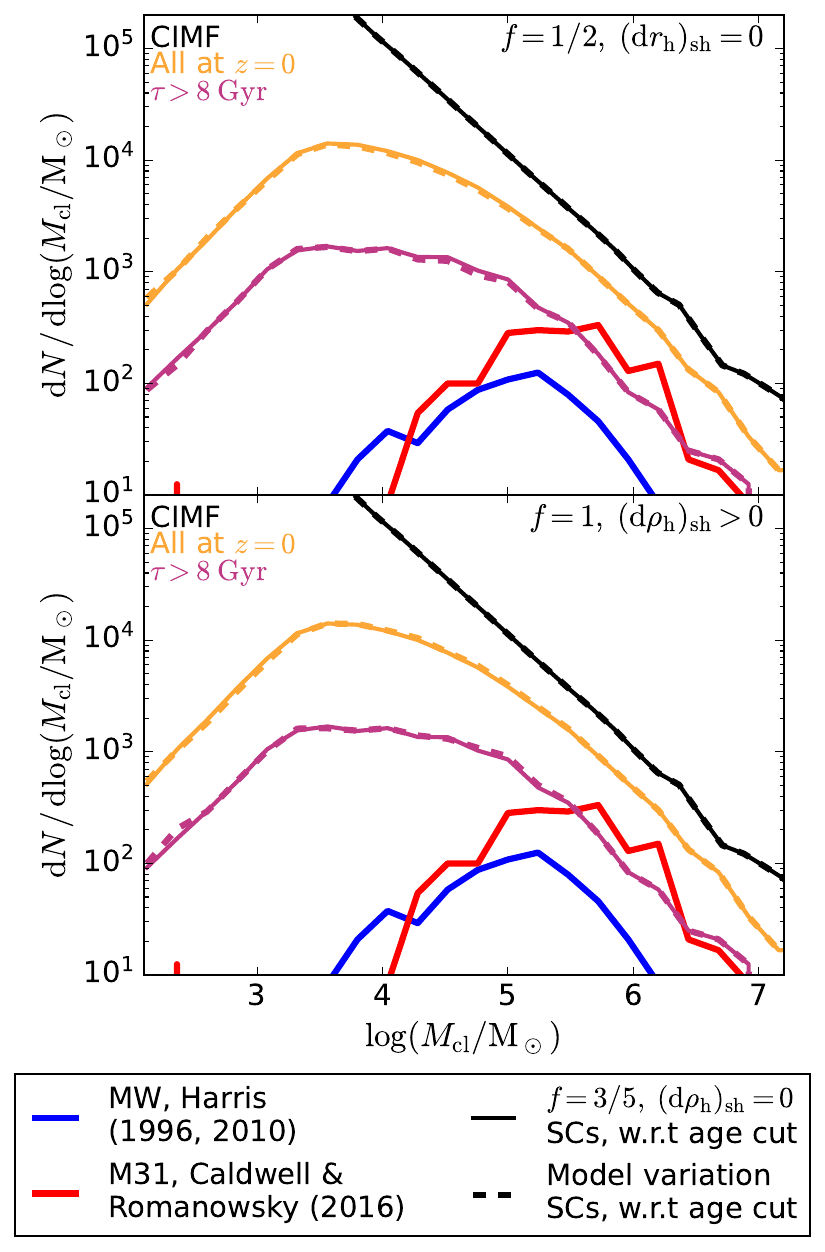}}
    \caption{Mass functions for the simulated SC populations for different choices of $f$ in the dynamical half-mass radius evolution.}
    \label{fig:fparamtest}
\end{figure}

The fiducial model allows the half-mass radius to change in correspondence to the mass loss that the SCs experience following equation \ref{eq:size_evo}. As mentioned in Sect.~\ref{sec2:globes}, $f$ describes the increase of energy of the remaining bound stars after a tidal shock mass loss event, and how it is related to the density at the half-mass radius, $\rho_\mathrm{h}$. The particular $f=3/5$ choice describes the change of $r_\mathrm{h}$ after tidal shocks mass loss such that $\rho_\mathrm{h}$ remains constant. Here we test the sensibility of the model to this particular choice by running two variations of the model with choices of $f=1$, and $f=1/2$. These choices correspond to $\rho_\mathrm{h}$ increasing after shock events, potentially shielding SCs from subsequent mass loss, and to no $r_\mathrm{h}$ change from shock events, experiencing only the overall expansion from stellar evolution and two-body relaxation, respectively. The MFs obtained from these variations, and how they compare to the fiducial choice, are shown in Fig.~\ref{fig:fparamtest}. The MFs are remarkably similar with very small differences with respect to the fiducial $f=3/5$ choice. The reason behind this behaviour could be two-fold: 
\begin{itemize}
    \item The strength of the tidal shocks is not sufficiently high to be able to distinguishing among these parameters.
    \item The change in the half-mass radius due to tidal shocks is subdominant with respect to the change from the enhanced two-body relaxation.
\end{itemize}
Nevertheless, we kept the $f=3/5$ choice as a safe spot on trying to keep most physics as can be possible with this model.

\FloatBarrier
\section{Shocks enhancement variations}
\label{app:shocks}
\begin{figure}
    \centering
    \resizebox{0.95\hsize}{!}{\includegraphics{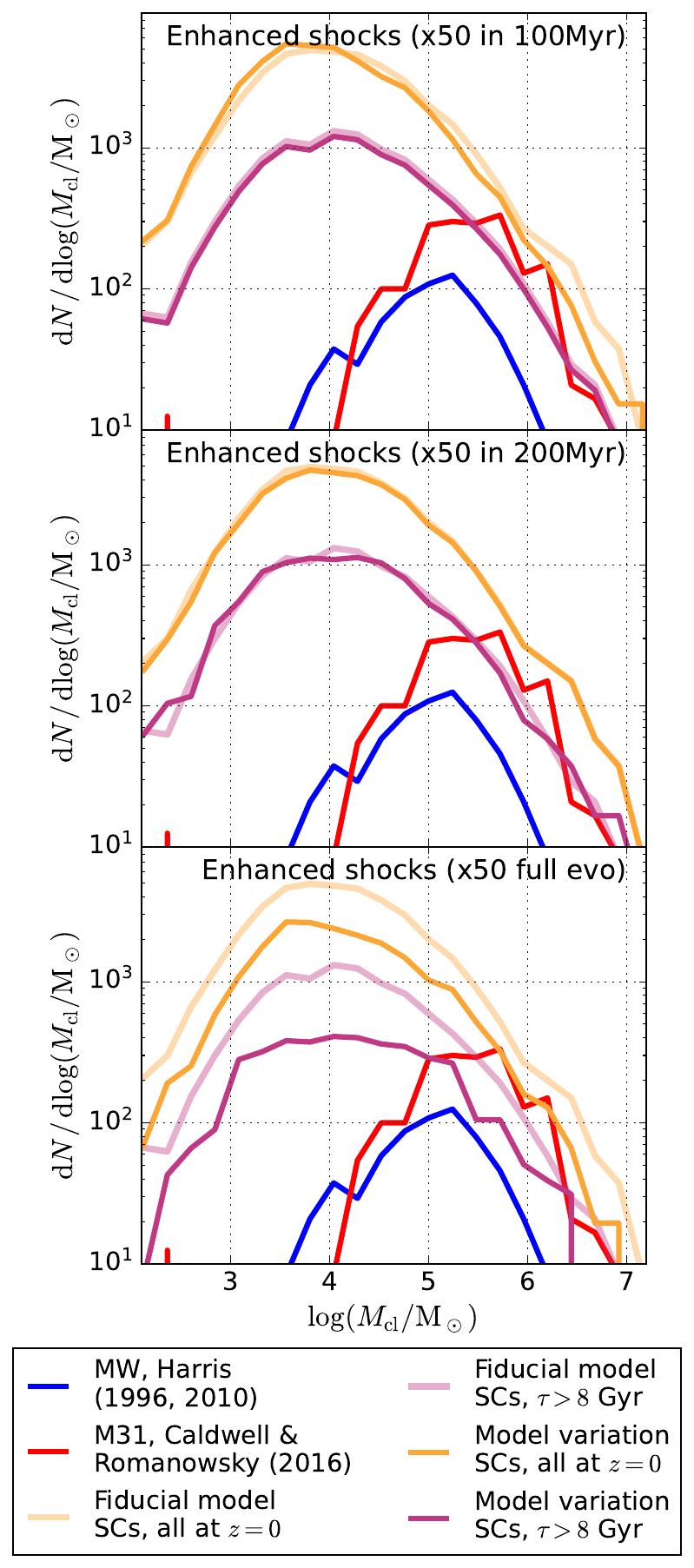}}
    \caption{Mass functions for the simulated SC populations for the additional shocks enhancement variations.}
    \label{fig:cce_test}
\end{figure}

We further explore the tidal shocks disruption in our model by performing additional variations enhancing the shock-driven mass loss with a higher factor of 50 over different time periods. The MFs produced in this variations are presented in Fig.~\ref{fig:cce_test}. The top and central panels explore the effect of this enhancement only during the first stages of the SCs evolution, with the aim to test how the model is capturing the early disruption from the natal environment \citep{Kruijssen_2012a}. The duration of the enhancement is based on the typical timescale of this rapid disruption phase quoted in \citet{Kruijssen_2015} and \citet{De_Lucia_2024}, i.e. $t_\mathrm{cce}=175$~Myr. The effect of this enhancement during these short periods is negligible. This suggests that the SCs that are disrupted in short timescales, i.e. $\tau<1$~Gyr, in the \texttt{fiducial} model just accelerate their disruption while the effect on the rest of the SCs is minimal given that their lifetime is likely large, i.e. $\tau>1$~Gyr.

Moving now to keeping this enhancement over a larger period, the MFs in the bottom panel reproduce the same behaviour discussed in Sect.~\ref{results:variation_mfs}. The normalisation is lowered but the peak of the MFs is mostly unchanged with an important population present at the low mass tail of the old ($\tau>8$~Gyr) population. We interpret here that the size evolution of the SCs acts as an agent suppressing the disruption. In particular, this stronger disruption, expected from e.g. the inclusion of cold phase of the ISM, shields the SCs through their change of half-mass radius. 

\FloatBarrier
\section{Individual halo SFR, CFR, and age distributions}
\label{app:age_distribution}
\begin{figure*}
    \centering
    \includegraphics[width=17cm]{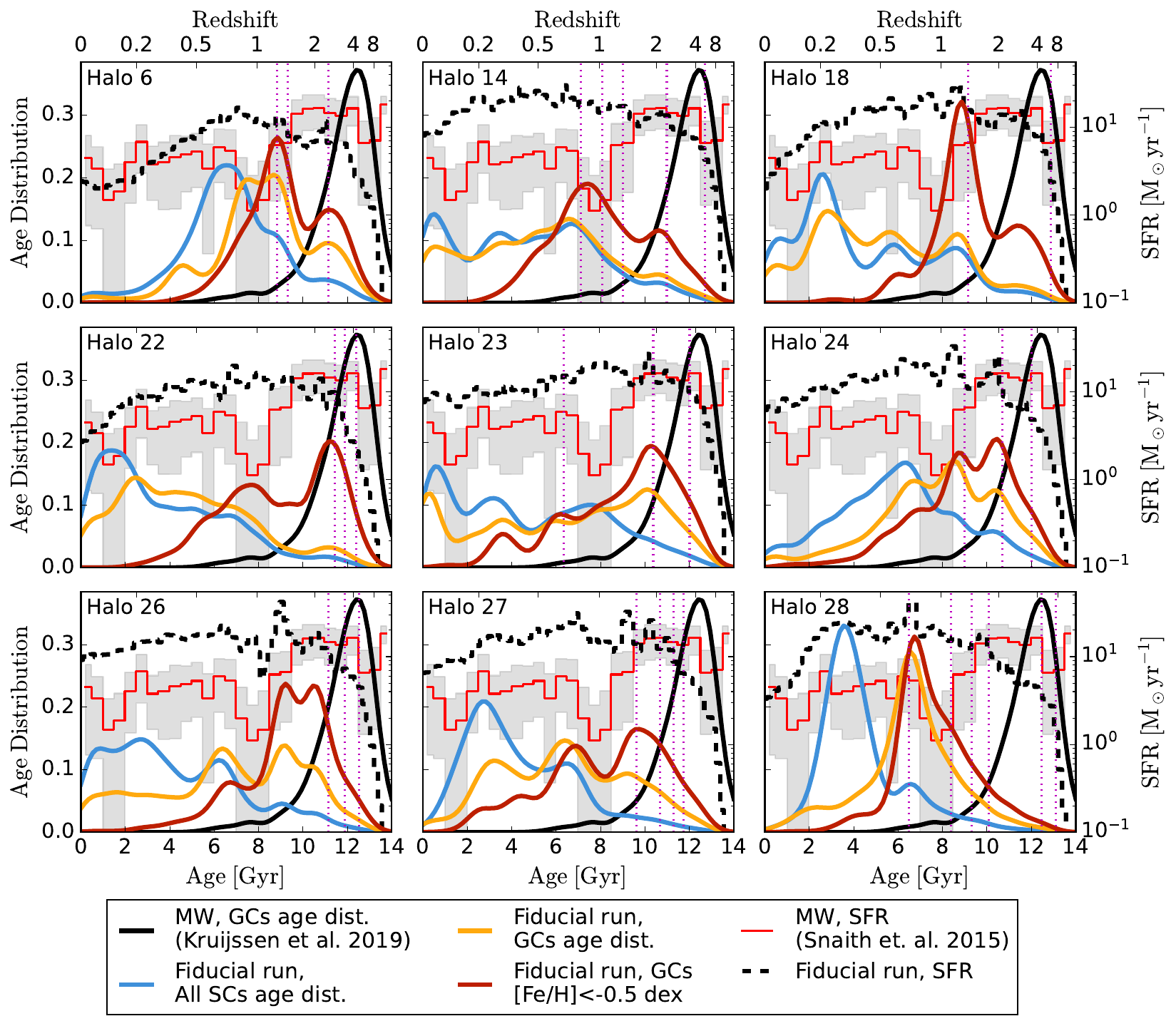}
    \caption{Age distribution of the simulated cluster populations for the individual halos. Different age distributions are shown for all SCs, GC selection, and GC selection with a metallicity cut. For comparison, the MW GC age distribution is included. The individual halo SFR is shown and compared to that of the MW from \citet{Snaith_2015}. The vertical dashed lines denote mergers with a baryon mass ratio of at least 1:12 at infall time.}
    \label{fig:halo_age_dist}
\end{figure*}

Finally, we present here a detailed comparison of the different SC age distributions of the simulated halos (bottom panel of Fig.~\ref{fig:cfr_age}) and their relation with their SFHs. We show in Fig.~\ref{fig:halo_age_dist} the individual halo SFH, and SC and GC age distributions in comparison to the SFH of the MW from \citet{Snaith_2015}, and its GC age distribution as compiled by \citet{Kruijssen_2019c}. We also include the age distribution of the simulated GCs applying the same metallicity cut as in Fig.~\ref{fig:tsfr_age} to target the old populations of each halo. The infall times of the main merger events for the halo are included as vertical dotted lines in each panel. These are identified as having a baryon mass ratio of at least 1:12 at the time of infall, i.e. when the merging system crosses $R_{200}$. Looking at all the individual SFRs, no simulation has the high early SFR such as seen in the case of the MW: $\sim 10 \textrm{--} 20 \,\mathrm{M}_\odot \, \mathrm{yr}^{-1}$ during the first $\sim 4$~Gyr of evolution and forming half of its stellar mass \citep{Snaith_2015}. Only after the first $\sim 2 \textrm{--} 4$~Gyr the simulated SFRs reach values comparable to the estimated SFR of the MW. Furthermore, past this time between $z\simeq 1\textrm{--}2$, the SFRs of the simulated halos are around an order of magnitude higher than that of the MW. For example, the halos with the oldest peaks in the GC age distribution, and closer to that of the MW, are halo 23 and 24. For these halos, the SFR quickly grows to match the MW SFR at $\approx 13$~Gyr lookback time and maintains this high SFR until peaking at $z\simeq2$. The GC age distribution of halo 23 reflects this behaviour and peaks at the same $z\simeq2$ redshift, which is $\approx 2$~Gyr later than the peak of the Galactic GCs age distribution. Halo 24, similarly shows this peak as well as another at $z\simeq1$ that correlates with its SFR. This behaviour heavily impacts any attempt to reproduce the observed GC age distribution when any of the halos is able to reproduce the MW SFH. In particular, this does not constitute a limitation of the formation prescription, but simply follows from the parent simulation/galaxy formation model.

When looking at GC formation at high redshift, there is a general correlation with merger induced starbursts. This is evident from the GC age distributions with the $[\mathrm{Fe/H}]<-0.5$ metallicity cut. Here, the distributions are consistent with older populations whose peaks follow early accretion events, and are generally closer to the observational one for the MW. In particular, most of the distributions after applying the metallicity cut are correlated with early mergers. This is evidenced with the age distributions peaks coinciding with the infall times of the mergers. We bring the attention to halo 22 here, as the halo that, after the metallicity cut, gives the closest age distribution with respect to the MW. We defer a dedicated analysis of this correlation with the accretion history to future work.

\end{appendix}

\end{document}